\documentclass[preprint,aps,amssymb,amsmath]{revtex4}
\usepackage{amssymb}
\usepackage[dvips]{graphicx}
\begin{document}


. \vspace{2cm}

\centerline { \textbf{\large STRONG-COUPLING THEORY}} \centerline{
\textbf{\large OF HIGH TEMPERATURE SUPERCONDUCTIVITY}}
 \centerline{A.S. Alexandrov}
\centerline { Loughborough University, Loughborough LE11 3TU, United
 Kingdom}

\section*{INTRODUCTION}
 The seminal work by Bardeen, Cooper and Schrieffer (BCS)\cite{bcs}
 extended further by Eliashberg \cite{eli} to the intermediate coupling regime solved one of the major scientific
problems of Condensed Matter Physics in the last century. While the
BCS theory provides a qualitatively correct description of some
novel superconductors like magnesium diboride and doped fullerenes,
if the phonon dressing of carriers (i.e. polaron formation) is
properly taken into account,  high-temperature superconductivity
(HTS) of cuprates represents a challenge to the conventional theory.
Here I discuss a multi-polaron approach to the problem based on our
extension of the BCS theory to the strong-coupling regime
\cite{alebook1}. Attractive electron correlations,  prerequisite to
any HTS, are caused by an almost unretarded electron-phonon (e-ph)
interaction sufficient to overcome the direct Coulomb repulsion in
this regime. Low energy physics is that of small polarons and
bipolarons (real-space electron (hole) pairs dressed by phonons).
They are itinerant quasiparticles existing in the Bloch states at
temperatures below the characteristic phonon frequency. Since there
is almost no retardation (i.e. no Tolmachev-Morel-Anderson
logarithm) reducing the Coulomb repulsion, e-ph interactions should
be relatively strong to overcome the direct Coulomb repulsion, so
carriers \emph{must} be polaronic to form pairs in novel
superconductors. I identify  the
 Fr\"{o}hlich
electron-phonon interaction  as the most essential for pairing in
superconducting cuprates. Many experimental observations have been
satisfactorily understood in the framework of the bipolaron theory
\cite{alebook1}
 providing evidence for a
novel state of electronic matter in layered cuprates, which is a
charged Bose-liquid of small mobile bipolarons.

Here  the band structure and essential interactions in oxide
superconductors are discussed in section 1, and  the
"Fr\"ohlich-Coulomb" model of HTS is introduced in section 2,
including discussions of single-polaron (2.1, 2.2, 2.3, 2.4, 2.5)
and multipolaron (2.4, 2.6, 3.1, 3.2) problems, low-energy
structures (3.3),
 and the phase diagram of cuprates (3.4).
"Individual" versus Cooper pairing (3.5), normal state properties
(section 4), in particular in-plane resistivity, the Hall effect,
magnetic susceptibility and the Lorenz number (4.1), the  Nernst
effect (4.2), diamagnetism (4.3), spin and charge pseudogaps, and
c-axis transport (4.4) are also discussed. I present a
 parameter-free evaluation of
$T_c$ (5.1), and an explanation of  isotope effects (5.2), specific
heat anomaly (5.3), upper critical fields (5.4), symmetries and
space modulations of the order parameter (5.5), and a model of
overdoped cuprates
 as mixtures  of  mobile bipolarons and degenerate lattice
polarons (section 6).


\section{ Band structure and essential
interactions in cuprates} A significant fraction of theoretical
research in the field of HTS has suggested that the interaction in
novel superconductors is essentially repulsive and unretarded, and
it could provide high $T_{c}$ without  phonons. Indeed strong
on-site repulsive correlations (Hubbard $U$) are essential in
shaping the insulating state of  undoped (parent) compounds.
Different from conventional band-structure insulators with
completely filled and empty Bloch bands, the Mott insulator arises
from a potentially metallic half-filled band as a result of the
Coulomb blockade of electron tunnelling to neighboring sites
\cite{mott}.

In our approach to cuprate superconductors  we take the view that
cuprates and related transition metal oxides are charge-transfer
Mott-Hubbard insulators at $any$ relevant level of doping
\cite{alebook1}. The one-particle density-of-states (DOS) of
cuprates is schematically represented by Fig.1, as it has been
established in a number of site-selective experiments \cite{sel} and
in the first-principle numerical ("LDA+U") \cite{first} and
semi-analytical cluster \cite{ovc} band structure calculations
properly taking into account the strong on-site repulsion. Here
d-band of the transition metal (Cu) is split into the lower and
upper Hubbard bands by the on-site repulsive interaction $U$, while
the first band to be doped is an oxygen band within the Hubbard gap.
The oxygen band is less correlated and completely filled in parent
insulators, so a single oxygen hole has well defined quasi-particle
properties in the absence of interactions with phonons and with spin
fluctuations of d-band electrons.

Unfortunately,  the Hubbard $U$ model shares an inherent difficulty
in determining the order when the Mott-Hubbard insulator is doped.
While some groups have claimed that it describes high-$T_{c}$
superconductivity at finite doping, other authors could not find any
superconducting instability. Therefore it has been concluded that
models of this kind are highly conflicting and confuse the issue by
exaggerating the magnetism rather than clarifying it \cite{lau}. The
Hubbard-$U$ model of HTS and its strong-coupling $"t-J"$
approximation \cite{pla}   are  also refutable  on experimental
ground. A characteristic magnetic interaction, which is allegedly
responsible for  pairing in the model, is the spin-exchange
interaction, $J=4t^{2}/U$, of the order of $0.1$ eV (here $t$ is the
hopping integral). On the other hand, a simple parameter-free
estimate of the Fr\"ohlich electron-phonon interaction (routinely
neglected within the Hubbard $U$ approach) yields the effective
attraction as high as $1$ eV \cite{alebook1}. This estimate  is
obtained using the familiar expression for the polaron level shift,
$E_{p},$  the high-frequency, $\epsilon _{\infty }$, and the static,
$\epsilon _{0},$ dielectric constants of the host insulator,
measured experimentally \cite{alebra1},
\begin{equation}
E_{p}={\frac{1}{{2\kappa }}}\int_{BZ}{\frac{d^{3}q}{{(2\pi )^{3}}}}{\frac{%
4\pi e^{2}}{{q^{2}}}},
\end{equation}
where $\kappa ^{-1}=\epsilon _{\infty }^{-1}-\epsilon _{0}^{-1}$ and
the size of the integration region is the Brillouin zone (BZ).
\begin{figure}[tbp]
\begin{center}
\includegraphics[angle=-90,width=0.50\textwidth]{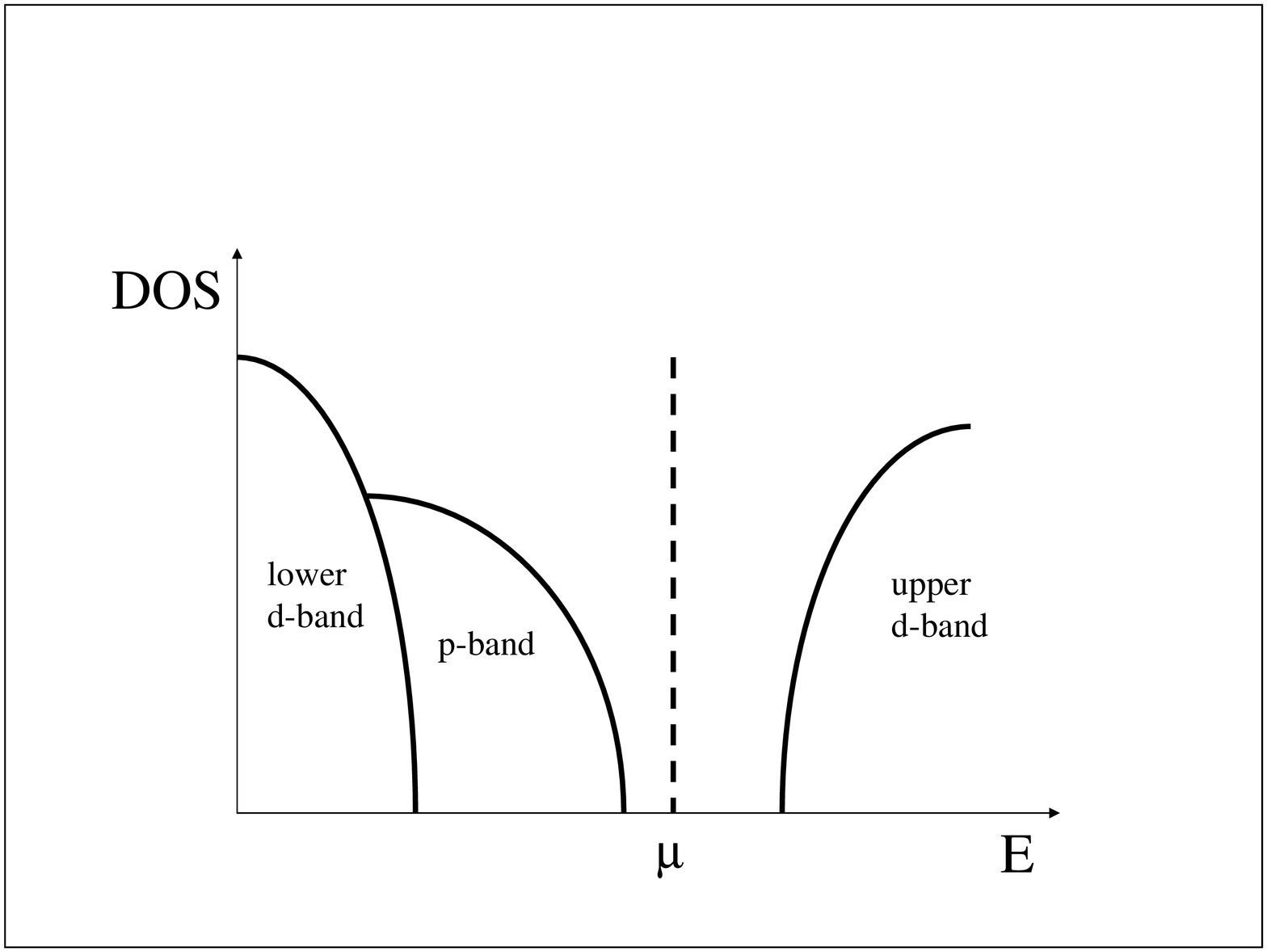}
\end{center}
\caption{ DOS in cuprates. The chemical potential $\mu$ is inside
the charge transfer gap as observed in the tunnelling experiments
\cite{boz0} because of  bipolaron formation \cite{alebook1} . It
could enter the oxygen band in overdoped cuprates, if bipolarons
 coexist with unpaired  degenerate polarons (section 6).}
\end{figure}
Since $\epsilon _{\infty }=5$ and $\epsilon_0=30$ in La$_2$CuO$_4$
one obtains $E_p=0.65$ eV. Hence the attraction, which is about
$2E_p$, induced by the long-range lattice deformation in parent
cuprates
 is one order of magnitude larger than the exchange
magnetic interaction. There is virtually no screening of
 e-ph interactions with $c-$axis polarized optical phonons in
doped cuprates because the upper limit for an out-of-plane plasmon
frequency ($< 200$ cm$^{-1}$)\cite{mar1} is well below
characteristic phonon frequencies, $\omega\approx$ 400 - 1000 cm
$^{-1}$ . Hence the Fr\"ohlich interaction remains the most
essential pairing interaction at any doping.

Further compelling evidence for the strong e-ph interaction has come
from isotope effects  \cite{zhao}, more recent high resolution angle
resolved photoemission spectroscopies (ARPES)
 \cite{LAN}, and a number of earlier optical \cite{mic1,ita,tal,tim} and
 neutron-scattering \cite{ega} studies of cuprates.   The strong
coupling with optical phonons, unambiguously established in all
high-temperature superconductors,  transforms holes into lattice
mobile polarons and mobile superconducting bipolarons as has been
proposed \cite{alerun}  prior the discovery \cite{mul,chu}.

When  the e-ph interaction  binds holes into intersite oxygen
bipolarons \cite{alebook1},  the chemical potential remains pinned
inside the charge transfer gap. It is found
 at a half of  the bipolaron binding energy, Fig.1, above the oxygen band edge
 shifted by the polaron level shift $E_{p}$, as clearly observed
in the tunnelling experiments by Bozovic et al. in optimally doped
La$_{1.85}$Sr$_{0.15}$ Cu O$_4$ \cite{boz0}. The bipolaron binding
energy as well as the singlet-triplet bipolaron exchange energy
(section 3) are thought to be the origin of normal state charge and
spin pseudogaps, respectively, as  has been proposed by us
\cite{alegap} and later found  experimentally \cite{kabmic}. In
overdoped samples carriers screen part of the e-ph interaction with
low frequency phonons. Hence, the bipolaron binding energy decreases
\cite{alekabmot} and the hole bandwidth increases with doping. As a
result, the chemical potential could enter the oxygen band in
overdoped samples  because of an overlap of the bipolaron and
polaron bands, so a Fermi-level crossing could be seen in ARPES
(section 6).

\section{"Fr\"ohlich-Coulomb" model of HTS}

Experimental facts tell us that any realistic
 description of
 high temperature superconductivity
 should treat the long-range Coulomb and
\emph{unscreened} e-ph interactions on an equal footing. In the past
decade we have developed a "Fr\"ohlich-Coulomb" model (FCM)
\cite{ale5,alebook1,alekor} to deal with the strong long-range
Coulomb and the strong long-range
 e-ph interactions in cuprates and other related compounds.
The model Hamiltonian explicitly includes a long-range
electron-phonon and the Coulomb interactions as well as the kinetic
and deformation energies.  The implicitly present large Hubbard $U$
term prohibits double occupancy and removes the need to distinguish
fermionic spins since the exchange interaction is negligible
compared with the direct Coulomb and the electron-phonon
interactions.

 Introducing spinless fermionic, $c_{\bf n}$, and
phononic, $d_{{\bf m}\alpha }$, operators the Hamiltonian of the
model  is written as
\begin{eqnarray}
H = & - & \sum_{\bf n \neq n'} \left[ T({\bf n-n'}) c_{\bf
n}^{\dagger } c_{\bf n'} - V_{c}({\bf n-n'}) c_{\bf n}^{\dagger}
c_{\bf n}c_{\bf n'}^{\dagger } c_{\bf n'} \right]  \nonumber \\
& - &  \sum_{\alpha,\bf n m} \omega_{\alpha} g_{\alpha}({\bf m-n})
({\bf e}_{\alpha } \cdot {\bf u}_{\bf m-n}) c_{\bf n}^{\dagger }
c_{\bf n}
(d_{{\bf m}\alpha}^{\dagger}+d_{{\bf m}\alpha }) \nonumber \\
& + &
 \sum_{{\bf m}\alpha} \omega_{\alpha}\left( d_{{\bf m}\alpha
}^{\dagger} d_{{\bf m}\alpha }+1/2 \right),
\end{eqnarray}
where $T({\bf n})$ is the hopping integral in a rigid lattice, ${\bf
e}_{ \alpha}$ is the polarization vector of the $\alpha$th vibration
coordinate, ${\bf u}_{\bf m-n} \equiv ({\bf m-n})/|{\bf m-n}|$ is
the unit vector in the direction from electron ${\bf n}$ to  ion
${\bf m}$, $g_{\alpha}({\bf m-n)}$ is the dimensionless e-ph
coupling function, and $V_{c}({\bf n-n'})$ is the inter-site Coulomb
repulsion. $g_{\alpha}({\bf m-n)}$ is proportional to the {\em
force} acting between the electron on site ${\bf n}$ and the ion on
${\bf m}$. For simplicity, we assume that all the phonon modes are
non-dispersive with the frequency $\omega_{\alpha}$. We also use
$\hbar =k_B=c=1$.

If the electron-phonon interaction is strong, i.e. the conventional
e-ph coupling constant of the BCS theory is large, $\lambda >1$,
then the weak-coupling BCS \cite{bcs} and the intermediate-coupling
Migdal-Eliashberg \cite{mig,eli} approaches cannot be applied
\cite{alebreak}. Nevertheless the Hamiltonian, Eq.(2), can be solved
analytically by using the $"1/\lambda"$ multi-polaron expansion
technique \cite{alebook1}, if $\lambda =E_p/zT(a) >1$. Here the
polaron level shift is
\begin{equation}
 E_{p} = \sum_{{\bf n}
\alpha} \omega_{\alpha} g_{\alpha}^{2}({\bf n}) ({\bf
e}_{\alpha}\cdot {\bf u}_{\bf n})^{2} ,
\end{equation}
and $zT(a)$ is about the half-bandwidth in a rigid lattice. As I
discuss below, the model shows a rich phase diagram depending on the
ratio of the inter-site Coulomb repulsion $V_{c}$ and the polaron
level shift $E_{p}$ \cite{alekor}. The ground state of FCM is a
\emph{polaronic} Fermi liquid when the Coulomb repulsion is large, a
\emph{bipolaronic} high-temperature superconductor at intermediate
Coulomb repulsions, and a charge-segregated insulator if the
repulsion is weak. FCM predicts \emph{superlight } polarons and
bipolarons in cuprates with a remarkably  high superconducting
critical temperature. Cuprate bipolarons are relatively light
because they are \emph{inter-site} rather than  \emph{on-site} pairs
due to the strong on-site repulsion,  and because mainly $c$-axis
polarized optical phonons are responsible for the in-plane mass
renormalization. The relatively small mass renormalization of
polaronic and bipolaronic carries in FCM has been confirmed
numerically using the exact QMC \cite{Korn2}, cluster
diagonalization \cite{feh3} and variational \cite{bon2} simulations.

(Bi)polarons describe many properties of
 cuprates \cite{alebook1}, in particular  normal-state transport (section 4),
 including  in-plane and out-of-plane resistivity, the Hall effect, spin susceptibility,
 thermal conductivity,
 normal state pseudogaps, the
  Nernst effect,   normal state diamagnetism, superconducting
  transition, including  high values of $T_c$, isotope effects,  unusual
  upper critical fields,   symmetries and real-space modulations
  of the superconducting order parameter (section 5).

\subsection {Single lattice polaron}
Let us first discuss a single lattice-polaron problem. Conducting
electrons in inorganic and organic matter interact with vibrating
ions. If phonon frequencies are sufficiently low, the local
deformation of ions, caused by electron itself,  creates a potential
well, which traps the electron even in a perfect crystal lattice.
This {\it self-trapping }phenomenon was predicted by Landau
\cite{land}. It was studied in greater detail by Pekar \cite{pek},
Fr\"{o}hlich \cite{fro}, Feynman \cite{fey}, Rashba \cite{ras},
Devreese\cite{dev} and other authors in the effective mass
approximation for the electron placed in a continuous polarizable
medium, which leads to a so-called {\it large} or {\it continuous}
polaron. Large polaron wave functions and corresponding lattice
distortions spread over many lattice sites.  The trapping is never
complete in the perfect lattice. Due to  finite phonon frequencies
ion polarizations
 follow polaron motion if the motion is  sufficiently slow. Hence,
large polarons with a low kinetic energy propagate through the
lattice as  free electrons but with an enhanced effective mass.

When  the electron-phonon (e-ph) interaction energy $E_{p}$ is
compared with  the electron energy-bandwidth, all electrons in the
Bloch bands of the crystal are ``dressed'' by phonons.  In this
strong-coupling regime, $\lambda=E_p/D
>1,$ the finite bandwidth $2D$ becomes important, so the continuous approximation
cannot be applied.  The main features of  small polarons were
understood by Tjablikov \cite{tja}, Yamashita and Kurosava
\cite{yam} , Sewell \cite{sew}, Holstein \cite{hol}  and his school,
Lang and Firsov \cite{lan}, Eagles \cite{eag} , and by other
researches and described in several review papers and textbooks
\cite{small}.
 The polaron shift of the atomic level and an exponential reduction of the
bandwidth (see below) at large values of $\lambda $  are among those
features. The shift can be easily understood using a toy model of an
electron localized on site \textbf{n} and interacting with a single
ion vibrating near site \textbf{m} in the direction connecting
\textbf{n }and \textbf{m}, Fig.2. The vibration part of the
Hamiltonian in this toy  model is
\begin{equation}
H_{ph}= -{1\over {2M}}{\partial^{2}\over{\partial x^{2}}}+
 {kx^{2}\over{2}},
\end{equation}
where  $M$ is the ion mass, $k=M\omega^2$ is the spring constant,
and $x$ is the ion displacement. The electron potential energy  due
to its Coulomb interaction with the ion is approximately
\begin{equation}
V= V_0 (1-x/a),
\end{equation}
where $V_0= -Ze^2/a$ is the Coulomb energy in a rigid lattice (an
analog of the crystal field potential), and $a$ is the average
distance between sites. Hence the Hamiltonian of the model is given
by
\begin{equation}
H= E_a \hat{n}+fx\hat{n}-{1\over {2M}}{\partial^{2}\over{\partial
x^{2}}}+
 {kx^{2}\over{2}},
\end{equation}
where $E_a$ is the atomic level at site ${\bf m}$ in the rigid
lattice, which includes the crystal field,  $f=Ze^2/a^2$ is the
Coulomb force, and $\hat{n}=c^{\dagger}c$ is the occupation number
operator on site ${\bf n}$ expressed in terms of the electron
annihilation $c$ and creation $c^{\dagger}$ operators. This
Hamiltonian can be readily diagonals  using a displacement
transformation of the vibration coordinate $x$,
\begin{equation}
x=y-\hat{n}f/k.
\end{equation}
The transformed Hamiltonian has no electron-phonon coupling,
\begin{equation}
\tilde{H}=(E_a-E_p) \hat{n}-{1\over {2M}}{\partial^{2}\over{\partial
y^{2}}}+
 {ky^{2}\over{2}},
 \end{equation}
where we used $\hat{n}^2=\hat{n}$ because of the Fermi statistics.
It describes a small polaron   at the atomic level,  shifted by the
\emph{polaron level shift} $E_p=f^2/2k$, and entirely decoupled from
ion vibrations. The ion vibrates near a new equilibrium, shifted by
$f/k$, with the "old" frequency $\omega$. As a result of the local
ion deformation, the total energy of the whole system decreases by
$E_p$ since a decrease of the electron energy by $-2E_p$ overruns an
increase of the deformation energy $E_p$.
\begin{figure}[tbp]
\begin{center}
\includegraphics[angle=-90,width=0.80\textwidth]{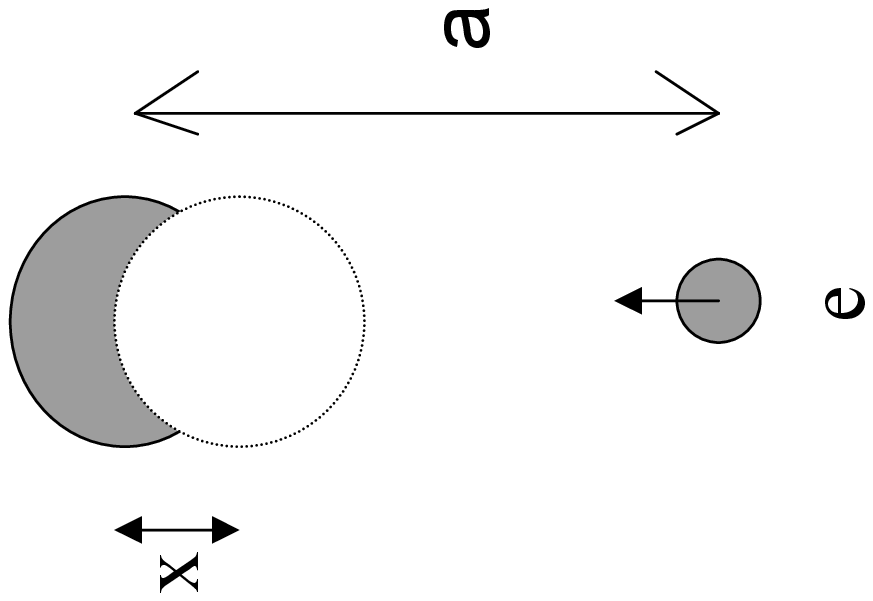}
\vskip -0.5mm
\end{center}
\caption{Localized electron shifts the equilibrium position of an
ion and lowers the atomic energy level.}
\end{figure}

The tunnelling of  small polarons in the lattice can be understood
within a simple Holstein model  \cite{hol} consisting of  two
molecules and a single electron. Here I slightly simplify the
original
 Holstein model replacing two molecules  by two rigid sites $1$ ("left") and $2$
("right") with the hopping amplitude $t$ between them. The electron
interacts  with a vibrational mode of an ion, placed at some
distance in between, Fig.3, rather than with the intra-molecular
vibrations:
\begin{equation}
H=  t(c_1^{\dagger}c_2+c_2^{\dagger}c_1)+H_{ph}+H_{e-ph},
\end{equation}
where $H_{e-ph}$ depends on the polarization of vibrations, and
$E_a=0$ is taken.  If the ion vibrates along the perpendicular
direction to the hopping (in "c"-direction) we have
\begin{equation}
 H_{e-ph}=f_cx(c_1^{\dagger}c_1+c_2^{\dagger}c_2),
 \end{equation}
and
\begin{equation}
 H_{e-ph}=f_ax(c_1^{\dagger}c_1-c_2^{\dagger}c_2),
 \end{equation}
 if the ion vibrates along the hopping ("a" direction).

 The wave-function of the electron and the ion is a linear superposition
of two terms describing the electron on the "left" and on the
"right" site, respectively,
\begin{equation}
\psi=[u(x)c_1^{\dagger} +v(x) c_2^{\dagger}]|0\rangle,
\end{equation}
where $|0\rangle$ is the vacuum state describing a rigid lattice
without the extra electron. Substituting $\psi$ into the
Schr\"odinger equation, $H\psi=E\psi$, we obtain two coupled
equations for the amplitudes,
\begin{equation}
(E-f_{a,c}x-H_{ph})u(x)=tv(x),
\end{equation}
\begin{equation}
(E\pm f_{a,c}x-H_{ph})v(x)=tu(x).
\end{equation}
\begin{figure}[tbp]
\begin{center}
\includegraphics[angle=-90,width=0.70\textwidth]{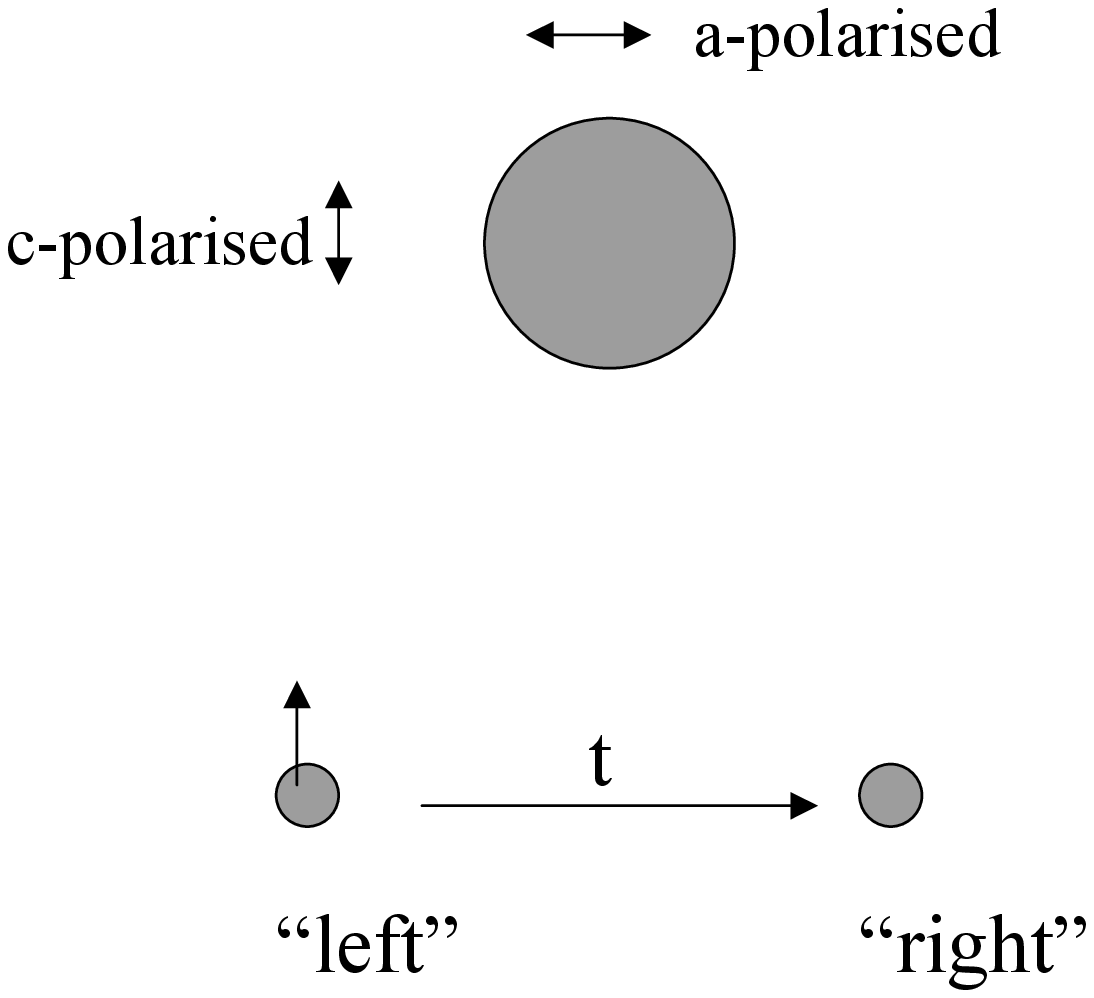}
\vskip -0.5mm
\end{center}
\caption{ Electron tunnels between sites $1$ ("left") and $2$
("right") with the amplitude $t$ and interacts with   c-axis or
a-axis polarized vibrational modes of the ion, placed in between.}
\end{figure}

 There is the exact solution for  the "c"-axis polarization,
 when a change in the ion position leads to the same shift of the electron energy on the left and on the right sites,
 \begin{equation}
 u(x)=u \chi_n(x), \nonumber
 \end{equation}
 \begin{equation}
 v(x)=v\chi_n(x),
  \end{equation}
  where $u$ and $v$ are constants and
  \begin{equation}
\chi _{n}(x)=\left( \frac{M\omega }{\pi (2^{n}n!)^2 }\right) ^{1/4}
H_{n}[(x-f_c/k)(M \omega )^{1/2}] \exp[-M(x-f_c/k)^{2}\omega /2],
\end{equation}
is the harmonic oscillator wave-function.  There are two ladders of
levels given by
\begin{equation}
E^{\pm}_{n}=-E_p\pm t +\omega(n+1/2)
\end{equation}
with $E_p=f_c^2/2k$. Here
\begin{equation}
H_{n}(\xi )=(-1)^{n}e^{\xi ^{2}}\frac{d^{n}e^{-\xi ^{2}}}{d\xi ^{n}}
\end{equation}
are the Hermite polynomials, and $n=0,1,2,3,...$ . Hence the c-axis
single-ion deformation  leads to the polaron level shift but without
any renormalization of the hopping integral $t$. In contrast,
$a$-polarized vibrations with the opposite shift of the electron
energy on the left and on the right sites,  strongly renormalize the
hopping integral. There is no simple general  solution of the
Holstein model in this case, but one can find it in two limiting
cases, $nonadiabatic$, when $t\ll \omega$ and $adiabatic$, when
$t\gg \omega$.

\subsection{Non-adiabatic small polaron}
 In the  non-adiabatic regime the ion vibrations are  fast and the
 electron hopping is slow. Hence one can apply a perturbation theory in powers of $t$ to solve
 \begin{equation}
\left(
  \begin{array}{cc}
   E-f_ax-H_{ph}& -t \\
   -t &  E+f_ax-H_{ph} \\
  \end{array}
\right)\left(
         \begin{array}{c}
           u(x) \\
           v(x) \\
         \end{array}
       \right)=0.
\end{equation}
 We take $t=0$ in zero order, and obtain
 a two-fold \emph{degenerate} ground state $[u^{l,r}(x),v^{l,r}(x)]$,
 corresponding to the polaron  localized on the left ($l$) or on the right ($r$) sites,
 \begin{eqnarray}
 u^{l}(x)=exp\left[-{M\omega\over{2}}(x+f_a/k)^{2} \right],\cr
  v^{l}(x)=0
  \end{eqnarray}
  and
  \begin{eqnarray}
   u^{r}(x)=0, \cr
    v^{r}(x)=exp\left[-{M\omega\over{2}}(x-f_a/k)^{2} \right]
  \end{eqnarray}
with the energy $E_0=-E_p+\omega/2$, where $E_p=f_a^2/2k$. The
eigenstates  are found as  linear superpositions of two unperturbed
states,
 \begin{equation}
\left(
  \begin{array}{c}
    u(x) \\
    v(x) \\
  \end{array}
\right)=\alpha\left(
                \begin{array}{c}
                 u^l(x) \\
                  0 \\
                \end{array}
              \right)+\beta\left(
                             \begin{array}{c}
                               0 \\
                               v^{r}(x)\\
                             \end{array}
                           \right).
\end{equation}
Here the  coefficients $\alpha$ and $\beta$ are  independent of $x$.
The conventional secular equation for $E$ is obtained,  multiplying
the first row by  $u^{l}(x)$ and the second row by $v^r(x)$,
 and integrating over the vibration
coordinate, $x$, each of two equations of the system. The result is
\begin{equation}
\det \left(
  \begin{array}{cc}
    E-E_0 & \tilde{t} \\
    \tilde{t}& E-E_0 \\
  \end{array}
\right)=0
\end{equation}
with the renormalized hopping integral
\begin{equation}
{\tilde{t}\over{t}}= {\int^{\infty}_{-\infty} dx u^{l}(x)
 v^{r}(x)\over{\int^{\infty}_{-\infty}
 dx |u^{l}(x)|^{2}}}.
 \end{equation}
 The corresponding eigenvalues, $E_{\pm}$ are
 \begin{equation}
 E_{\pm}=\omega/2-E_{p} \pm \tilde{t}.
 \end{equation}
 The hopping integral splits the degenerate level, as in the rigid lattice, but  an
 effective `bandwidth' $2\tilde{t}$ is significantly reduced compared with
 the bare one
 \begin{equation}
 \tilde{t}=t \exp (-2E_p/\omega).
 \end{equation}
This polaron band narrowing  originates in a small
 overlap integral of two displaced oscillator wave functions $u^{l}(x)$
 and $v^{r}(x)$.

 \subsection{Adiabatic small polaron}
In the adiabatic regime, when $t\gg \omega$, the electron tunnelling
is fast compared with the ion motion. Hence one can apply the
Born-Oppenheimer adiabatic approximation taking
 the wave function in the form
 \begin{equation}
\left(
  \begin{array}{c}
    u(x) \\
   v(x) \\
  \end{array}
\right)=\chi(x)\left(
                 \begin{array}{c}
                   u_a(x) \\
                  v_a(x) \\
                 \end{array}
               \right).
\end{equation}
 Here $u_a(x)$ and $v_a(x)$ are  the electron wave functions obeying the Schr\"odinger equation with
 the \emph{frozen }ion deformation $x$, i.e.
\begin{equation}
\left(
  \begin{array}{cc}
   E_a(x)-f_ax& -t \\
    -t & E_a(x)+f_ax\\
  \end{array}
\right)\left(\begin{array}{c}
u_a(x)\\
v_a(x) \\
\end{array}
\right)=0.
\end{equation}
The lowest energy level  is found as
\begin{equation}
E_a(x)=- \sqrt{(f_ax)^2+t^{2}}.
\end{equation}
 $E_a(x)$ together with $kx^2/2$ play the role of a potential energy term
in the equation for the `vibration' wave function, $\chi(x)$,
\begin{equation}
\left[-{1\over {2M}}{\partial^{2}\over{\partial x^{2}}}+
 {kx^{2}\over{2}}- \sqrt{(f_ax)^2+t^{2}}\right] \chi(x)=E\chi(x).
\end{equation}
Terms with the first and second derivatives of the electron
wave-functions $u_a(x)$ and $v_a(x)$ are small compared with the
corresponding
 derivatives of $\chi(x)$  in the adiabatic approximation, so they can be neglected in Eq.(31).
As a result we arrive with the familiar double-well potential
problem, where the potential energy  $U(x)=kx^{2}/2-
\sqrt{(f_ax)^2+t^{2}}$ has two symmetric minima, separated by a
barrier.
  Minima
are located approximately at
\begin{equation}
x_{m}=\pm f_a/k
\end{equation}
in the strong-coupling limit, $E_p\gg t$, and the potential energy
near the bottom of each potential well is about
\begin{equation}
U(x)=-E_{p}+{k(|x|-f_a/k)^2\over{2}}.
\end{equation}
If the barrier were impenetrable, there would be the ground state
energy level $E_{0}=-E_p+\omega/2$, the same for both wells. The
tunnelling under the barrier results in a splitting of this level
$2\tilde{t}$, which corresponds to a polaron band in the lattice. It
can be estimated using the  quasi-classical approximation as
\begin{equation}
\tilde{t}\propto \exp\left[-2\int_{0}^{x_{m}}p(x)dx\right],
\end{equation}
where $ p(x)=\sqrt{2M[U(x)-E_0]}\approx (Mk)^{1/2}|x-f_a/k|$ is the
classical momentum

 Calculating the integral   one finds the exponential reduction of the "bandwidth",
\begin{equation}
\tilde{t}\propto \exp(-2E_p/\omega),
\end{equation}
which is the same as in the nonadiabatic regime. Holstein found also
corrections to this expression up to terms of the order of
$1/\lambda^{2}$, which allowed him to estimate  the pre-exponential
factor as
\begin{equation}
\tilde{t}\approx \sqrt {E_{p}\omega} \exp(-2E_p/\omega).
\end{equation}

  The term in front of the exponent
   differs from
  $t$ of the non-adiabatic case.  It is thus apparent that the
 perturbation  approach covers only part of the entire lattice
 polaron region, $\lambda \geq 1$. The upper limit of applicability of
 the perturbation theory is given by
  $t<\sqrt {E_{p} \omega}$.
For the remainder of the region the adiabatic approximation is more
accurate.

 \subsection{$``1/\lambda $'' expansion technique: polaron band}
The kinetic energy is small compared with  the interaction energy as long as $%
\lambda >1.$ Hence an analytical approach to the multi-polaron
problem is possible with the $``1/\lambda $'' expansion technique
\cite{alebook1}, which treats the kinetic energy as a perturbation.
The technique is based on the fact, known for a long time, that
there is an analytical exact solution of a $single$ polaron problem
in the strong-coupling limit $\lambda \rightarrow \infty $.
Following Lang and Firsov \cite{lan} we apply the canonical
transformation $e^{S}$   diagonalizing the Hamiltonian, Eq.(2). The
diagonalization is exact, if $T({\bf m})=0$ (or $\lambda =\infty $).
In the Wannier representation for electrons and phonons,
\[
S=\sum_{{\bf m\neq n,}\alpha }g_{\alpha }({\bf m-n})({\bf e}_{\alpha }\cdot {\bf u}_{%
{\bf m-n}})c_{{\bf n}}^{\dagger }c_{{\bf n}}(d_{{\bf m}\alpha }^{\dagger }-d_{%
{\bf m}\alpha }).
\]

The transformed Hamiltonian is
\begin{eqnarray}
\tilde{H} &=&e^{-S}He^{S}=\sum_{{\bf n\neq n^{\prime }}}\hat{\sigma}_{{\bf %
nn^{\prime }}}c_{{\bf n}}^{\dagger }c_{{\bf n^{\prime }}}+\omega\sum_{%
{\bf m}\alpha }\left( d_{{\bf m}\alpha}^{\dagger }d_{{\bf m}\alpha}+\frac{1}{2}%
\right) + \\
&&\sum_{{\bf n\neq n^{\prime }}}v({\bf n-n^{\prime }})c_{{\bf
n}}^{\dagger
}c_{{\bf n}}c_{{\bf n^{\prime }}}^{\dagger }c_{{\bf n^{\prime }}}-E_{p}\sum_{%
{\bf n}}c_{{\bf n}}^{\dagger }c_{{\bf n}} \nonumber,
\end{eqnarray}
where for simplicity we take $\omega_{\alpha}=\omega$. The last term
describes the energy gained by polarons due to the e-ph interaction.
The third term on the right-hand side  is the polaron-polaron
interaction,
\begin{equation}
v({\bf n-n^{\prime }})=V_{c}({\bf n-n^{\prime }})-V_{ph}({\bf n-n^{\prime }}%
),
\end{equation}
where
\begin{eqnarray*}
V_{ph}({\bf n-n^{\prime }}) &=&2\omega \sum_{{\bf m,}\alpha }g_{\alpha }({\bf %
m-n})g_{\alpha }({\bf m-n^{\prime }})\times \\
&&({\bf e}_{\alpha}\cdot {\bf u}_{{\bf m-n}})({\bf e}_{\alpha }\cdot {\bf u}_{%
{\bf m-n^{\prime }}}).
\end{eqnarray*}
The phonon-induced interaction $V_{ph}$ is due to displacements of
common
ions caused by two electrons. Finally, the transformed hopping operator $\hat{\sigma%
}_{{\bf nn^{\prime }}}$  is given by
\begin{eqnarray}
\hat{\sigma}_{{\bf nn^{\prime }}} &=&T({\bf n-n^{\prime }})\exp \left[ \sum_{%
{\bf m,}\alpha }\left[ g_{\alpha}({\bf m-n})({\bf e}_{\alpha }\cdot {\bf u}_{{\bf m-n%
}})\right. \right. \\
&&-\left. \left. g_{\alpha }({\bf m-n^{\prime }})({\bf e}_{\alpha }\cdot {\bf u}_{%
{\bf m-n^{\prime }}})\right] (d_{{\bf m}\alpha }^{\dagger }-d_{{\bf
m}\alpha })\right] \nonumber.
\end{eqnarray}
This term is  perturbation at a large $\lambda $.  It accounts for
the polaron and \emph{bipolaron} tunnelling and high temperature
superconductivity \cite{alebook1}. In  particular crystal structures
like perovskites, a bipolaron tunnelling appears already in the
first order in $T({\bf n})$ (see below), so that $\hat{\sigma}_{{\bf
nn^{\prime }}}$ can be averaged over  phonon vacuum, if the
temperature is low enough, $T\ll \omega$. The result is
\begin{equation}
t({\bf n-n^{\prime }})\equiv \left\langle \left\langle \hat{\sigma}_{{\bf %
nn^{\prime }}}\right\rangle \right\rangle _{ph}=T({\bf n-n^{\prime
}})\exp [-g^{2}({\bf n-n^{\prime }})],
\end{equation}
where
\begin{eqnarray*}
g^{2}({\bf n-n^{\prime }}) &=&\sum_{{\bf m},\alpha }g_{\alpha }({\bf m-n})({\bf e}%
_{\alpha }\cdot {\bf u}_{{\bf m-n}})\times \\
&&\left[ g_{\alpha}({\bf m-n})({\bf e}_{\alpha }\cdot {\bf u}_{{\bf
m-n}})-g_{\alpha
}({\bf m-n^{\prime }})({\bf e}_{\alpha }\cdot {\bf u}_{{\bf m-n^{\prime }}})%
\right] .
\end{eqnarray*}
By comparing Eqs.(40) and Eqs.(38,3), the bandwidth renormalization
exponent can be expressed via $E_{p}$ and $V_{ph}$ as follows
\begin{equation}
g^{2}({\bf n-n^{\prime }})=\frac{1}{\omega}\left[ E_{p}-\frac{1}{2}%
V_{ph}({\bf n-n^{\prime }})\right] .
\end{equation}
In zero order with respect to the hopping the Hamiltonian, Eq.(37)
describes localized polarons and independent phonons, which are
vibrations of ions around new equilibrium positions depending on the
polaron occupation numbers. The phonon frequencies remain unchanged
in this limit. The middle of the electron band falls by the polaron
level-shift $E_{p}$ due to a potential well created by lattice
deformation. The finite hopping term leads to the polaron tunnelling
because of degeneracy of the zero order Hamiltonian with respect to
 site positions of the polaron.

\subsection{From continuous  to small Holstein and small Fr\"{o}hlich polarons:
QMC simulation} The narrowing of the band and the polaron effective
mass strongly depend on the radius of the electron-phonon
interaction \cite{ale5}. Let us compare the small Holstein polaron
(SHP) formed by a short-range e-ph interaction and a small polaron
formed by a long-range (Fr\"{o}hlich) interaction, which we refer to
as the small Fr\"{o}hlich polaron (SFP). For simplicity we consider
the interaction with a single phonon branch. In general, there is no
simple relation between the polaron level-shift $E_{p}$ and the
exponent $g^{2}$. This relation depends on the form of the
electron-phonon interaction. In the nearest-neighbor approximation
the effective mass renormalization is given by
\[
m^{\ast }/m=e^{g^{2}},
\]
where $m$ is the bare band mass and $g^2\equiv g^2({\bf a})$.
\begin{figure}[tbp]
\begin{center}
\includegraphics[angle=-90,width=0.55\textwidth]{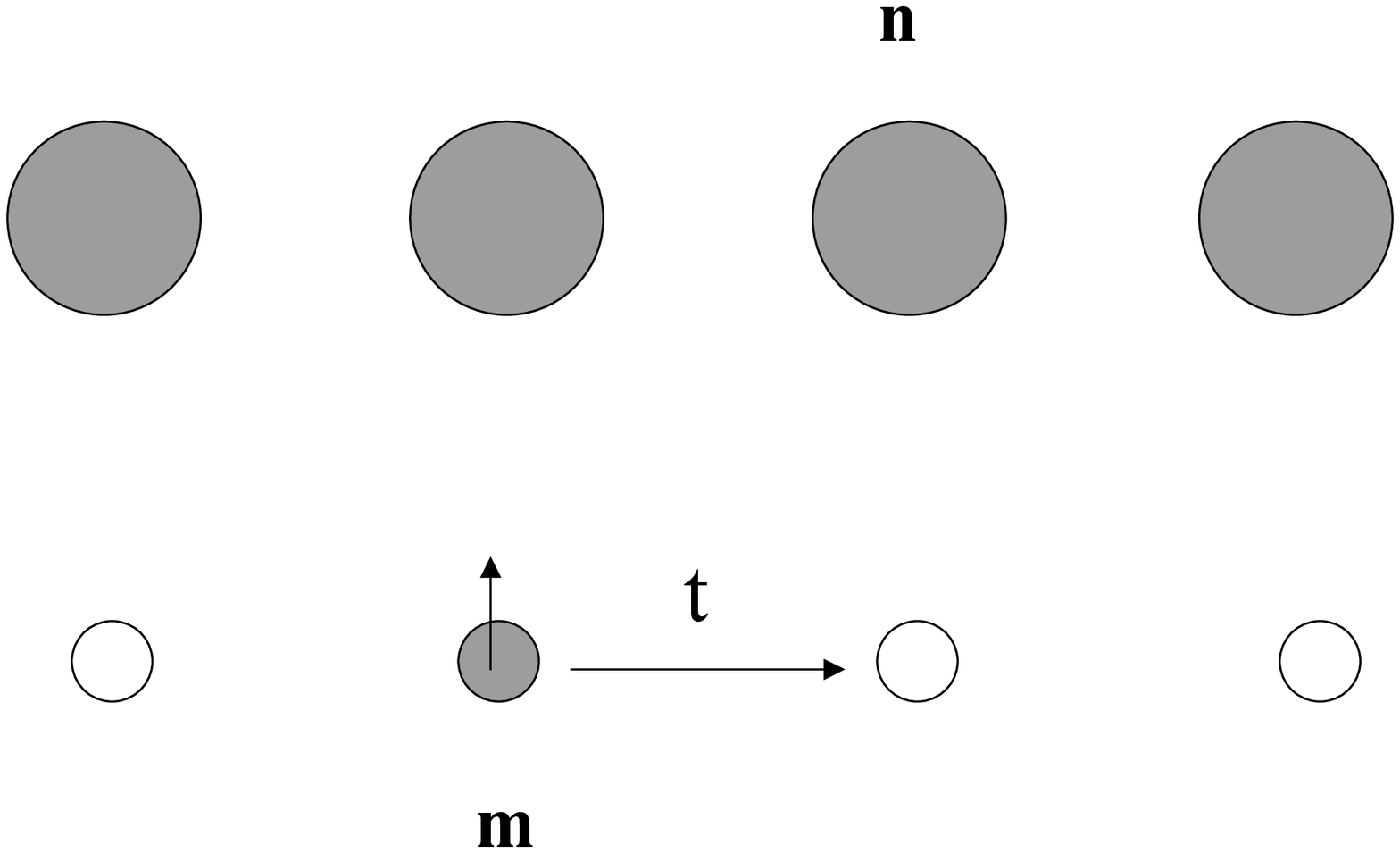}
\vskip -0.5mm
\end{center}
\caption{One-dimensional model of the lattice polaron on the chain
interacting with displacements of all ions of another chain.}
\end{figure}

If the interaction is short-ranged, $g_{\alpha}({\bf m})=\kappa
\delta _{{\bf m},0}$ (the Holstein model), then $g^{2}=E_{p}/\omega
$. Here $\kappa $ is a constant. In general, we have $g^{2}=\gamma
E_{p}/\omega $ with the numerical coefficient $\gamma$ less than
$1$. To estimate $\gamma $ let us consider a one-dimensional chain
model with the long-range Coulomb interaction between the electron
on chain ($\times )$ and ion vibrations of another chain ($\circ ),$
polarized in the direction perpendicular to the chains \cite{Korn2},
Fig.4. The corresponding force is given by
\begin{equation}
g_{\alpha }({\bf m-n})({\bf e}%
_{\alpha }\cdot {\bf u}_{{\bf m-n}})=\frac{\kappa }{(|{\bf m}-{\bf
n}|^{2}+1)^{3/2}}. \label{ten}
\end{equation}
Here the distance along the chains $|{\bf m}-{\bf n}|$ is measured
in units of the lattice constant $a$, the inter-chain distance is
also $a$, and we take $a=1$. For this long-range interaction we
obtain $\gamma=0.39$. Hence the effective mass renormalization is
much smaller than in the Holstein model, roughly as $m_{SFP}^{\ast
}\propto (m_{SHP}^{\ast })^{1/2}$.

Not only does the small polaron mass strongly depend on the radius
of the electron-phonon interaction, but also does the range of the
applicability of the analytical $1/\lambda $ expansion theory. The
theory appears almost exact in a wide region of parameters for the
Fr\"{o}hlich interaction. The exact polaron mass in a wide region of
the adiabatic parameter $\omega /T(a)$ and coupling was calculated
with the continuous-time path-integral Quantum Monte Carlo (QMC)
algorithm \cite{Korn2}. This method is free from any systematic
finite-size, finite-time-step and finite-temperature errors and
allows for an {\em exact} (in the QMC sense) calculation of the
ground-state energy and the effective mass of the lattice polaron
for any electron-phonon interaction.
\begin{figure}[tbp]
\begin{center}
\includegraphics[angle=0,width=0.75\textwidth]{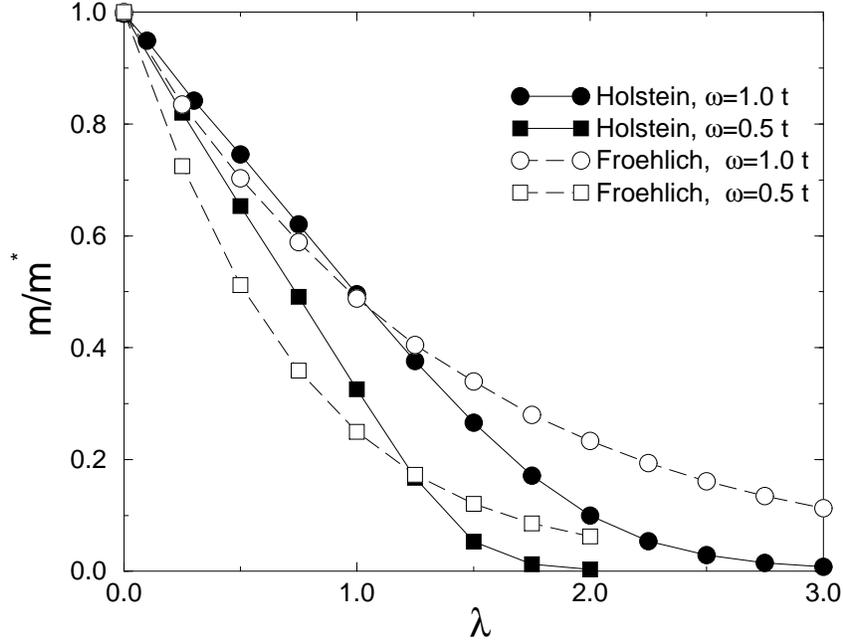}
\vskip -0.5mm
\end{center}
\caption{Inverse effective polaron mass in units of $1/m=2T(a)a^2$
($T(a)=t$ on the graph) \cite{Korn2}}
\end{figure}
At large $\lambda $ ($>1.5$) SFP was found to be much lighter than
SHP, while the large Fr\"{o}hlich polaron (i.e. at $\lambda <1$) was
$heavier$ than the large Holstein polaron with the same binding
energy, Fig.5. The
mass ratio $m_{FP}^{\ast }/m_{HP}^{\ast }$ is a non-monotonic function of $%
\lambda $. The effective mass of the Fr\"{o}hlich polaron,
$m_{FP}^{\ast }(\lambda )$ is well fitted by a single exponent,
which is $e^{0.73\lambda }$ for $\omega=T(a)$ and $e^{1.4\lambda }$
for $\omega =0.5\,T(a)$. The exponents are remarkably close to those
obtained with the Lang-Firsov transformation, $e^{0.78\lambda }$ and
$e^{1.56\lambda }$, respectively. Hence, in the case of the
Fr\"{o}hlich interaction the transformation is perfectly accurate
even in the moderate adiabatic regime, $\omega /T(a)\leq 1 $ for
$any$ coupling strength. It is not the case for the Holstein
polaron. If the interaction is short-ranged, the same analytical
technique is applied only in the nonadiabatic regime $\omega
/T(a)>1.$

Another interesting point is that the size of SFP and the length,
over which the distortion spreads, are $different$. In the
strong-coupling limit the polaron is almost localized on one site
${\bf m}$. Hence, the size of its wave function is the atomic size.
On the other hand, the ion displacements, proportional to the
displacement force $g_{\alpha}({\bf m-n})$, spread over a large
distance. Their amplitude at a site ${\bf n}$ falls with the distance as $|%
{\bf m-n}|^{-3}$ in our one-dimensional model. The polaron cloud
(i.e. lattice distortion) is more extended than the polaron itself.
Such polaron tunnels with a larger probability than the Holstein
polaron due to a smaller $relative$ lattice distortion around two
neighboring sites. For the short-range e-ph interaction the {\em
entire} lattice deformation disappears at one site and then forms at
its neighbor, when the polaron tunnels from site to site. Therefore
$\gamma =1$ and the polaron is very heavy already at $\lambda
\approx 1$. On the contrary, if the interaction is long-ranged, only
a fraction of the total deformation
changes every time the polaron tunnels from one site to its neighbor, and $%
\gamma $ is smaller than $1.$  A lighter mass of SFP compared with
the nondispersive SHP is a generic feature of any dispersive
electron-phonon interaction.

\subsection{Attractive correlations of small polarons}
Lattice deformation also strongly affects the interaction between
electrons. At large distances polarons repel each other in ionic
crystals, but their Coulomb repulsion is substantially reduced
due to  ion polarization. Nevertheless two $%
large$ polarons can be bound into a $large$ bipolaron by an exchange
interaction even with no additional e-ph interaction but the
Fr\"{o}hlich one \cite{vin,sup,ada,emin,bas,ver0}.

When a short-range deformation potential and molecular e-ph
interactions (e.g. of the Jahn-Teller type \cite{mul00})  are taken
into account together with the long-range Fr\"{o}hlich interaction,
they can overcome the Coulomb repulsion \cite{ale5}. The resulting
interaction becomes attractive at a short distance of about a
lattice constant. Then two  small polarons readily form a bound
state, i.e. a $small$ bipolaron \cite{ander,cha,alerun,aubr},
because their band is  narrow. Consideration of particular lattice
structures shows that small bipolarons are mobile even when the
electron-phonon coupling is strong and the bipolaron binding energy
is large \cite{ale5}(see below).  Here we  encounter a novel
electronic state of matter, a charged Bose liquid of electron
molecules with double elementary charge 2e, qualitatively different
from  normal Fermi-liquids in ordinary metals and from the
Bardeen-Cooper-Schrieffer (BCS) superfluids in conventional
superconductors.

The origin of the attractive force between two small polarons can be
readily understood from about the same toy model as in Fig.2, but
with two electrons on neighbor sites \textbf{1,2} interacting with
an ion in between \textbf{3}, Fig.6. For generality we now assume
that the ion is a three-dimensional oscillator described by a
displacement vector $\bf u$, rather than by a single-component
displacement $x$ as in Fig.2.
\begin{figure}[tbp]
\begin{center}
\includegraphics[angle=-90,width=0.80\textwidth]{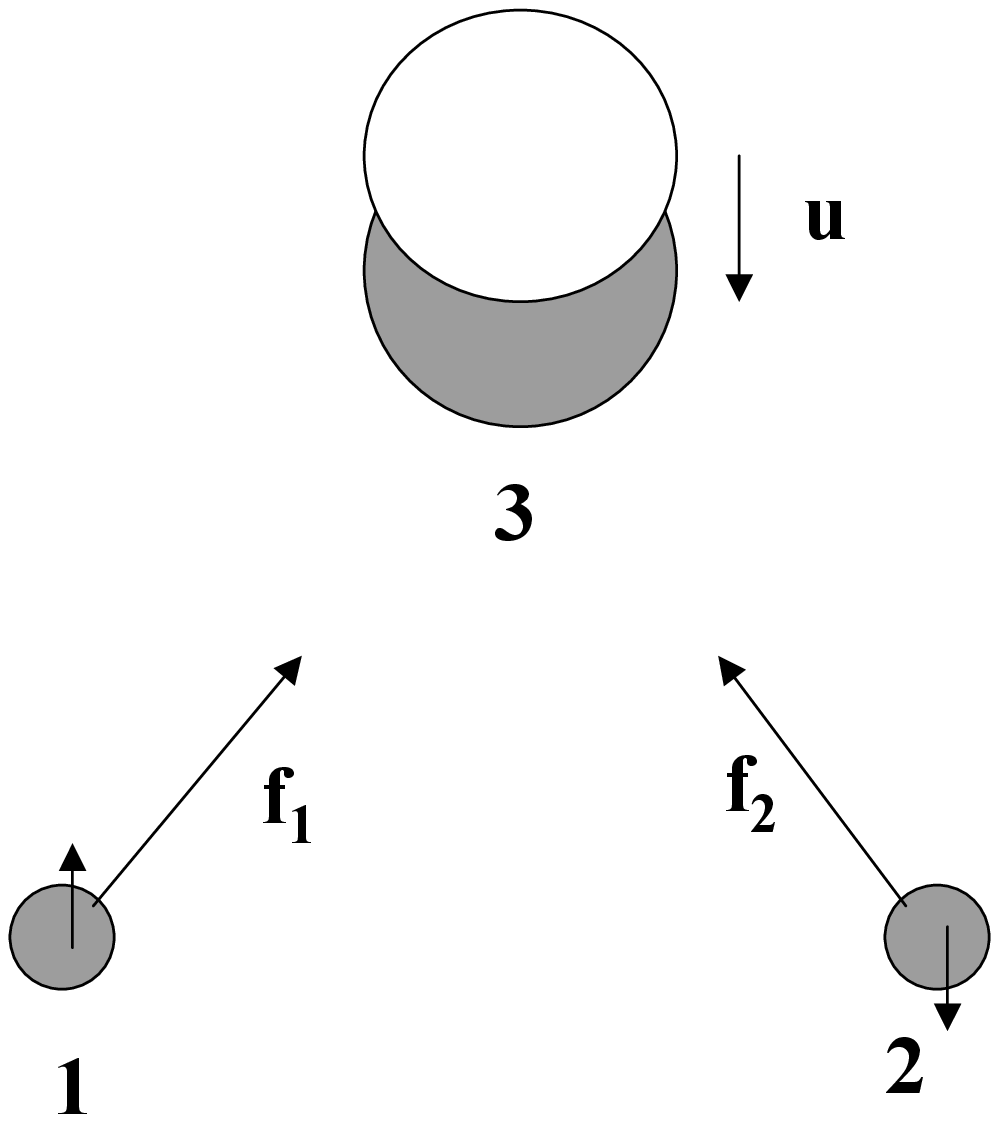}
\vskip -0.5mm
\end{center}
\caption{Two localized electrons shift the equilibrium position of
the ion (${\bf 3}$). As a result two electrons on neighboring sites
${\bf 1}$  and ${\bf 2}$  attract each other.}
\end{figure}
The vibration part of the Hamiltonian in  the model is
\begin{equation}
H_{ph}= -{1\over {2M}}\left({\partial\over{\partial {\bf
u}}}\right)^{2}+
 {ku^{2}\over{2}},
\end{equation}
 Electron potential energies  due to the Coulomb interaction with the ion are approximately
\begin{equation}
V_{1,2}= V_0 (1-{\bf u}\cdot {\bf e_{1,2}}/a),
\end{equation}
where  ${\bf e_{1,2}}$ are units vectors connecting sites ${\bf
1,2}$ and site ${\bf 3}$, respectively. Hence the Hamiltonian of the
model is given by
\begin{equation}
H= E_a (\hat{n}_{1} +\hat{n}_{2})+ {\bf u} \cdot ({\bf
f_{1}}\hat{n}_{1}+{\bf f_{2}}\hat{n}_{2})-{1\over
{2M}}\left({\partial\over{\partial {\bf u}}}\right)^{2}+
 {ku^{2}\over{2}},
\end{equation}
where   ${\bf f_{1,2}}=Ze^2{\bf e_{1,2}}/a^2$ is the Coulomb force,
and $\hat{n}_{1,2}$ are  occupation number operators at every site.
This Hamiltonian is also readily diagonalized  by the same
displacement transformation of the vibronic coordinate ${\bf u}$ as
above,
\begin{equation}
{\bf u}={\bf v}-\left({\bf f_{1}}\hat{n}_{1}+{\bf
f_{2}}\hat{n}_{2}\right)/k.
\end{equation}
The transformed Hamiltonian has no electron-phonon coupling,
\begin{equation}
\tilde{H}=(E_a-E_p) (\hat{n}_{1}+\hat{n}_{2})
+V_{ph}\hat{n}_{1}\hat{n}_{2}-{1\over
{2M}}\left({\partial\over{\partial {\bf v}}}\right)^{2}+
 {kv^{2}\over{2}},
 \end{equation}
 and describes two small polarons   at their atomic levels  shifted by the polaron level shift $E_p=f_{1,2}^2/2k$, which are entirely decoupled from ion vibrations.  As a result, the lattice deformation   caused by two electrons leads to  an effective interaction between them, $V_{ph}$, which should be added to their Coulomb repulsion, $V_{c}$,
\begin{equation}
V_{ph}= - {\bf f_{1}} \cdot {\bf f_{2}}/k.
 \end{equation}
When $V_{ph}$ is negative and larger by magnitude than the positive
$V_{c}$ the resulting interaction becomes attractive.

Applying the  polaron canonical transformation to a generic
``Fr\"{o}hlich-Coulomb'' Hamiltonian, allows us explicitly calculate
the effective attraction of small polarons \cite{alekor}, Eq.(38),
and elaborate more physics behind the lattice sums in Eq.(3,38) and
Eq.(41). If a carrier (electron or hole) acts on an ion with a force
${\bf f}$, it displaces the ion by some vector ${\bf x}={\bf f}/k$.
Here $k$ is the ion's
force constant. The total energy of the carrier-ion pair is $-{\bf f}%
^{2}/(2k)$. This is precisely the summand in Eq.(3) expressed via
dimensionless coupling constants. Now consider two carriers
interacting with
the {\em same} ion, see Fig.7. The ion displacement is ${\bf x}=({\bf f}%
_{1}+{\bf f}_{2})/k$ and the energy is $-{\bf f}_{1}^{2}/(2k)-{\bf f}%
_{2}^{2}/(2k)-({\bf f}_{1}\cdot {\bf f}_{2})/k$. Here the last term
should be interpreted as an ion-mediated interaction between the two
carriers. It depends on the scalar product of ${\bf f}_{1}$ and
${\bf f}_{2}$ and consequently on the relative positions of the
carriers with respect to the ion. If the ion is an isotropic
harmonic oscillator, as we assume here, then the following simple
rule applies. If the angle $\phi $ between ${\bf f}_{1}$ and ${\bf
f}_{2}$ is less than $\pi /2$ the polaron-polaron interaction will
be attractive, if otherwise it will be repulsive. In general, some
ions will generate attraction, and some repulsion between polarons,
Fig.7.
\begin{figure}[tbp]
\begin{center}
\includegraphics[angle=-0,width=0.57\textwidth]{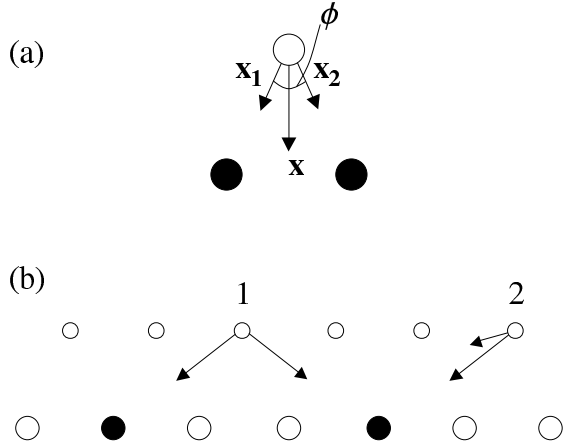} \vskip -0.5mm
\end{center}
\caption{Mechanism of the polaron-polaron interaction. (a) Together,
 two polarons (solid circles) deform the lattice more effectively
than separately. An effective attraction occurs when the angle $\phi
$ is less than $\pi /2$ . (b) A mixed situation. Ion 1 results in
repulsion between two polarons while ion 2 results in attraction. }
\end{figure}
The overall sign and magnitude of the interaction is given by the
lattice sum in Eq.(38), the evaluation of which is elementary. One
should also note that according to Eq.(41) an attractive interaction
reduces the polaron mass (and consequently the bipolaron mass),
while repulsive interaction enhances the mass.

\section{Superlight  bipolarons in high-$T_{c}$ cuprates}

Consideration of particular lattice structures shows that small
inter-site bipolarons are perfectly mobile even when the
electron-phonon coupling is strong and the bipolaron binding energy
is large. Let us analyze the important case of copper-based
high-$T_{c}$ oxides. As discussed in the introduction they are doped
charged-transfer ionic insulators with narrow electron bands.
Therefore, the interaction between holes can be analyzed using
computer simulation techniques based on a minimization of the ground
state energy of an ionic insulator with two holes, the lattice
deformations and the Coulomb repulsion fully taken into account, but
neglecting the kinetic energy terms. Using these techniques net
inter-site interactions of the in-plane oxygen hole with the $apex$
hole, Fig.8, and of two in-plane oxygen holes, Fig.10, were found to
be attractive in $La_{2}CuO_{4}$ \cite{cat} with the binding
energies $\Delta =119meV$ and $\Delta =60meV$, respectively.  All
other interactions were found to be repulsive.

\subsection{Apex bipolarons}
Both apex and in-plane bipolarons can tunnel from one
unit cell to another via the {\it %
\ single}-polaron tunnelling from one apex  oxygen to its apex
neighbor in case of the apex bipolaron \cite{ale5}, Fig.8, or via
the next-neighbor hopping in case of the in-plane bipolaron
\cite{alekor}, Fig.10.
\begin{figure}[tbp]
\begin{center}
\includegraphics[angle=-90,width=0.47\textwidth]{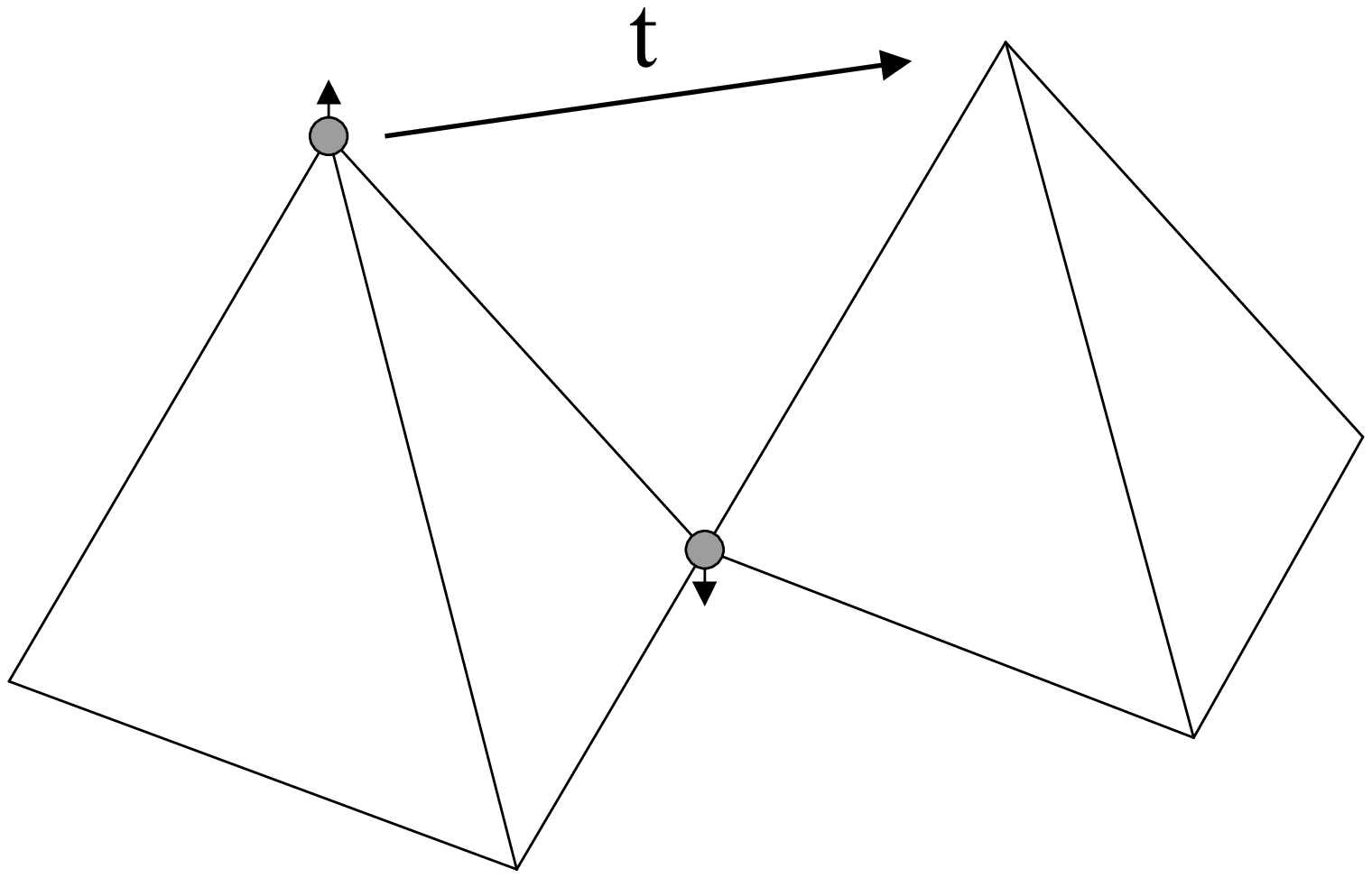} \vskip -0.5mm
\end{center}
\caption{Apex bipolaron tunnelling in perovskites }
\end{figure}
The Bloch bands of these bipolarons  are obtained using the
canonical transformation, described above,  projecting the
transformed Hamiltonian, Eq.(37), onto a reduced Hilbert space
containing only empty or doubly occupied elementary cells, and
averaging the result with respect to phonons \cite{alebook1}. The
wave function of the apex bipolaron localized, let us say, in the
cell ${\bf m}$ is written as
\begin{equation}
|{\bf m}\rangle =\sum_{i=1}^{4}A_{i}c_{i}^{\dagger
}c_{apex}^{\dagger }|0\rangle ,
\end{equation}
where $i$ denotes the $p_{x,y}$ orbitals and spins of the four plane
oxygen ions in the cell, Fig.8 and $c_{apex}^{\dagger }$ is the
creation operator for the hole on one of the three apex oxygen
orbitals with the spin, which is the same or opposite of the spin of
the in-plane hole depending on the total spin of the bipolaron. The
probability amplitudes $A_{i}$ are
normalized by the condition $|A_{i}|=1/2,$ if four plane orbitals $%
p_{x1},p_{y2},p_{x3}$ and $p_{y4}$ are involved, or by
$|A_{i}|=1/\sqrt{2}$ if only two of them are relevant. Then a matrix
element of the Hamiltonian Eq.(37)  describing the bipolaron
tunnelling to the nearest neighbor cell ${\bf m+a}$ is found as
\begin{equation}
t=\langle {\bf m}|\tilde{H}|{\bf m}+{\bf a}\rangle
=|A_{i}|^{2}T_{pp^{\prime }}^{apex}e^{-g^{2}},
\end{equation}
where $T_{pp^{\prime }}^{apex}e^{-g^{2}}$ is a \emph{single polaron}
hopping integral between two apex ions. The inter-site bipolaron
tunnelling appears already in the first order with respect to the
single-hole transfer  $T_{pp^{\prime }}^{apex}$, and the bipolaron
energy
spectrum consists of two bands $E^{x,y}({\bf K),}$ formed by the overlap of $%
p_{x}$ and $p_{y}$ $apex$ oxygen  orbitals, respectively:
\begin{eqnarray}
E^{x}({\bf K}) &=&t\cos (K_{x})-t^{\prime }\cos (K_{y}), \\
E^{y}({\bf K}) &=&-t^{\prime }\cos (K_{x})+t\cos (K_{y}). \nonumber
\end{eqnarray}
\begin{figure}[tbp]
\begin{center}
\includegraphics[angle=-0,width=0.47\textwidth]{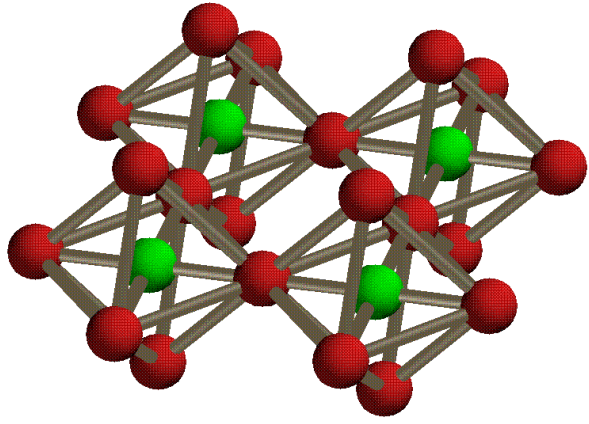} \vskip -0.5mm
\end{center}
\caption{Simplified model of the copper-oxygen perovskite layer. }
\end{figure}
They transform into one another under $\pi /2$ rotation. If
$t,t^{\prime
}>0, $ ``$x"$ bipolaron band has its minima at ${\bf K=}(\pm \pi ,0)$ and $y$%
-band at ${\bf K}=(0,\pm \pi )$. In these equations $t$ is the
renormalized hopping integral between $p$ orbitals of the same
symmetry elongated in the direction of the hopping ($pp\sigma $) and
$t^{\prime }$ is the renormalized hopping integral in the
perpendicular direction ($pp\pi $). Their ratio $t/t^{\prime
}=T_{pp^{\prime }}^{apex}/T^{\prime }{}_{pp^{\prime }}^{apex}=4$ as
follows from the tables of hopping integrals in solids. Two
different bands are not mixed because $T_{p_{x},p_{y}^{\prime
}}^{apex}=0$ for the nearest neighbors. A random potential does not
mix them either, if it varies smoothly on the lattice scale. Hence,
we can distinguish `$x$' and `$y$' bipolarons with a lighter
effective mass in $x$ or $y$ direction, respectively. The apex $z$
bipolaron, if formed, is $ca.$ four times less mobile than $x$ and
$y$ bipolarons. The bipolaron bandwidth is of the same order as the
polaron one, which is a specific feature of  inter-site
bipolarons. For a large part of the Brillouin zone near $%
(0,\pi )$ for `$x$' and $(\pi ,0)$ for `$y$' bipolarons, one can
adopt the effective mass approximation
\begin{equation}
E^{x,y}({\bf K})={\frac{K_{x}^{2}}{{2m_{x,y}^{\ast \ast }}}}+{\frac{%
K_{y}^{2}}{{2m_{y,x}^{\ast \ast }}}}
\end{equation}
with $K_{x,y}$ taken relative to the band bottom positions and
$m_{x}^{\ast \ast }=1/t$, $m_{y}^{\ast \ast }=4m_{x}^{\ast \ast }$.

\subsection{In-plane bipolarons}
Now let us consider  in-plane bipolarons in a two-dimensional
lattice of ideal octahedra that can be regarded as a simplified
model of the copper-oxygen perovskite layer, Fig.9 \cite{alekor}.
The lattice period is $a=1$ and the distance between the apical
sites and the central plane is $h=a/2=0.5$. For mathematical
transparency we assume that all in-plane atoms, both copper and
oxygen, are static but apex oxygens are independent
three-dimensional isotropic harmonic oscillators. Due to poor
screening, the hole-apex interaction is purely coulombic,
\[
g_{\alpha }({\bf m-n})=\frac{\kappa _{\alpha }}{|{\bf m-n}|^{2}},
\]
where $\alpha =x,y,z$. To account for the experimental fact that
$z$-polarized
phonons couple to the holes stronger than others \cite{tim}, we choose $%
\kappa _{x}=\kappa _{y}=\kappa _{z}/\sqrt{2}$. The direct hole-hole
repulsion is
\[
V_{c}({\bf n-n^{\prime }})=\frac{V_{c}}{\sqrt{2}|{\bf n-n^{\prime
}}|}
\]
so that the repulsion between two holes in the nearest neighbor (NN)
configuration is $V_{c}$. We also include the bare NN hopping
$T_{NN}$, the next nearest neighbor (NNN) hopping across copper
$T_{NNN}$ and the NNN hopping between the pyramids $T_{NNN}^{\prime
}$.

\begin{figure}[tbp]
\begin{center}
\includegraphics[angle=-0,width=0.50\textwidth]{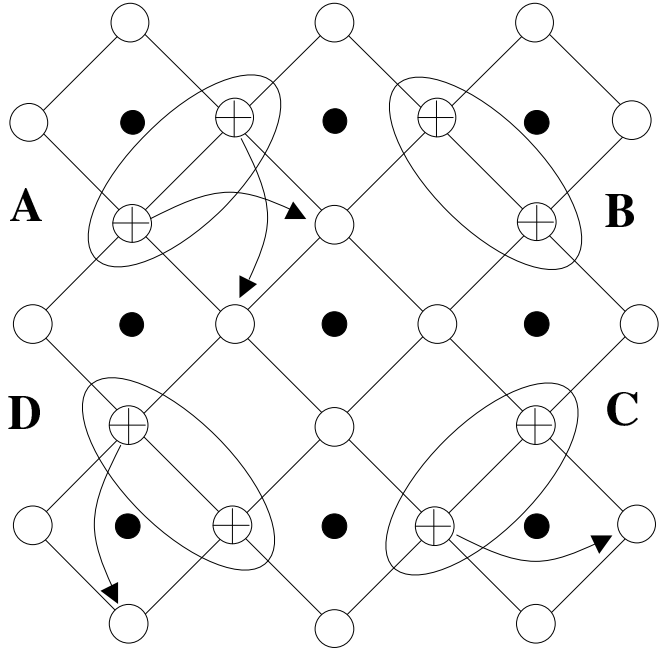} \vskip -0.5mm
\end{center}
\caption{Four degenerate in-plane bipolaron configurations A, B, C,
and D . Some single-polaron hoppings are indicated by arrows. }
\end{figure}

The polaron shift is given by the lattice sum Eq.(3), which after
summation over polarizations yields
\begin{equation}
E_{p}=2\kappa _{x}^{2}\omega _{0}\sum_{{\bf m}}\left( \frac{1}{|{\bf m-n}%
|^{4}}+\frac{h^{2}}{|{\bf m-n}|^{6}}\right) =31.15\kappa
_{x}^{2}\omega _{0}, \label{televen}
\end{equation}
where the factor $2$ accounts for two layers of apical sites. For
reference,
the Cartesian coordinates are ${\bf n}=(n_{x}+1/2,n_{y}+1/2,0)$, ${\bf m}%
=(m_{x},m_{y},h)$, and $n_{x},n_{y},m_{x},m_{y}$ are integers. The
polaron-polaron attraction is
\begin{equation}
V_{ph}({\bf n-n^{\prime }})=4\omega \kappa _{x}^{2}\sum_{{\bf m}}\frac{%
h^{2}+({\bf m-n^{\prime }})\cdot ({\bf m-n})}{|{\bf m-n^{\prime }}|^{3}|{\bf %
m-n}|^{3}}.  \label{ttwelve}
\end{equation}
Performing the lattice summations for the NN, NNN, and NNN'
configurations one finds $V_{ph}=1.23\,E_{p},$ $0.80\,E_{p}$, and
$0.82\,E_{p}$,
respectively. As a result, we obtain a net inter-polaron interaction as $%
v_{NN}=V_{c}-1.23\,E_{p}$, $v_{NNN}=\frac{V_{c}}{\sqrt{2}}-0.80\,E_{p}$, $%
v_{NNN}^{\prime }=\frac{V_{c}}{\sqrt{2}}-0.82\,E_{p}$, and the mass
renormalization exponents as $g_{NN}^{2}=0.38(E_{p}/\omega)$, $%
g_{NNN}^{2}=0.60(E_{p}/\omega)$ and $(g^{\prime
}{}_{NNN})^{2}=0.59(E_{p}/\omega)$.

Let us now discuss different regimes of the model. At
$V_{c}>1.23\,E_{p}$, no bipolarons are formed and the system is a
polaronic Fermi liquid. Polarons tunnel in the {\em square} lattice
with  $t=T_{NN}\exp (-0.38E_{p}/\omega)$ and  $t^{\prime
}=T_{NNN}\exp (-0.60E_{p}/\omega)$ for NN and NNN hoppings,
respectively. Since $g_{NNN}^{2}\approx (g_{NNN}^{\prime })^{2} $
one can neglect the difference between NNN hoppings within and
between the octahedra. A single polaron spectrum is therefore
\begin{equation}
E_{1}({\bf k})=-E_{p}-2t^{\prime }[\cos k_{x}+\cos k_{y}]-4t\cos
(k_{x}/2)\cos (k_{y}/2).  \label{tfifteen}
\end{equation}
The polaron mass is $m^{\ast }=1/(t+2t^{\prime })$. Since in general $%
t>t^{\prime }$, the mass is mostly determined by the NN hopping amplitude $t$%
.

If $V_{c}<1.23\,E_{p}$ then intersite NN bipolarons form. The
bipolarons
tunnel in the plane via four resonating (degenerate) configurations $A$, $B$%
, $C$, and $D$, as shown in Fig.10. In the first order of the
renormalized hopping integral, one should retain only these lowest
energy configurations and discard all the processes that involve
configurations with higher energies. The result of such a projection
is the bipolaron Hamiltonian
\begin{eqnarray}
H_{b} &=&(V_{c}-3.23\,E_{p})\sum_{{\bf l}}[A_{{\bf l}}^{\dagger }A_{{\bf l}%
}+B_{{\bf l}}^{\dagger }B_{{\bf l}}+C_{{\bf l}}^{\dagger }C_{{\bf l}}+D_{%
{\bf l}}^{\dagger }D_{{\bf l}}] \\
&&-t^{\prime }\sum_{{\bf l}}[A_{{\bf l}}^{\dagger }B_{{\bf l}}+B_{{\bf l}%
}^{\dagger }C_{{\bf l}}+C_{{\bf l}}^{\dagger }D_{{\bf l}}+D_{{\bf l}%
}^{\dagger }A_{{\bf l}}+H.c.]  \nonumber \\
&&-t^{\prime }\sum_{{\bf n}}[A_{{\bf l-x}}^{\dagger }B_{{\bf l}}+B_{{\bf l+y}%
}^{\dagger }C_{{\bf l}}+C_{{\bf l+x}}^{\dagger }D_{{\bf l}}+D_{{\bf l-y}%
}^{\dagger }A_{{\bf l}}+H.c.],  \nonumber
\end{eqnarray}
where ${\bf l}$ numbers octahedra rather than individual sites, ${\bf x}%
=(1,0)$, and ${\bf y}=(0,1)$. A Fourier transformation and
diagonalization of a $4\times 4$ matrix yields the bipolaron
spectrum:
\begin{equation}
E_{2}({\bf K})=V_{c}-3.23E_{p}\pm 2t^{\prime }[\cos (K_{x}/2)\pm
\cos (K_{y}/2)].  \label{tseventeen}
\end{equation}
There are four bipolaronic subbands combined in the band of the width $%
8t^{\prime }$. The effective mass of the lowest band is $m^{\ast
\ast }=2/t^{\prime }$. The bipolaron binding energy is $\Delta
\approx 1.23E_{p}-V_{c}.$ Inter-site bipolarons already move in the
{\em first} order of the single polaron hopping. This remarkable
property is entirely due to the strong on-site repulsion and
long-range electron-phonon interactions that leads to a non-trivial
connectivity of the lattice. This fact combines with a weak
renormalization of $t^{\prime }$ yielding a {\em superlight}
bipolaron with the mass $m^{\ast \ast }\propto \exp
(0.60\,E_{p}/\omega )$. We recall that in the Holstein model
$m^{\ast \ast }\propto \exp (2E_{p}/\omega )$ \cite{alerun}. Thus
the mass of the Fr\"{o}hlich bipolaron in the perovskite layer
scales approximately as a {\em cubic root} of that of the Holstein
bipolaron.
\begin{figure}[tbp]
\begin{center}
\includegraphics[angle=-0,width=0.40\textwidth]{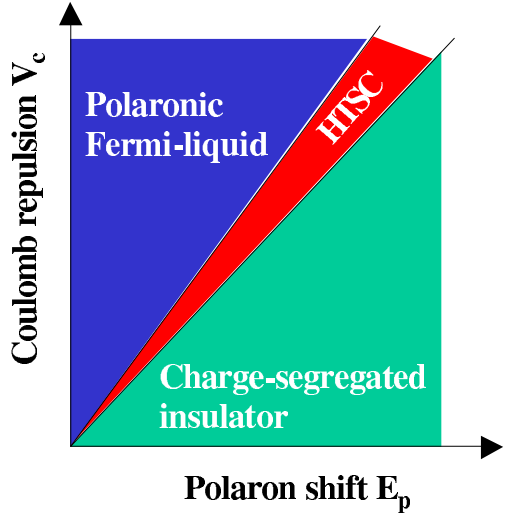} \vskip -0.5mm
\end{center}
\caption{FCM phase diagram.}
\end{figure}
At even stronger e-ph interaction, $V_{c}<1.16E_{p}$, NNN bipolarons
become stable. More importantly, holes can now form 3- and
4-particle clusters. The dominance of the potential energy over
kinetic in the transformed Hamiltonian enables us to readily
investigate these many-polaron cases. Three holes placed within one
oxygen square have four degenerate states with the energy
$2(V_{c}-1.23E_{p})+V_{c}/\sqrt{2}-0.80E_{p}$. The first-order
polaron hopping processes mix the states resulting in a ground state
linear
combination with the energy $E_{3}=2.71V_{c}-3.26E_{p}-\sqrt{%
4t^{2}+t^{\prime }{}^{2}}$. It is essential that between the squares
such triads could move only in higher orders of polaron hopping. In
the first order, they are immobile. A cluster of four holes has only
one state within
a square of oxygen atoms. Its energy is $E_{4}=4(V_{c}-1.23E_{p})+2(V_{c}/%
\sqrt{2}-0.80E_{p})=5.41V_{c}-6.52E_{p}$. This cluster, as well as
all bigger ones, are also immobile in the first order of polaron
hopping. We would like to stress that at distances much larger than
the lattice constant the polaron-polaron interaction is always
repulsive, and the formation of infinite clusters, stripes or
strings is  prohibited. We conclude that at $V_{c}<1.16E_{p}$ the
system quickly becomes a charge segregated insulator, Fig.11.

The fact that within the window, $1.16E_{p}<V_{c}<1.23E_{p}$, there
are no three or more polaron bound states, means that bipolarons
repel each other. The system is effectively a charged Bose-gas,
which is a superconductor. The superconductivity window, that we
have found, is quite narrow. This indicates that the superconducting
state in cuprates requires a rather fine balance between electronic
and ionic interactions, Fig.11.

\subsection{Low-energy (bi)polaron energy structure of cuprates}
The considerations set out above lead us to a simple model of
cuprates  \cite{alemot3,alebook1}. The main assumption is that $all$
$electrons$ are bound into small singlet and triplet $intersite$
bipolarons stabilized by e-ph interactions. As the undoped plane has
a half-filled $Cu3d^{9}$ band there are no empty states for
bipolarons to move if they are inter-site. Their Brillouin zone is
half of the original electron BZ and is completely filled with
hard-core bosons. $Hole$ pairs, which appear with doping, have
enough empty states to move, and they are responsible for low-energy
kinetics. Above $T_{c}$ a material such as $YBa_{2}Cu_{3}O_{6+x}$
contains a non-degenerate gas of hole bipolarons in singlet and in
triplet states. Triplets are separated from singlets by a spin gap
$J$ and have a lower mass due to a lower binding energy, Fig.12. The
main part of the electron-electron correlation energy (Hubbard $U$
and the inter-site Coulomb repulsion) and the electron-phonon
interaction are taken into account in the binding energy of
bipolarons $\Delta ,$ and in their band-width renormalization as
described above. When the hole  density is small, $n_{b}\ll 1$ (as
in cuprates), their bipolaronic operators are almost bosonic. The
hard-core interaction does not play any role in this dilute limit,
so only the long-distance Coulomb repulsion is relevant. This
repulsion is
significantly reduced due to a large static dielectric constant in oxides, $%
\epsilon _{0}\gg 1$. Hence, carriers are almost free charged bosons
coexisting with thermally excited nondegenerate single fermions, so
that the conventional Boltzmann kinetics (see below) and the
Bogoliubov transformation \cite{bog} for a charged Bose gas  are
perfectly applied in the normal and in the superconducting state,
respectively.

\begin{figure}
\begin{center}
\includegraphics[angle=-0,width=0.60\textwidth]{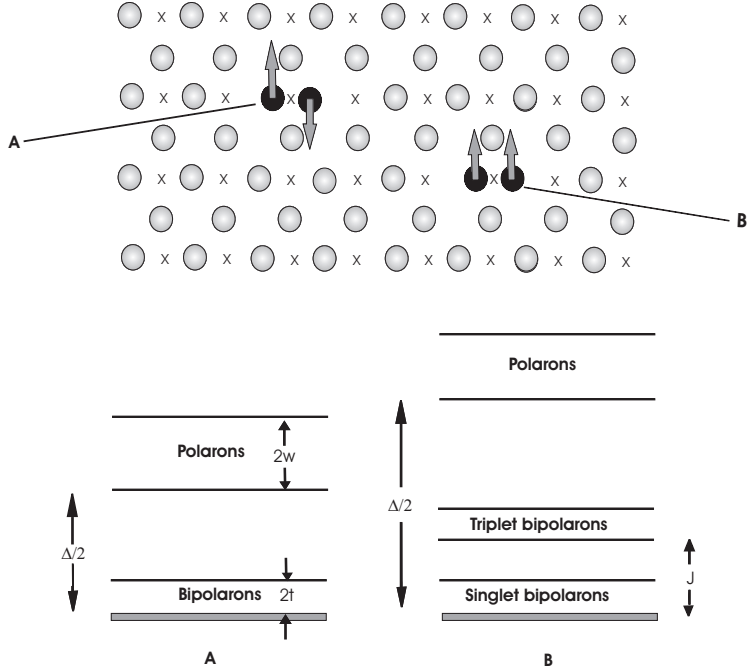}
\vskip -0.5mm \caption{Bipolaron picture of high temperature
superconductors. $A$ corresponds to a singlet intersite bipolaron.
$B$ is a triplet intersite bipolaron, which naturally includes the
addition of an extra excitation band. The crosses are copper sites
and the circles are oxygen sites, $w$ is a half bandwidth of the
polaron band, $t$ is a half bandwidth of the bipolaronic band,
$\Delta/2$ is the bipolaron binding energy per polaron and $J$ is
the exchange energy per bipolaron.}
\end{center}
\end{figure}
The population of singlet, $n_{s}$, triplet $n_{t}$, and polaron,
$n_{p}$ bands is determined by the chemical potential $\mu \equiv
T\ln y$, where $y$ is found using the thermal equilibrium of singlet
and triplet bipolarons and polarons,
\begin{equation}
2n_{s}+2n_{t}+n_{p}=x.
\end{equation}
Applying the effective-mass approximation for quasi-two-dimensional
energy spectra of all particles we obtain for $0< y <1$
\begin{equation}
-m_{s}^{\ast \ast }\ln \left( {1-y}\right) -3m_{t}^{\ast \ast }\ln \left( {%
1-ye^{-J/T}}\right) +m^{\ast }\ln \left( {1+y}^{1/2}{e^{-\Delta /(2T)}}%
\right) ={\frac{\pi x}{{T}}},
\end{equation}
in the normal state, and $y=1$ in the superconducting state. Here
$x$ is the total number of holes per unit area. If the polaron
energy spectrum is (quasi)one-dimensional, an additional $T^{-1/2}$
appears in front of $\ln$ in the third term on the left hand side of
Eq.(59).

\subsection{Role of disorder and the phase diagram of cuprates}
 We should also take into account localization of holes by
the random potential, because doping inevitably introduces some
disorder. The Coulomb repulsion restricts the number of charged
bosons in each localized state, so that the distribution function
will show a mobility edge $E_{c}$ \cite{alebramot}. The number of
bosons in a single potential well is
determined by the competition between their long-range Coulomb repulsion c.a. $%
 {4e^{2}/(\epsilon_0 \xi) }$ and the binding energy $E_{c}-\epsilon $. If
the
localization length diverges with the critical exponent $\nu <1$ $(\xi \sim {%
(E_{c}-\epsilon )^{-\nu }})$, we can apply a `single well-single
particle' approximation assuming that there is only one boson in
each potential well. Within this approximation $localized$ $charged$
bosons effectively obey the Fermi-Dirac statistics, so that their
density is given by
\begin{equation}
n_{L}(T)=\int_{-\infty }^{E_{c}}{\frac{N_{L}(E)dE}{y^{-1}{\exp
(E/T)+1}}},
\end{equation}
where $N_{L}(E)$ is the density of localized states. Near the
mobility edge it remains constant $N_{L}(E)\approx {n_{L}/\Gamma ,}$
where ${\Gamma }$ is of the order of the binding energy in a single
random potential well, and
$n_{L}$ is the number of localized states per unit area. The number of $%
empty $ localized states turns out to be linear as a function of
temperature in a wide temperature range $T < \Gamma $ from Eq.(60).
Then the conservation of the total number of carriers yields for the
chemical potential:
\begin{eqnarray}
\frac{\pi (x-2n_{L})}{T} &=&-m_{s}^{\ast \ast }\ln \left(
{1-y}\right)
-3m_{t}^{\ast \ast }\ln \left( {1-ye^{-J/T}}\right) + \\
&&m^{\ast }\ln \left( {1+y}^{1/2}{e^{-\Delta /(2T)}}\right)
-{\frac{2\pi n_{L}}{{\Gamma }}}\ln (1+y^{-1}).  \nonumber
\end{eqnarray}
If the number of localized states is about the same as the number of pairs, $%
n_{L}\approx x/2$, a solution of this equation does not depend on
temperature in a wide temperature range above $T_{c}$. With $y$ to
be a constant ($y \approx 0.6$ in a wide range of parameters in
Eq.(61)), the number of singlet bipolarons in the Bloch states is
linear in temperature,
\begin{equation}
n_{s}(T)=(x/2-n_{L})+T{\frac{n_{L}}{{\Gamma }}}\ln (1+y^{-1}).
\end{equation}
The numbers of triplet pairs and single polarons are exponentially
small at low temperatures, $T\ll J,\Delta/2.$

\begin{figure}
\begin{center}
\includegraphics[angle=-90,width=0.80\textwidth]{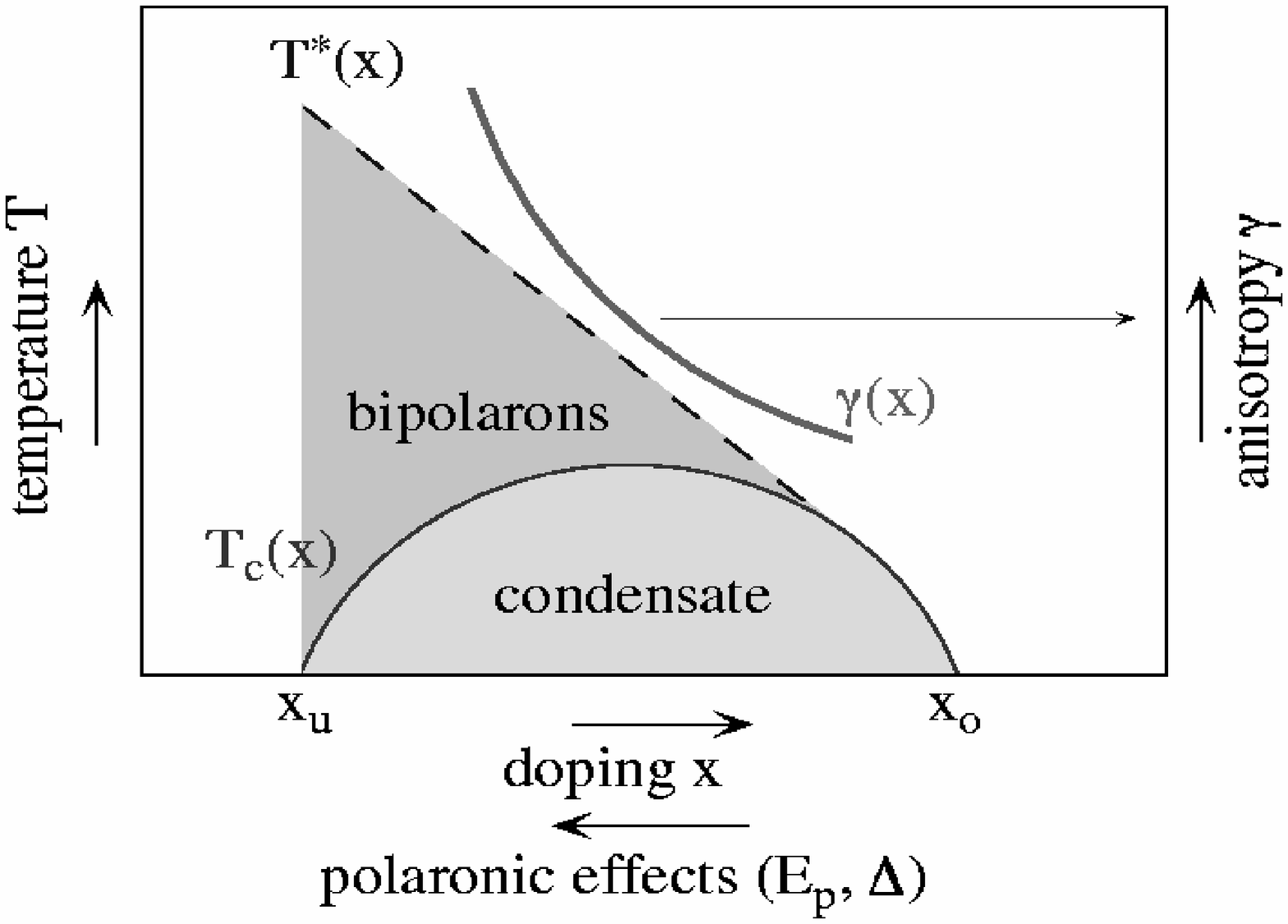}
\vskip -0.5mm
 \caption{Phase diagram of superconducting cuprates in
the bipolaron theory (courtesy of J. Hofer).}
\end{center}
\end{figure}
The model suggests a phase diagram of the cuprates as shown in
Fig.13. This phase diagram is based on the assumption that to
account for the high values of $T_{c}$ in cuprates one has to
consider electron-phonon interactions larger than those used in the
BSC-Eliashberg theory of superconductivity. Regardless of the
adiabatic ratio, the Migdal-Eliashberg theory of superconductivity
and the Fermi-liquid theory break at $\lambda \approx 1$. The
many-electron system collapses into the small (bi)polaron regime at
$\lambda \geq 1$ with well separated vibrational and charge-carrier
degrees of freedom. Even though it seems that these carriers should
have a mass too large to be mobile, the inclusion of the on-site
Coulomb repulsion and a poor screening of the long-range
electron-phonon interaction leads to $mobile$ intersite bipolarons
as discussed above. In the normal state the Bose gas of the
bipolarons is non-degenerate and below $T_{c}$ their phase coherence
sets in and hence superfluidity of the double-charged $2e$ bosons
occurs. There are also thermally excited single polarons and
triplets in the model, Fig.12, which are responsible   for the
crossover regime at $T^{\ast }$ and  normal state charge and spin
(pseudo)gaps in cuprates. These pseudogaps were predicted as half of
the binding energy $\Delta $ and the singlet-triplet exchange energy
$J$ of preformed bipolarons, respectively \cite{alegap}. $T^{\ast }$
is a temperature, where the polaron density compares with the
bipolaron one.

\subsection{Low Fermi energy: individual pairing in cuprates}
 Clear evidence for bipolarons comes from a
parameter-free estimate \cite{alefermi} of the renormalized
Fermi-energy $\epsilon _{F}$, which yields a very small value in
cuprates, where the band structure is quasi-two-dimensional with a
few degenerate hole pockets. Applying the parabolic approximation
for the band dispersion we obtain the {\it renormalized} Fermi
energy as
\begin{equation}
\epsilon _{F}={\frac{\pi n_{i}d}{{m_{i}^{\ast }}}},
\end{equation}
where $d$ is the interplane distance, and $n_{i},m_{i}^{\ast }$ are
the density of holes and their effective mass in each of the hole
subbands $i$ renormalized by the electron-phonon interaction. One
can express the renormalized band-structure parameters through the
in-plane magnetic-field penetration depth at $T\approx 0$, measured
experimentally:
\begin{equation}
\lambda _{H}^{-2}=4\pi e^{2}\sum_{i}\frac{n_{i}}{m_{i}^{\ast }}.
\end{equation}
As a result, we obtain a {\em parameter-free }expression for the
``true'' (i.e. renormalized) Fermi energy as
\begin{equation}
\epsilon _{F}={\frac{d}{{4ge^{2}\lambda _{H}^{2}}}},
\end{equation}
where $g$ is the degeneracy of the spectrum, which  may depend on
doping in cuprates. One expects $4$ hole pockets inside  BZ due to
the Mott-Hubbard gap in underdoped cuprates. If the hole band minima
are shifted with doping to BZ boundaries, all their wave vectors
would belong to the stars with two or more prongs. The groups of
wave vectors for these stars have only 1D representations. It means
that the spectrum will be degenerate with respect to the number of
prongs which
the star has, i.e $g \geq 2$. The only exception is the minimum at ${\bf %
k}=(\pi ,\pi )$ with one prong and $g=1$. Hence, in cuprates the
degeneracy is $1\leq g \leq 4$. Because Eq.(65) does not contain any
other band-structure parameters, the estimate of $\epsilon _{F}$
using this equation does not depend very much on the parabolic
approximation for the band dispersion.

Generally, the ratios $n/m^{\ast }$ in Eq.(63) and in Eq.(64) are
not necessarily the same. The `superfluid' density in Eq.(64) might
be different from the total density of delocalized carriers in
Eq.(63). However, in a translation invariant system they must be the
same \cite{pop}. This is also true even in the extreme case of a
pure two-dimensional superfluid, where quantum fluctuations are
important. One can, however, obtain a reduced value of the zero
temperature superfluid density in the dirty limit, $l\ll \xi (0)$,
where $\xi (0)$ is the zero-temperature coherence length. The latter
was measured directly in cuprates as the size of the vortex core. It
is about 10 $\AA $ or even less. On the contrary, the mean free path
was found surprisingly large at low temperatures, $l\sim $ 100-1000
$\AA $. Hence, it is rather probable that all cuprate
superconductors are in the clean limit, $l\gg \xi (0)$, so  the
parameter-free expression for $\epsilon _{F}$, Eq.(65), is perfectly
applicable.

\begin{table}[tbp] \caption{The Fermi energy
(multiplied by the degeneracy) of cuprates}
\begin{tabular}[t]{llllllll}
Compound & $T_{c}$ (K) & $\lambda _{H,ab}$ $(\AA )$ & d$(\AA )$ & $%
g\epsilon_{F}$ (meV) &  &  &  \\ \hline
$La_{1.8}Sr_{0.2}CuO_{4}$ & 36.2 & 2000 & 6.6 & 112 &  &  &  \\
$La_{1.78}Sr_{0.22}CuO_{4}$ & 27.5 & 1980 & 6.6 & 114 &  &  &  \\
$La_{1.76}Sr_{0.24}CuO_{4}$ & 20.0 & 2050 & 6.6 & 106 &  &  &  \\
$La_{1.85}Sr_{0.15}CuO_{4}$ & 37.0 & 2400 & 6.6 & 77 &  &  &  \\
$La_{1.9}Sr_{0.1}CuO_{4}$ & 30.0 & 3200 & 6.6 & 44 &  &  &  \\
$La_{1.75}Sr_{0.25}CuO_{4}$ & 24.0 & 2800 & 6.6 & 57 &  &  &  \\
$YBa_{2}Cu_{3}O_{7}$ & 92.5 & 1400 & 4.29 & 148 &  &  &  \\
$YBa_{2}Cu_{3}O_{6.7}$ & 66.0 & 2100 & 4.29 & 66 &  &  &  \\
$YBa_{2}Cu_{3}O_{6.57}$ & 56.0 & 2900 & 4.29 & 34 &  &  &  \\
$YBa_{2}Cu_{3}O_{6.92}$ & 91.5 & 1861 & 4.29 & 84 &  &  &  \\
$YBa_{2}Cu_{3}O_{6.88}$ & 87.9 & 1864 & 4.29 & 84 &  &  &  \\
$YBa_{2}Cu_{3}O_{6.84}$ & 83.7 & 1771 & 4.29 & 92 &  &  &  \\
$YBa_{2}Cu_{3}O_{6.79}$ & 73.4 & 2156 & 4.29 & 62 &  &  &  \\
$YBa_{2}Cu_{3}O_{6.77}$ & 67.9 & 2150 & 4.29 & 63 &  &  &  \\
$YBa_{2}Cu_{3}O_{6.74}$ & 63.8 & 2022 & 4.29 & 71 &  &  &  \\
$YBa_{2}Cu_{3}O_{6.7}$ & 60.0 & 2096 & 4.29 & 66 &  &  &  \\
$YBa_{2}Cu_{3}O_{6.65}$ & 58.0 & 2035 & 4.29 & 70 &  &  &  \\
$YBa_{2}Cu_{3}O_{6.6}$ & 56.0 & 2285 & 4.29 & 56 &  &  &  \\
$HgBa_{2}CuO_{4.049}$ & 70.0 & 2160 & 9.5 & 138 &  &  &  \\
$HgBa_{2}CuO_{4.055}$ & 78.2 & 1610 & 9.5 & 248 &  &  &  \\
$HgBa_{2}CuO_{4.055}$ & 78.5 & 2000 & 9.5 & 161 &  &  &  \\
$HgBa_{2}CuO_{4.066}$ & 88.5 & 1530 & 9.5 & 274 &  &  &  \\
\hline
\end{tabular}
\end{table}

A parameter-free estimate of the Fermi energy obtained by using
Eq.(65) is presented in Table 1. The renormalized Fermi energy  in
many cuprates is less than $100$ $meV$, if the degeneracy $g\geq 2$
is taken into account. That should be compared with the
characteristic phonon frequency, which is estimated as the plasma
frequency of oxygen ions,
\begin{equation}
\omega =\sqrt{\frac{4\pi Z^{2}e^{2}N}{M}}.
\end{equation}
One obtains  $\omega $=$84meV$ with $Z=2$, $N=6/V_{cell}$, $%
M=16a.u.$ for $YBa_{2}Cu_{3}O_{6}$. Here $V_{cell}$ is the volume of
the (chemical) unit cell. The estimate agrees with the measured
phonon spectra. As established experimentally in cuprates (see the
introduction)  high-frequency optical phonons are strongly coupled
with holes. A low Fermi energy is a serious problem for the
Migdal-Eliashberg approach. Since the Fermi energy is small and
phonon frequencies are high, the Coulomb pseudopotential $\mu
_{c}^{\ast }$ is of the order of the bare Coulomb repulsion, $\mu
_{c}^{\ast }\simeq \mu _{c}\simeq 1$ since the
Tolmachev-Morel-Anderson logarithm is ineffective. Hence, to get a
pairing, one has to have a strong coupling, $\lambda >\mu _{c}$.
However, one cannot increase $\lambda $ without accounting for the
polaron collapse of the band. Even in the region of the
applicability of the BCS-Eliashberg theory (i.e. at $\lambda \leq
0.5$), the non-crossing diagrams cannot be treated as vertex
$corrections$ like in Ref.\cite{pie}, since they are comparable to
the standard terms, when $\omega/\epsilon _{F}\geq 1$.

In many cases (Table 1) the renormalized Fermi energy is so small
that pairing is certainly individual, i.e. the bipolaron size is
smaller than the inter-carrier distance. Indeed, this is the case,
if
\begin{equation}
\epsilon _{F}\leq \pi \Delta .
\end{equation}
If the bipolaron binding energy is  twice of the pseudogap
experimentally measured in the normal state of many cuprates \cite{kabmic}, $%
\Delta > 100meV,$  Eq.(67) is well satisfied in underdoped and even
in a few optimally and overdoped cuprates. One should notice that
the coherence length in a charged Bose gas has nothing to do with
the size of the boson. It depends on the interparticle distance and
the mean-free path \cite{alebook1}, and might be as large as in the
BCS superconductor. Hence, it would be incorrect to apply the ratio
of the coherence length to the inter-carrier distance as a criterium
of the BCS-Bose liquid crossover. The correct criterium is given by
Eq.(67).

\section{Normal state properties of cuprates in FCM}

The low-energy FCM electronic structure  of  cuprates  is shown in
Fig.12 \cite{alemot3}. Polaronic p-holes  are bound in lattice
inter-site singlets (A) or in singlets and  triplets (B) (if spins
are included in Eq.(2)) at any temperature. Above T$_{c}$ a charged
bipolaronic  Bose liquid is non-degenerate and below $T_{c}$ phase
coherence (ODLRO) of the preformed bosons sets in. The state above
$T_{c}$ is perfectly "normal" in the sense that the off-diagonal
order parameter (i.e. the Bogoliubov-Gor'kov anomalous average
$\cal{F}(\mathbf{r,r^{\prime }})=\langle \psi_{\downarrow
}(\mathbf{{r})\psi _{\uparrow }({r^{\prime }}\rangle}$) is zero
above the resistive transition temperature $T_{c}$. Here
$\psi_{\downarrow,\uparrow }(\mathbf{r})$ annihilates  electrons
with spin $\downarrow, \uparrow$ at point ${\bf r}$. (Bi)polarons
and thermally excited phonons  are well decoupled in the
strong-coupling regime of the electron-phonon interaction, so
 the conventional Boltzmann kinetics for mobile polaronic and bipolaronic carries is
applied.

\subsection{Normal state in-plane resistivity, the Hall effect, magnetic susceptibility and
the Lorenz number } A nonlinear temperature dependence of the
$in$-plane resistivity below $T^{\ast }$, a temperature-dependent
paramagnetic susceptibility, and a peculiar maximum in the Hall
ratio well above $T_c$ have remained  long-standing problems of
cuprate physics. The bipolaron model provides their quantitative
description \cite{alebramot,jung,alezavdzu}. Here we use a `minimum'
bipolaron model Fig.12A, which includes the singlet bipolaron band
and the spin 1/2 polaron band separated by $T^{\ast }$, and the
$\tau -$approximation   in weak electric ${\bf E}$ and  magnetic
fields, $\bf B\perp E$ \cite{alezavdzu}. Bipolaron and
single-polaron non-equilibrium distributions are found as
\begin{equation}
f({\bf k})=f_{0}(E)+\tau \frac{\partial f_{0}}{\partial E}{\bf
v}\cdot \left\{ {\bf F}+\Theta {\bf n}\times {\bf F}\right\} ,
\label{4}
\end{equation}
where ${\bf v=}\partial E/\partial {\bf k,}$ ${\bf
F}={\vec{\bf\nabla}}(\mu -2e\phi )$, $f_{0}(E)=[y^{-1}{\exp
(E/T)-1]}^{-1}$ and the Hall angle $\Theta =\Theta _{b}=2eB\tau
_{b}/m_{b}$ for bipolarons with the energy $E=k^{2}/(2m_{b})$, and
${\bf F}={\vec{\bf\nabla}}(\mu /2-e\phi ) $,
$f_{0}(E)=\{y^{-1/2}{\exp [(E+T^{\ast })/T]+1\}}^{-1}$,
$E=k^{2}/(2m_{p})$, and $\Theta =\Theta _{p}=eB\tau _{p}/m_{p}$ for
thermally excited polarons. Here $m_{b}$ and $m_p$ are the bipolaron
and polaron mass, respectively, $y=\exp
(\mu /T),$ $\mu $ is the chemical potential,  and ${\bf %
n=B/}B$ is a unit vector in the direction of the magnetic field.
Eq.(68) is used to calculate the electrical resistivity and the Hall
ratio as
\begin{eqnarray}
\rho &=&\frac{m_{b}}{4e^{2}\tau _{b}n_{b}(1+An_{p}/n_{b})}, \\
R_{H} &=&\frac{1+2A^{2}n_{p}/n_{b}}{2en_{b}(1+An_{p}/n_{b})^{2}},
\end{eqnarray}
where $A=\tau _{p}m_{b}/(4\tau _{b}m_{p})$. The atomic densities of
quasi two-dimensional carriers are found as
\begin{eqnarray}
n_{b}=\frac{m_{b}T}{2\pi }|\ln (1-y)|, \\
n_{p}=\frac{m_{p}T}{\pi }\ln \left[ 1+y^{1/2}\exp \left( -T^{\ast }/T\right) %
\right] .
\end{eqnarray}
and the chemical potential is determined by doping $x$ using
$2n_{b}+n_{p}=x-n_{L}$, where $n_{L}$ is the number of carriers
localized by disorder (here we take the lattice constant $a=1$).
\begin{figure}
\begin{center}
\includegraphics[angle=-0,width=0.40\textwidth]{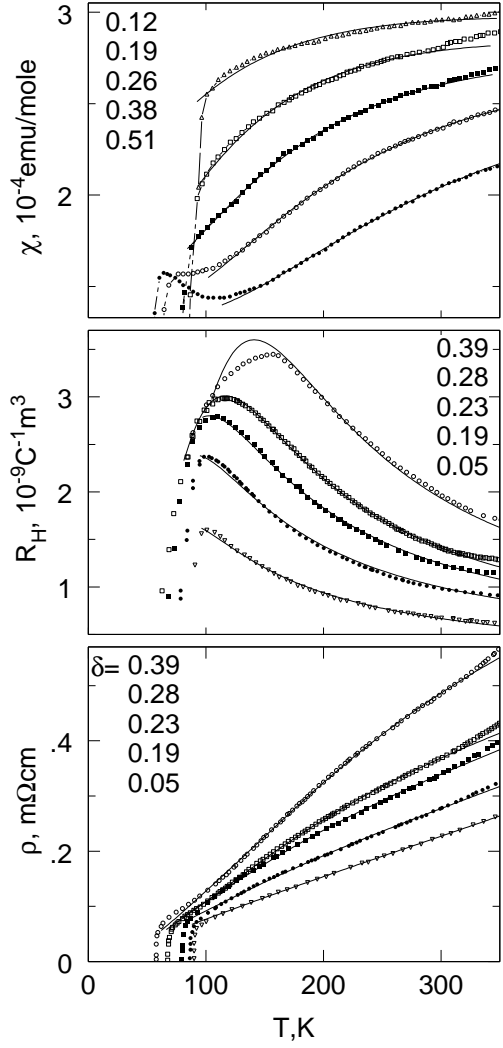}
\vskip -0.5mm \caption{Uniform magnetic susceptibility, $\chi(T)$,
Hall ratio, $R_H(T)$ and resistivity, $\rho(T)$, of underdoped
$YBa_{2}Cu_{3}O_{7-\delta}$ \cite{coop} fitted by the theory; see
the Table below for parameters. }
\end{center}
\end{figure}
Polarons are not degenerate. Their number  remains small compared
with twice the number of bipolarons, $n_p/(2n_b)<0.2$, in the
relevant temperature range $T < T^{\ast }$, so that
\begin{equation}
 y\approx 1-\exp(-T_0/T),
\end{equation}
where $T_0=\pi (x-n_L)/m_b \approx T_c$ is about the superconducting
critical temperature of the (quasi)two-dimensional Bose gas. Because
of this reason, the experimental $T_c$ was taken as  $T_0$ in our
fits. Using Eqs.(70,71,72) we obtain
\begin{equation}
R_{H}(T)=R_{H0}\frac{1+2A^2y^{1/2}(T/T_{c})\exp \left( -T^{\ast
      }/T\right) }{
[1+A(T/T_{c})y^{1/2}\exp \left( -T^{\ast }/T\right) ]^{2}},
\end{equation}
where $R_{H0}=[e(x-n_{L})]^{-1}$.  If we assume that the number of
localized carriers depends only weakly on temperature in underdoped
cuprates since their average  ionization energy is sufficiently
large, then $R_{H0}$ is temperature independent at $T<T^{\ast }$. As
proposed in Ref.\cite{alebramot} the scattering rate  at relatively
high temperatures is due to inelastic
 collisions of itinerant carriers with those localized by
 disorder, so it is proportional to $T^2$. We also have to take into account the residual
 scattering of polarons by optical
 phonons, so that
$\tau^{-1}\propto (T/T_{1})^{2}+\exp \left( -\omega /T\right)$, if
the temperature is low compared with the characteristic phonon
energy $\omega$. The relaxation times of each type of carriers
scales with their charge $e^\ast$ and mass as $\tau_{p,b} \propto
m_{p,b}^{-3/2}(e^\ast)^{-2}$, so we estimate $A=(m_b/m_p)^{5/2}
\approx 6$ if we take $m_{b}\approx 2m_p$ . As a result  the
in-plane resistivity is given by
\begin{equation}
\rho (T)=\rho _{0}\frac{(T/T_{1})^{2}+\exp \left( -\omega /T\right)
}{[1+A(T/T_{c})y^{1/2}\exp \left( -T^{\ast }/T\right) ]},
\end{equation}
where $\rho _{0}\sim m_{b}/[2e^{2}(x-n_L)]$ and $T_{1}$ are
temperature independent. Finally, one can easily obtain the uniform
magnetic susceptibility due to nondegenerate spin 1/2 polarons as
\cite{AKM}
\begin{equation}
\chi (T)=By^{1/2}\exp \left( -T^{\ast }/T\right) +\chi _{0},
\end{equation}
where $B=(\mu _{B}^{2}m_{p}/\pi )$, and $\chi_{0} $ is the magnetic
susceptibility of the parent Mott insulator.
\begin{table}[tbp]
\begin{tabular}{|c|c||c|c|c||c|c|c|c|c|}
$\delta$ &$T_c$& $\rho_0$&$R_{H0}$ &$10^4B$ & $10^4\chi_0$ & $T^*$& $\omega$  & $T_1$   \\
& K & $m\Omega cm$&$\frac{10^{-9}m^3}{C}$& $\frac{emu}{mole}$ &
$\frac{emu}{mole}$ &K&K  &K
\\ \hline
  0.05 &  90.7 & 1.8  & 0.45&&& 144&447 & 332
\\ \hline
  0.12 &   93.7 &&&  2.6 &2.1&155 &&
\\ \hline
  0.19 &  87 &3.4 & 0.63 &4.5&1.6& 180&477& 454
\\ \hline
  0.23 &  80.6& 5.7 & 0.74&&&210& 525  & 586
\\ \hline
  0.26 &   78 & & &5.4 &1.5&259&&
\\ \hline
  0.28 & 68.6 & 8.9& 0.81&&&259& 594  & 786
\\ \hline
  0.38 & 61.9 && &7.2 &1.4& 348&&
\\ \hline
  0.39 & 58.1& 17.8& 0.96&&&344& 747 & 1088
\\ \hline
  0.51 & 55 && & 9.1 &1.3& 494&&
\\ \hline
\end{tabular}
\end{table}
 Our model numerically fits the Hall ratio,
$R_H(T)$, the in-plane resistivity, $\rho(T)$, and the magnetic
susceptibility $\chi(T)$ of $YBa_{2}Cu_{3}O_{7-\delta }$ within the
physically relevant range of all parameters (see Fig.14 and the
Table). The ratio of polaron and bipolaron mobilities $A=7$ used in
all fits is close to the above estimate,  and $\chi _{0}\approx
1.5\times 10^{-4}emu/mole$ is very close to the susceptibility of a
slightly doped insulator \cite{coop}. The maximum of $R_{H}(T)$ is
due to the contribution of thermally excited polarons into
transport, and the temperature dependence of the in-plane
resistivity below $T^{\ast }$ is due to this contribution and the
combination of the carrier-carrier and carrier-phonon scattering.
The characteristic phonon frequency from the resistivity fit (see
the Table) decreases with doping and the pseudogap $T^{\ast }$ shows
the doping behavior as observed in other independent experiments
\cite{kabmic}.

Notwithstanding our explanation of the Hall ratio, the in-plane
resistivity and the bulk magnetic susceptibility might be not so
convincing as a direct measurement of the double
 charge $2e$ on carriers in the normal state. In
1993, we  discussed the thermal conductivity of preformed bosons
\cite{NEV}. The contribution from  carriers to the thermal transport
provided by the Wiedemann-Franz law depends strongly on the
elementary charge as $\sim(e^{\ast })^{-2}$ and should be
significantly suppressed if $e^{\ast}=2e$. The Lorenz number, $L$,
has been directly measured in $YBa_{2}Cu_{3}O_{6.95}$ by Zhang et
al. \cite{zha} using the thermal Hall conductivity. Remarkably, the
measured value of $L$ just above $T_{c}$ was found just the same as
predicted by the bipolaron model \cite{NEV}, $L\approx 0.15L_{e}$,
where $L_{e}$ is the conventional Fermi-liquid Lorenz number. The
breakdown of the Wiedemann-Franz law has been also explained in the
framework of the bipolaron model \cite{leeale}.

\subsection{Normal-state Nernst effect}
In disagreement with the
weak-coupling BCS  and the strong-coupling
 bipolaron theories a significant fraction of research in the field
of high-temperature superconductivity suggests that the
superconducting transition is only a phase ordering while the
superconducting order parameter $\cal{F}(\mathbf{r,r^{\prime }})$
remains nonzero above the resistive $T_c$. One of the key
experiments supporting this viewpoint is the large Nernst signal
observed  in the normal (i.e. resistive) state of cuprates  (see
Ref. \cite{xu,cap,cap2} and references therein). Some authors
\cite{xu,ong}  claim that numerous resistive determinations of the
upper critical field, $H_{c2}(T)$ in cuprates have been misleading
since the Nernst signal \cite{xu} and the diamagnetic magnetization
\cite{ong} imply that  $H_{c2}(T)$  remains large at $T_c$ and
above. They propose a "vortex scenario", where the long-range phase
coherence is destroyed by mobile vortices, but the amplitude of the
off-diagonal order parameter remains finite and  the Cooper pairing
with a large binding energy exists  well above $T_c$ supporting the
so-called  "preformed Cooper-pair" or "phase fluctuation"  model
\cite{kiv}. The model is  based on the assumption that the
superfluid density is small compared with  the normal carrier
density in cuprates.  These interpretations  seriously undermine
many theoretical and experimental  works on superconducting
cuprates, which consider the state above $T_c$ as perfectly normal
with no off-diagonal order, either long or short.

We believe that  the vortex (or phase fluctuation) scenario
contradicts  straightforward resistive and other measurements, and
it is theoretically inconsistent. This scenario is impossible to
reconcile with the extremely sharp resistive transitions at $T_c$ in
high-quality underdoped, optimally doped  and overdoped cuprates.
For example, the in-plane and out-of-plane resistivity of $Bi-2212$,
where the anomalous Nernst signal has been measured \cite{xu}, is
perfectly "normal" above $T_c$, Fig.15, showing only a few percent
positive or negative magnetoresistance \cite{zavale}, explained with
bipolarons \cite{zavalemos}.
\begin{figure}
\begin{center}
\includegraphics[angle=-0,width=0.75\textwidth]{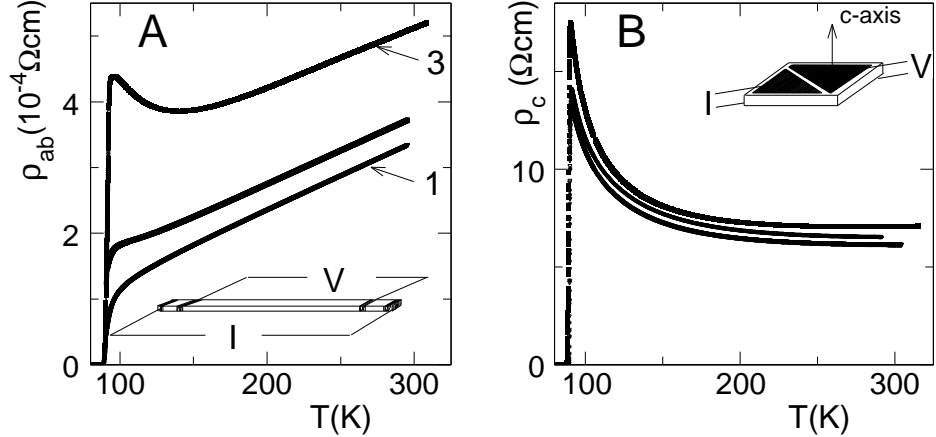}
\vskip -0.5mm \caption{In-plane (A) and out-of-plane (B) resistivity
of 3   single crystals of Bi$_2$Sr$_2$CaCu$_2$O$_8$ \cite{zavale}
showing no signature of phase fluctuations  above the resistive
transition. }
\end{center}
\end{figure}
Both in-plane \cite{buc,mac0,boz,lawrie,gan} and out-of-plane \cite
{alezavnev,hof2,zve} resistive transitions  of high-quality samples
are sharp and remain sharp in the magnetic field providing a
reliable determination of the genuine $H_{c2}(T)$. The vortex
entropy \cite{cap} estimated from the Nernst signal is an order of
magnitude smaller than the difference between the entropy of the
superconducting state and the extrapolated entropy of the normal
state obtained from the specific heat. The preformed Cooper-pair
model \cite{kiv}  is incompatible with a great number of
thermodynamic, magnetic, and kinetic measurements, which show that
only holes (density x), doped into a parent insulator are carriers
\emph{both} in  the normal and the superconducting states of
cuprates. The assumption \cite{kiv} that the superfluid density is
small compared with the normal-state carrier density is also
inconsistent with the theorem \cite{pop}, which proves that the
number of supercarriers at $T=0$K  should be the same as the number
of normal-state carriers in any  clean superfluid.

Recently we described the unusual Nernst signal in cuprates in a
different manner as the  normal state phenomenon \cite{alezav}. We
have also extended this description to  cuprates with very low
doping level accounting  for their Nernst signal, the thermopower
and the insulating-like in-plane low temperature resistance
\cite{alecon} as observed \cite{xu,cap,cap2}.

Thermomagnetic effects appear in conductors subjected to a
longitudinal temperature gradient $\nabla _{x}T$ in $x$ direction
and a perpendicular magnetic field  in $z$ direction. The transverse
Nernst-Ettingshausen effect \cite{nernst}  (here the Nernst effect)
is the appearance of a transverse electric field $E_y$ in the third
direction. When bipolarons are formed in the strong-coupling regime,
the chemical potential is  negative, Eq.(73). It is found in the
impurity band just below the mobility edge at $T>T_c$. Carriers,
localized  below the mobility edge contribute to the longitudinal
transport together with the itinerant carriers in extended states
above the mobility edge. Importantly the contribution of localized
carriers of any statistics to the \emph{ transverse} transport is
normally small \cite{ell} since a microscopic Hall voltage will only
develop at junctions in the intersections of the percolation paths,
and it is expected that these are few for the case of hopping
conduction among disorder-localized states \cite{mott2}. Even if
this contribution is not negligible, it adds  to the contribution of
itinerant carriers to produce a large Nernst signal,
$e_{y}(T,B)\equiv -E_{y}/\nabla _{x}T$, while it reduces the
thermopower $S$ and the Hall angle $\Theta$. This unusual "symmetry
breaking" is completely at variance with  ordinary metals where the
familiar "Sondheimer" cancelation \cite{sond} makes  $e_{y}$ much
smaller than $S\tan \Theta$ because of the electron-hole symmetry
near the Fermi level. Such  behavior originates in the "sign" (or
"$p-n$") anomaly of the Hall conductivity of localized carriers. The
sign of their Hall effect  is often $opposite$ to that of the
thermopower as observed in many amorphous semiconductors \cite{ell}
and described theoretically \cite{fri}.

The Nernst signal is expressed in terms of the kinetic coefficients
$\sigma _{ij}$ and $\alpha _{ij}$ as
\begin{equation}
e_{y}={\frac{{\sigma _{xx}\alpha _{yx}-\sigma _{yx}\alpha
_{xx}}}{{\sigma _{xx}^{2}+\sigma _{xy}^{2}}}},
\end{equation}
where the current density  is given by $j_{i}=\sigma
_{ij}E_{j}+\alpha _{ij}\nabla _{j}T$.
 When the chemical potential $\mu$ is at the mobility edge,
 localized carriers contribute to the transport,
 so  $\sigma _{ij}$ and $\alpha _{ij}$ in Eq.(77) can be expressed
as $\sigma^{ext} _{ij}+\sigma^{l}_{ij}$ and $\alpha^{ext}
_{ij}+\alpha^{l}{ij}$, respectively. Since the Hall mobility of
carriers localized below $\mu$, $\sigma^{l}_{yx}$, has the  sign
opposite to that of carries in the extended states above $\mu$,
$\sigma^{ext}_{yx}$, the sign of the off-diagonal Peltier
conductivity $\alpha^{l}_{yx}$ should be the same as the sign of
$\alpha^{ext}_{yx}$. Then  neglecting the magneto-orbital effects in
the resistivity (since $\Theta \ll 1$ \cite{xu}) we obtain
\begin{equation}
S\tan \Theta \equiv {\sigma _{yx}\alpha _{xx}\over{\sigma
_{xx}^{2}+\sigma _{xy}^{2}}} \approx\rho (\alpha ^{ext}_{xx}-|\alpha
^{l}_{xx}|) (\Theta^{ext}-|\Theta^{l}|)
\end{equation}
and
\begin{equation}
e_{y}\approx\rho (\alpha^{ext} _{yx}+|\alpha^{l} _{yx}|)-S\tan
\Theta,
\end{equation}
where $\Theta^{ext}\equiv \sigma^{ext}_{yx}/\sigma_{xx}$,
$\Theta^{l}\equiv \sigma^{l}_{yx}/\sigma_{xx}$, and
$\rho=1/\sigma_{xx}$ is the resistivity.

\begin{figure}
\begin{center}
\includegraphics[angle=-90,width=0.45\textwidth]{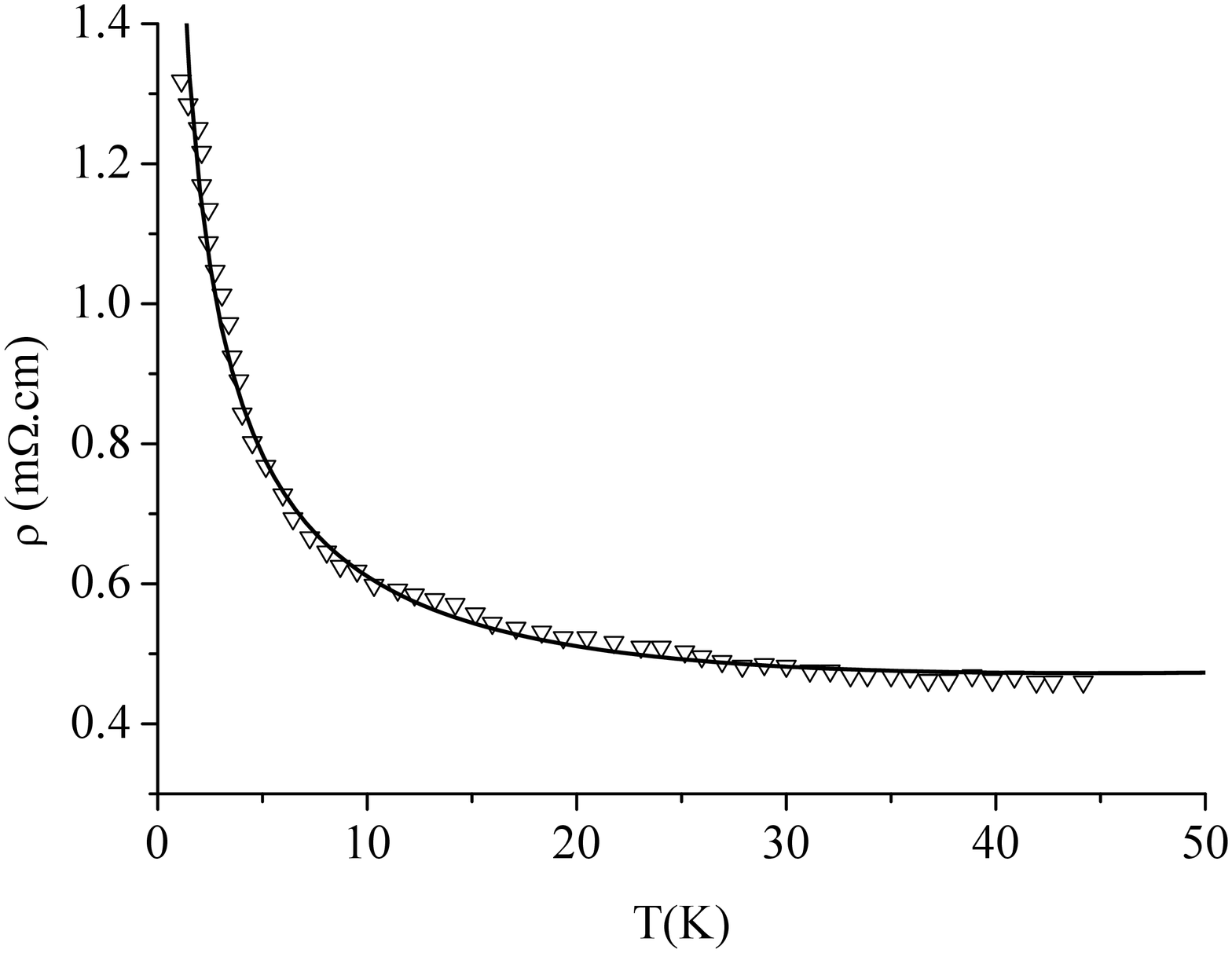}
\vskip -0.5mm \caption{Normal state in-plane resistivity of
underdoped La$_{1.94}$ Sr$_{0.06}$CuO$_4$ (triangles \cite{cap}) as
revealed in the field $B=12$ Tesla  and compared with the bipolaron
theory, Eq.(82) (solid line).}
\end{center}
\end{figure}
Clearly the model, Eqs.(78,79) can account for a low value of
$S\tan\Theta $ compared with a large value of $e_y$ in some
underdoped cuprates \cite{xu,cap2} due to the sign anomaly. Even in
the case when  localized bosons contribute little to the
conductivity   their contribution to the thermopower $S=\rho (\alpha
^{ext}_{xx}-|\alpha ^{l}_{xx}|))$ could almost cancel  the opposite
sign contribution of itinerant carriers \cite{alezav}. Indeed the
longitudinal conductivity of itinerant two-dimensional bosons,
$\sigma^{ext} \propto \int_0 dE E df(E)/dE$ diverges logarithmical
when $\mu$ in the Bose-Einstein distribution function
$f(E)=[\exp((E-\mu)/T)-1]^{-1}$ goes to zero and the relaxation time
$\tau$ is a constant. At the same time $\alpha^{ext}_{xx}\propto
\int_0 dE E(E-\mu) df(E)/dE$ remains finite, and it could have the
magnitude comparable   with  $\alpha^{l}_{xx}$. Statistics of
bipolarons  gradually changes from Bose to Fermi statistics with
lowering energy across the mobility edge because of the Coulomb
repulsion of bosons in localized states  \cite{alegile}. Hence one
can use the same expansion near the mobility edge as in  ordinary
amorphous semiconductors to obtain the familiar textbook result
$S=S_0T$ with a constant $S_0$ at low temperatures \cite{mott3}. The
model becomes particularly simple, if we   neglect the localized
carrier contribution to $\rho$, $\Theta$ and $\alpha_{xy}$, and take
into account that $\alpha^{ext}_{xy} \propto B/\rho^2$ and
$\Theta^{ext}\propto B/\rho$ in the Boltzmann theory. Then
Eqs.(78,79) yield
\begin{equation}
S\tan \Theta  \propto T/\rho
\end{equation}
and
\begin{equation}
e_{y}(T,B)\propto (1-T/T_1)/\rho.
\end{equation}
According to our earlier suggestion \cite{alelog} the
insulating-like low-temperature dependence of $\rho(T)$ in
underdoped cuprates  originates from the elastic scattering of
nondegenerate itinerant carriers off charged  impurities. We assume
here that the carrier density is temperature independent at low
temperatures in agreement  with the temperature-independent Hall
effect \cite{per}. The relaxation time of nondegenerate carriers
depends on temperature as $\tau \propto T^{-1/2}$ for scattering off
short-range deep potential wells, and as $T^{1/2}$ for very shallow
wells \cite{alelog}. Combining both scattering rates we obtain
\begin{equation}
\rho =\rho_0[(T/T_2)^{1/2}+(T_2/T)^{1/2}].
\end{equation}
Eq.(82) with $\rho_0=0.236$ m$\Omega\cdot$cm and $T_2=44.6$K fits
extremely well the experimental insulating-like normal state
resistivity of underdoped La$_{1.94}$ Sr$_{0.06}$CuO$_4$ in the
whole low-temperature range from  2K up to 50K, Fig.16,  as revealed
in the field $B=12$ Tesla \cite{cap,cap2}. Another high quality fit
can be  obtained combining the Brooks-Herring formula for the 3D
scattering off screened charged impurities, as proposed in
Ref.\cite{kast} for almost undoped $LSCO$, or the Coulomb scattering
in 2D ($\tau \propto T$) and a temperature independent scattering
rate off neutral impurities with the carrier exchange \cite{erg}
similar to the scattering of slow electrons by hydrogen atoms in
three dimensions. Hence the scale $T_2$, which determines the
crossover toward an insulating behavior, depends on the relative
strength of two scattering mechanisms. Importantly our expressions
(80,81) for  $S\tan \Theta$ and $e_y$ do not depend on the
particular scattering mechanism. Taking into account the excellent
fit of Eq.(82) to the experiment, they can be parameterized as
\begin{equation} S\tan \Theta = e_0
{(T/T_2)^{3/2}\over{1+T/T_2}},
\end{equation}
and
\begin{equation}
e_{y}(T,B)=e_0{(T_1-T) (T/T_2)^{1/2}\over{T_2+T}} ,
\end{equation}
where $T_1$ and $e_0$ are temperature independent.
\begin{figure}
\begin{center}
\includegraphics[angle=270,width=0.70\textwidth]{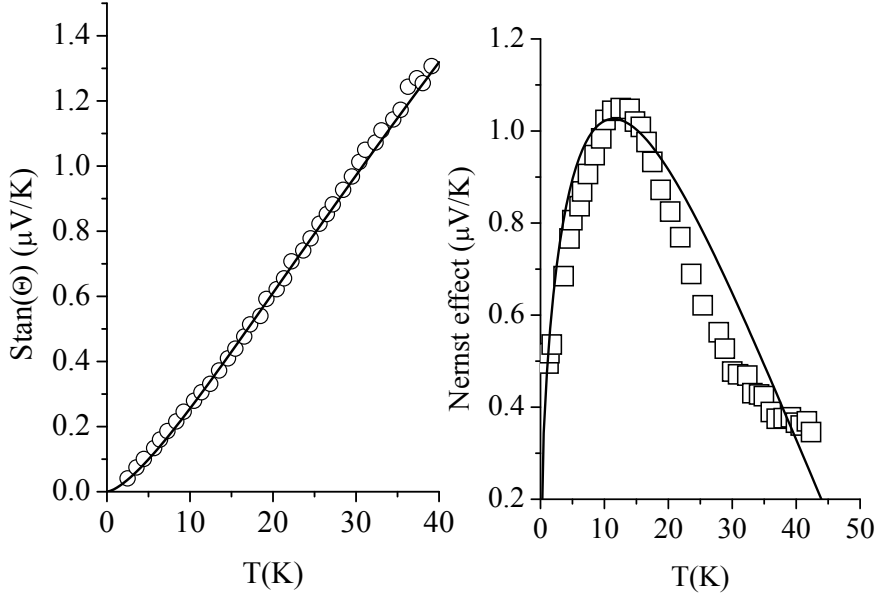}
\vskip -0.5mm \caption{$S\tan\Theta$ (circles \cite{cap2} )  and the
Nernst effect $e_y$  (squares \cite{cap})  of underdoped La$_{1.94}$
Sr$_{0.06}$CuO$_4$ at $B=12$ Tesla compared with the bipolaron
theory, Eqs.(18,19) (solid lines).}
\end{center}
\end{figure}

In spite of all simplifications, the model describes  remarkably
well both $S\tan \Theta$ and $e_y$  measured in La$_{1.94}$
Sr$_{0.06}$CuO$_4$ with a $single$ fitting parameter, $T_1=50$K
using the experimental $\rho(T)$. The constant  $e_0=2.95$ $\mu$V/K
scales the magnitudes of $S\tan \Theta$ and $e_y$.  The magnetic
field $B=12$ Tesla destroys the superconducting state of the
low-doped La$_{1.94}$ Sr$_{0.06}$CuO$_4$ down to $2$K, Fig.16, so
any residual superconducting order above $2$K is clearly ruled out,
while the Nernst signal, Fig.17, is remarkably large. The
coexistence of the large Nernst signal and a nonmetallic resistivity
is in sharp disagreement with the vortex scenario, but in agreement
with our model. Taking into account the field dependence of the
conductivity of localized carriers, the phonon-drug effect, and
their contribution to the transverse magnetotransport  can well
describe the magnetic field dependence of the Nernst signal
\cite{alezav} and improve the fit in Fig.17 at the expense of the
increasing number of fitting parameters.

\subsection{Normal state diamagnetism}
A number of experiments (see, for example,
\cite{mac,junM,hof,nau,igu,ong} and references therein), including
torque magnetometries, showed enhanced diamagnetism above $T_c$,
which has been explained as the fluctuation diamagnetism in quasi-2D
superconducting cuprates (see, for example Ref. \cite{hof}). The
data taken at relatively low magnetic fields (typically below 5
Tesla) revealed a crossing point in the magnetization $M(T,B)$ of
most anisotropic cuprates (e.g. $Bi-2212$), or in $M(T,B)/B^{1/2}$
of less anisotropic $YBCO$ \cite{junM}. The dependence of
magnetization (or $M/B^{1/2}$) on the magnetic field has been shown
to vanish at some characteristic temperature below $T_c$. However
the data taken in high magnetic fields (up to 30 Tesla) have shown
that the crossing point, anticipated for low-dimensional
superconductors and associated with superconducting fluctuations,
does not explicitly exist in magnetic fields above 5 Tesla
\cite{nau}.

Most surprisingly the torque magnetometery  \cite{mac,nau} uncovered
a diamagnetic signal somewhat above $T_c$ which increases in
magnitude with applied magnetic field. It has been  linked with the
Nernst signal and mobile vortexes   in the  normal state of cuprates
\cite{ong}. However, apart from the inconsistences mentioned above,
the vortex scenario of the normal-state diamagnetism is internally
inconsistent.  Accepting the vortex scenario and fitting  the
magnetization data in $Bi-2212$  with the conventional  logarithmic
field dependence \cite{ong}, one obtains surprisingly high upper
critical fields $H_{c2} > 120$ Tesla and a very large
Ginzburg-Landau parameter, $\kappa=\lambda/\xi >450$  even at
temperatures close to $T_c$. The in-plane low-temperature magnetic
field penetration depth is $\lambda=200$ nm in optimally doped
$Bi-2212$ (see, for example \cite{tallon}). Hence the zero
temperature coherence length $\xi$ turns out to be about  the
lattice constant, $\xi=0.45$nm, or even smaller. Such a small
coherence length rules out the "preformed Cooper pairs"  \cite{kiv},
since the pairs are virtually not overlapped at any size of the
Fermi surface in $Bi-2212$ . Moreover the magnetic field dependence
of $M(T,B)$ at and above $T_c$ is entirely inconsistent  with what
one expects from a vortex liquid.  While $-M(B)$  decreases
logarithmical at temperatures well below $T_c$, the  experimental
curves \cite{mac,nau,ong} clearly show that   $-M(B)$  increases
with the field at and  above $T_c$ , just opposite to what one could
expect in the vortex liquid.  This significant departure from the
London liquid behavior clearly indicates that the vortex liquid does
not appear above the resistive phase transition \cite{mac}.

Some time ago we  explained the anomalous diamagnetism in cuprates
as the Landau normal-state diamagnetism of preformed bosons
\cite{den}. The same model  predicted  the unusual upper critical
field \cite{aleH} observed in many superconducting cuprates
\cite{buc,mac0,boz,lawrie,gan,alezavnev,ZAV} (see below). Here we
extend the model to high magnetic fields taking into account the
magnetic pair-breaking of singlet bipolarons and the  anisotropy of
the energy spectrum. When the strong magnetic field is applied
perpendicular to the copper-oxygen plains the quasi-2D bipolaron
energy spectrum is quantized as
\begin{equation}
E_{\alpha}= \omega(n+1/2) +2t_c [1-\cos(K_zd)],
\end{equation}
where $\omega=2eB/m_b$, $n=0,1,2,...$, and $t_c$, $K_z$, $d$ are the
hopping integral, the momentum and the lattice period perpendicular
to the planes. Quantum numbers $\alpha$ also include  the momentum
along one of the in-plane directions. Expanding the Bose-Einstein
distribution function in powers of $exp[(\mu-E_\alpha)/T]$ with the
negative $\mu$ one can readily obtain (after summation over $n$) the
boson density
\begin{equation}
n_b={eB\over{\pi d}} \sum_{k=1}^{\infty} I_0(2t_c k/T) {\exp[
(\tilde{\mu} -2t_c)k/T]\over{1-\exp(-\omega k/T)}},
\end{equation}
and the magnetization
\begin{eqnarray}
M(T,B)&=&-n_b \mu_b+{eT\over{\pi d}} \sum_{k=1}^{\infty} I_0(2t_c
k/T) {\exp[ (\tilde{\mu} -2t_c)k/T]\over{1-\exp(-\omega k/T)}}\cr
&\times& \left({1\over{k}}-{\omega \exp(-\omega
k/T)\over{T[1-\exp(-\omega k/T)]}}\right),
\end{eqnarray}
where $\mu_b=e/m_b$, $\tilde{\mu}=\mu-\omega/2$ and $I_0(x)$ is the
modified Bessel function. At low temperatures $T \rightarrow 0$
Schafroth's result \cite{sha} is recovered, $M(0,B)= -n_b \mu_b$.
The magnetization of charged bosons is field-independent at low
temperatures. At high temperatures, $T \gg T_c$ the chemical
potential has a large  magnitude , so we can keep only terms with
$k=1$ in Eqs.(86,87) to obtain
\begin{equation}
M(T,B)=-n_b \mu_b+{Tn_b\over{B}}  \left(1-{\omega \exp(-\omega
/T)\over{T[1-\exp(-\omega/T)]}}\right).
\end{equation}
 The experimental conditions are such that $T \gg \omega$ when $T$ is of the order of $T_c$
 or higher, so that
 \begin{equation}
M(T,B)=-n_b \mu_b {\omega\over{6T}},
\end{equation}
 which is the Landau  orbital diamagnetism  of nondegenerate carriers.
The  bipolaron in-plane mass in cuprates is about $m_b\approx 10
m_e$ \cite{alebook1}. Using this mass yields $M(0,B) \approx 2000$
A/m with the bipolaron density $n_b=10 ^{21}$ cm$^{-3}$. Then the
magnitude and the field/temperature dependence of $M(T,B)$ near and
above $T_c$ are about the same as experimentally  observed in Refs
\cite{nau,ong}.

The pseudogap temperature $T^*$ depends on the magnetic field
predominantly because of the magnetic-field splitting of the
single-polaron band in Fig.12. As a result the bipolaron density
depends on the field (as well as on  temperature) near $T_c$ as
 \begin{equation}
 n_b(T,B)=n_b(T_c,0)\left[1+(T_c-T)/\tilde{T}_0 -(B/B_0)^\beta \right],
 \end{equation}
 where $\tilde{T}_0$ and $B_0$ are constants  depending on $T^*$, $\beta=2$
if the polaron spectrum is spin-degenerate, and $\beta=1$ if the
spin degeneracy is removed by the crystal field already in the
absence of the external field.

 Theoretical temperature and field dependencies of $M(T,B)$,  Eq.(87)
agree qualitatively  with the experimental curves in $Bi-2212$
\cite{nau,ong}, if the depletion of the bipolaron density, Eq.(90)
is taken into account. The  depletion of $n_b$ accounts for the
absence of the crossing point in $M(T,B)$ at high magnetic fields.
Nevertheless a quantitative fit to experimental $M(T,B)$ curves
using $\tilde{T}_0$ and $B_0$ as the fitting parameters is
premature. The experimental diamagnetic magnetization has been
extracted from the total magnetization assuming  that the normal
state paramagnetic contribution remains temperature-independent  at
all temperatures \cite{nau,ong}. This assumption is inconsistent
with a great number of NMR and the Knight shift measurements, and
even with the preformed Cooper-pair model itself. The Pauli
spin-susceptibility has been found temperature-dependent in these
experiments revealing  normal-state pseudogaps, contrary to the
assumption. Hence the experimental diamagnetic $M(T,B)$
\cite{nau,ong} has to be corrected by taking into account the
temperature dependence of the spin paramagnetism at relatively low
temperatures.

\subsection{Spin pseudogap, $c$-axis transport and charge pseudogap}
The pairing of holes into singlets well above $T_{c}$ should be seen
as a drop of  the nuclear magnetic relaxation rate $1/T_{1}$ with
temperature lowering. Indeed it is a common feature of the normal
state of many cuprates. The bipolaron model has described the
temperature dependence of $1/T_{1}$ \cite{alegap}. The conventional
contact hyperfine coupling of nuclear spin on a site $i$ with
electron spins is given in the site representation by
\begin{equation}
H_{i}=\hat{A}_{i}\sum_{j}c_{j\uparrow }^{\dagger }c_{j\downarrow
}+H.c.,
\end{equation}
where $\hat{A}_{i}$ is an operator acting on the nuclear spin, and
$j$ is its nearest neighbor sites. Performing projecting
transformations to  bipolarons as above we obtain the effective
spin-flip interaction of triplet bipolarons with the nuclear spin as
\begin{equation}
H_{i}\sim \sum_{j,l\neq l^{\prime }}b_{j,l}^{\dagger }b_{j,l^{\prime
}}+H.c.
\end{equation}
Here $l,l^{\prime }=0,\pm 1$ are z-components of spin $S=1$. The NMR
width due to the spin-flip scattering of triplet bipolarons on
nuclei is obtained using the Fermi-Dirac golden rule,
\begin{equation}
{\frac{1}{{T_{1}}}}=-\frac{B}{t^{2}}\int_{0}^{2t}dE\frac{\partial f(E)}{%
\partial E}
\end{equation}
where $f(E)=[\exp ((E+J-\mu )/T)-]^{-1}$ is the triplet distribution
function, and $2t$ is their bandwidth. For simplicity the triplet
DOS is taken as a constant ($=1/(2t)$). As a result, we obtain
\begin{equation}
{\frac{1}{{T_{1}}}}={\frac{BT\sinh (t/T)}{{t^{2}[\cosh [(t+J)/T-\ln
y]-\cosh (t/T)]}},}
\end{equation}
where $B$ is a temperature independent hyperfine coupling constant.

Eq.(94) describes all essential features of the nuclear spin
relaxation rate in copper-based oxides: the absence of the
Hebel-Slichter coherent peak
below $T_{c}$, the temperature-dependent Korringa ratio ($1/TT_{1}$) above $%
T_{c}$, and a large value of $1/T_{1}$ due to the small bandwidth
2$t$. It nicely fits the experimental data in $YBa_2Cu_4O_8$
\cite{tan} with reasonable values of the parameters, $t=250$K and
$J=150$K \cite{alegap}.  A similar unusual behavior of NMR was found
in underdoped $YBa_{2}Cu_{3}O_{6+x}$, and in many other cuprates.
The Knight shift, which measures the spin susceptibility of
carriers, also drops well above $T_{c}$ in many copper oxides, in
agreement with the bipolaron model. The `spin' gap has been observed
above and below $T_{c}$ in $YBa_{2}Cu_{3}O_{6+x}$ with unpolarized
\cite{ros}, and polarized \cite{moo} neutron scattering.

The bipolaron model has also quantitatively explained  $c$-axis
transport and the anisotropy of cuprates
\cite{alekabmot,in,hof2,zve}. The crucial point is that single
polarons dominate in  $c$-axis transport at finite
temperatures because they are much lighter than bipolarons in  $c$%
-direction. Bipolarons can propagate across the planes  due to a
simultaneous two-particle tunnelling alone, which is much less
probable than a single polaron tunnelling. Along the planes polarons
and inter-site bipolarons propagate with  comparable effective
masses, as shown above. Hence in the mixture of nondegenerate
quasi-two-dimensional bosons and thermally excited fermions, only
fermions  contribute to  $c$ -axis transport, if the temperature is
not very low, which leads to  thermally activated $c$ -axis
transport and to a fundamental relation between the anisotropy and
the uniform magnetic susceptibility of cuprates \cite{alekabmot}.

 The exponential temperature
dependence of c-axis resistivity and $"c"$ versus $"ab"$ anisotropy
was interpreted within the framework of the bipolaron model in many
cuprates, in particular in $La_{2-x}Sr_{x}CuO_{4}$
\cite{alekabmot,zhaomul},
$%
Bi_{2}Sr_{2}CaCu_{2}O_{8+\delta }$ \cite{in}, $YBa_{2}Cu_{3}O_{6+x}$
\cite{zve}, and $HgBa_{2}CuO_{4+\delta }$ \cite{hof2}. Importantly,
the uniform magnetic susceptibility above $T_{c}$ \emph{increases}
with doping. It proves once more that cuprates are doped insulators,
where low energy charge and spin degrees of freedom are due to holes
doped into a parent insulating matrix with no free carriers and no
free spins. A rather low magnetic susceptibility of parent
insulators in their paramagnetic phase is presumably due to a
singlet pairing of copper spins.

\section{Superconducting state of cuprates}

\subsection{Parameter-free evaluation of $T_{c}$: Bose-Einstein condensation versus
the Kosterlitz-Thouless transition} An ultimate goal of the theory
of superconductivity is to provide an expression for $T_{c}$ as a
function of some well-defined parameters characterizing the
material. In the framework of the BCS theory $T_{c}$ is fairly
approximated by the familiar McMillan's formula, which works well
for simple metals and their alloys. But applying a theory of this
kind to high-T$_{c}$ cuprates is problematic. Since bare electron
bands are narrow, strong correlations result in the Mott insulating
state of undoped parent compounds. As a result, $\mu ^{\ast }$ is
ill-defined in doped cuprates, and polaronic effects are important
as in many doped semiconductors. Hence, an estimate of $T_{c}$ in
cuprates within the BCS theory appears to be an exercise in
calculating the Coulomb pseudopotential $\mu ^{\ast }$ rather than
$T_{c}$ itself. One cannot increase $\lambda $ either without
accounting for a polaron collapse of the band as discussed above.
This appears at $\lambda \approx 1$.

On the other hand, the bipolaron theory provides a parameter-free
expression for $T_{c}$ \cite{alekabTc}, which fits the
experimentally measured $T_{c}$ in many cuprates for any level of
doping. $T_{c}$ is calculated using the density sum rule as the
Bose-Einstein condensation (BEC) temperature of $2e$ charged bosons
on a lattice. Just before the discovery \cite{mul} we predicted
$T_{c}$ as high as $\approx 100K$ using an estimate of the bipolaron
effective mass \cite{alekab0}. Uemura \cite{uem} established a
correlation of $T_{c}$ with the in-plane magnetic field penetration
depth measured by $\mu sR$ technique in many cuprates as $T_{c}\sim
1/\lambda _{ab}^{2}$. The technique is based on the implantation of
spin polarized muons. It monitors the time evolution of the muon
spin polarization. He concluded that cuprates are neither BCS nor
BEC superfluids but they are in a crossover region from one to the
other, because the experimental $T_{c}$ was found about $3$ or more
times below the BEC temperature.

Here we calculate $T_{c}$ of a bipolaronic superconductor taking
properly into account the microscopic band structure of bipolarons
in layered cuprates as derived in section 3. We arrive at a
parameter-free expression for $T_{c},$ which in contrast to Ref.
\cite{uem} involves not only the in-plane, $\lambda _{ab}$ but also
the out-of-plane, $\lambda _{c}$, magnetic field penetration depth,
and a normal state Hall ratio $R_{H}$ just above the transition. It
describes the experimental data for a few dozen different samples,
Fig.18, clearly indicating that many cuprates are in the BEC rather
than in the crossover regime.
\begin{figure}
\begin{center}
\includegraphics[angle=-90,width=0.45\textwidth]{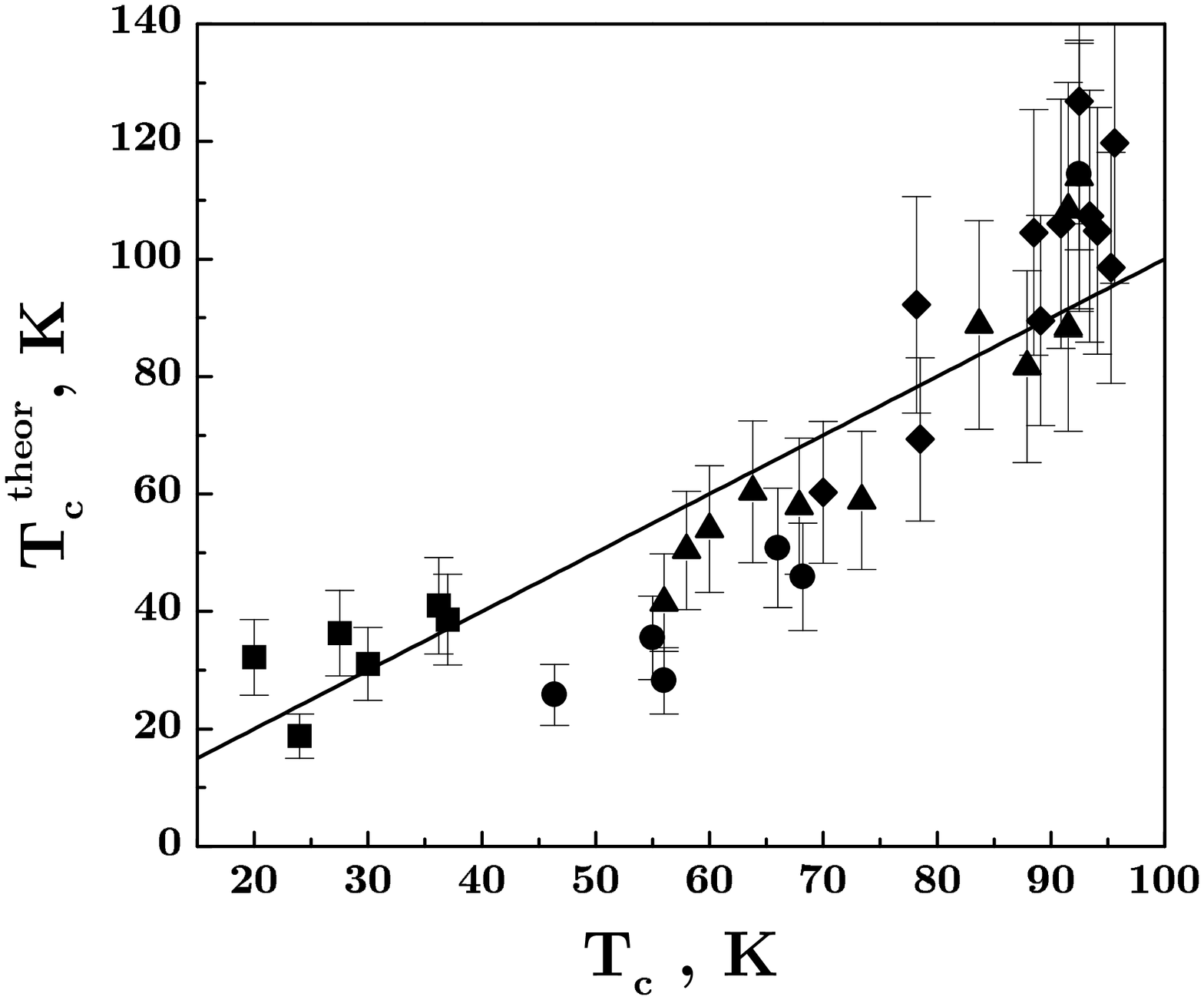}
\vskip -0.5mm \caption{Theoretical critical temperature compared
with the experiment (the theory is exact for samples on the straight line) for $%
LaSrCuO$ compounds (squares), for $Zn$ substituted $%
YBa_{2}Cu_{1-x}Zn_{x}O_{7}$ (circles), for $YBa_{2}Cu_{3}O_{7-%
\delta}$ (triangles), and for $HgBa_{2}CuO_{4+\delta }$ (diamonds).
Experimental data for the London penetration depth are taken from T.
Xiang $et$ $al$,  Int. J. Mod. Phys. B{\bf 12}, 1007 (1998) and
B. Janossy $et$ $al$,  Physica C{\bf 181}, 51 (1991) in $%
YBa_{2}Cu_{3}O_{7-\delta }$  and $YBa_{2}Cu_{1-x}Zn_{x}O_{7}$; from
V.G. Grebennik $et$ $al$, \emph{Hyperfine Interactions }{\bf 61},
1093 (1990) and C. Panagopoulos (private communication) in
underdoped and overdoped $La_{2-x}Sr_{x}CuO_{4}$, respectively, and
from J. Hofer $et$ $al$, Physica C, {\bf 297}, 103 (1998) in $%
HgBa_{2}CuO_{4+\delta}$. The Hall coefficient above $T_{c}$  is
taken from A. Carrington $et$ $al$,  Phys. Rev. B{\bf 48}, 13051
(1993) and J. R. Cooper (private communication)
($YBa_{2}Cu_{3}O_{7-\delta
}$  and $YBa_{2}Cu_{1-x}Zn_{x}O_{7}$) and from H.Y. Hwang $%
et$ $al$,  Phys. Rev. Lett. {\bf 72}, 2636 (1994)
($La_{2-x}Sr_{x}CuO_{4}$).}
\end{center}
\end{figure}
The energy spectrum of bipolarons is at least two-fold degenerate in
cuprates (section 3). One can apply the effective mass approximation
at $T\simeq T_{c}$, Eq.(52), because $T_{c}$ should be less than the
bipolaron bandwidth. Also  three-dimensional corrections to the
spectrum are important for the Bose-Einstein condensation. They are
well described by the tight-binding approximation as
\begin{equation}
E_{{\bf K}}^{x,y}={\frac{\hbar ^{2}K_{x,y}^{2}}{{2m_{x}^{\ast \ast }}}}+{%
\frac{\hbar ^{2}K_{y,x}^{2}}{{2m_{y}^{\ast \ast }}}}+2t_{c}[1-\cos
(K_{z}d)].
\end{equation}
 Substituting the spectrum,
Eq.(95) into the density sum rule,
\begin{equation}
\sum_{{\bf K},i=(x,y)}\left[ \exp (E_{{\bf K}}^{i}/T_{c})-1\right]
^{-1}=n_{b}
\end{equation}
one readily obtains $T_{c}$ as (in ordinary units)
\begin{equation}
k_{B}T_{c}=f\left( \frac{t_{c}}{{k_{B}T_{c}}}\right) \times {\frac{%
3.31\hbar ^{2}(n_{b}/2)^{2/3}}{{(m_{x}^{\ast \ast }m_{y}^{\ast \ast
}m_{c}^{\ast \ast })^{1/3}}}},
\end{equation}
where the coefficient $f(x)\approx 1$ is shown in Fig.19 as a
function of the anisotropy, $t_{c}/(k_{B}T_{c})$, and $m_{c}^{\ast
\ast }=\hbar ^{2}/(2|t_{c}|d^{2})$.

\begin{figure}
\begin{center}
\includegraphics[angle=-90,width=0.35\textwidth]{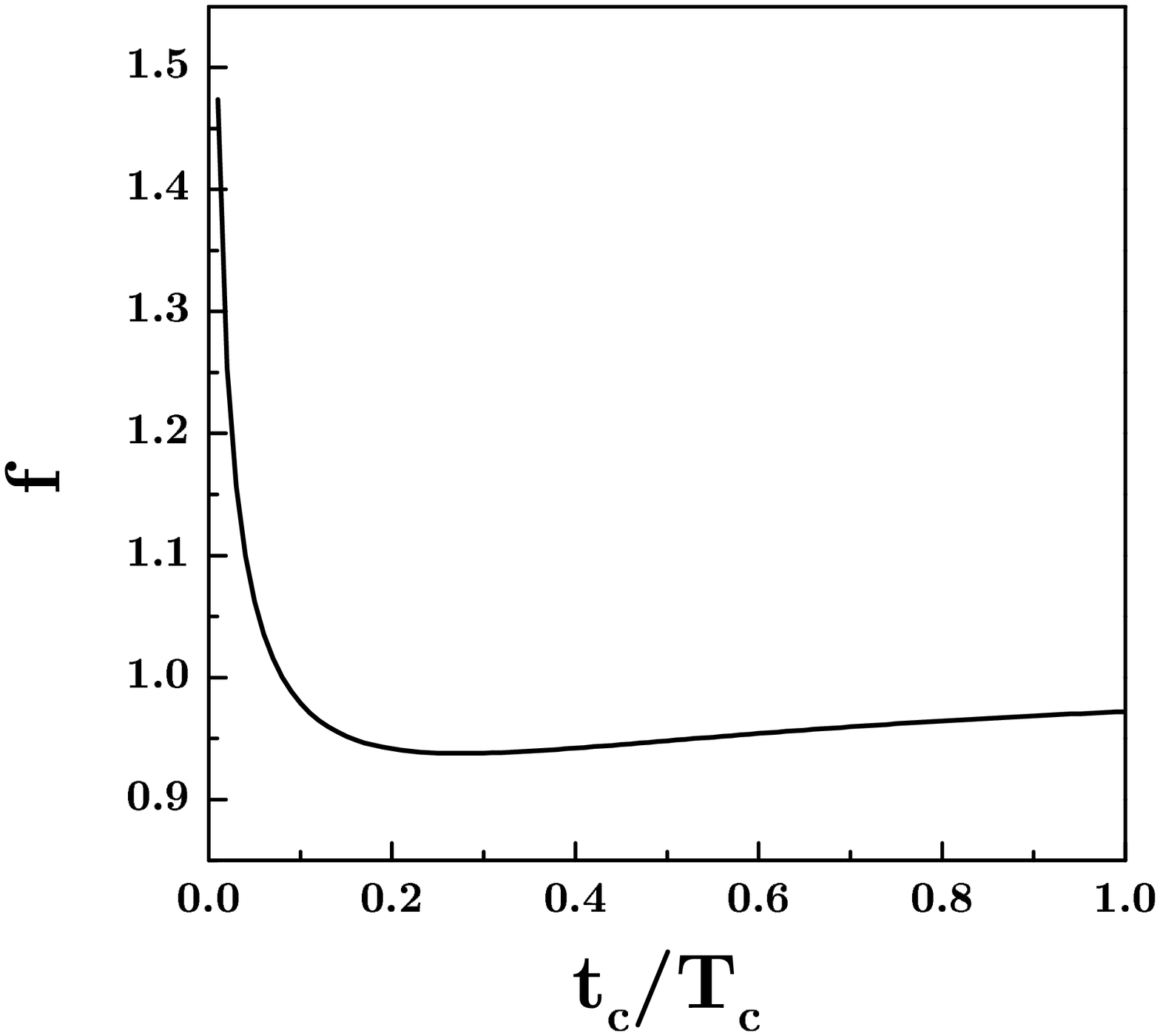}
\vskip -0.5mm \caption{Correction coefficient to the 3D
Bose-Einstein condensation temperature as a function of anisotropy.}
\end{center}
\end{figure}

This expression is rather ambiguous because the effective mass
tensor as well as the bipolaron density $n_{b}$ are not well known.
Fortunately, we can express the band-structure parameters via
in-plane,
\[
\lambda _{ab}=\left[ \frac{{m_{x}^{\ast \ast }m_{y}^{\ast \ast
}}}{8\pi n_{b}e^{2}({m_{x}^{\ast \ast }+m_{y}^{\ast \ast }})}\right]
^{1/2}
\]
and out-of-plane penetration depths,
\[
\lambda _{c}=\left[ \frac{{m_{c}^{\ast \ast }}}{16\pi
n_{b}e^{2}}\right] ^{1/2}
\]
(we use $c=1$). The bipolaron density is expressed through the
in-plane Hall ratio (above the transition) as
\begin{equation}
R_{H}={\frac{1}{{2en_{b}}}}\times {\frac{4{m_{x}^{\ast \ast
}m_{y}^{\ast \ast }}}{{(m_{x}^{\ast \ast }+m_{y}^{\ast \ast
})^{2}}}},
\end{equation}
which leads to
\begin{equation}
T_{c}=1.64f\left( \frac{{t_{c}}}{{k_{B}T_{c}}}\right) \left( {\frac{%
eR_{H}}{{\lambda _{ab}^{4}\lambda _{c}^{2}}}}\right) ^{1/3}.
\end{equation}
Here $T_{c}$ is measured in Kelvin, $eR_{H}$ in cm$^{3}$ and
$\lambda $ in
cm. The coefficient $f$ is about unity in a very wide range of $%
t_{c}/(k_{B}T_{c})\geq 0.01$, Fig.19. Hence, the bipolaron theory
yields a parameter-free expression, which unambiguously tells us how
near cuprates are to the BEC regime,
\begin{equation}
T_{c}\approx T_{c}(3D)=1.64\left( {\frac{eR_{H}}{{\lambda
_{ab}^{4}\lambda _{c}^{2}}}}\right) ^{{1/3}}.
\end{equation}
\small{
\begin{table}[tbp]
\caption{Experimental data on $T_{c}$(K), $ab$ and $c$ penetration depth($nm$%
), Hall coefficient ($10^{-3}(cm^{3}/C)$), and calculated values of
$T_{c}$
respectively for $La_{2-x}Sr_{x}CuO_{4}$ ($La$), $YBaCuO(x\%Zn)$ ($Zn$), $%
YBa_{2}Cu_{3}O_{7-x}$ ($Y$) and $HgBa_{2}CuO_{4+x}$ ($Hg$)
compounds}
\begin{tabular}[t]{llllllll}
Compound & $T^{exp}_{c}$ & $\lambda_{ab}$ & $\lambda_{c}$ & $R_{H}$, & $%
T_{c}(3D) $ & $T_{c}$ & $T_{KT}$ \\ \hline
$La$(0.2) & 36.2 & 200 & 2540 & 0.8 & 38 & 41 & 93 \\
$La$(0.22) & 27.5 & 198 & 2620 & 0.62 & 35 & 36 & 95 \\
$La$(0.24) & 20.0 & 205 & 2590 & 0.55 & 32 & 32 & 88 \\
$La$(0.15) & 37.0 & 240 & 3220 & 1.7 & 33 & 39 & 65 \\
$La$(0.1) & 30.0 & 320 & 4160 & 4.0 & 25 & 31 & 36 \\
$La$(0.25) & 24.0 & 280 & 3640 & 0.52 & 17 & 19 & 47 \\
$Zn$(0) & 92.5 & 140 & 1260 & 1.2 & 111 & 114 & 172 \\
$Zn$(2) & 68.2 & 260 & 1420 & 1.2 & 45 & 46 & 50 \\
$Zn$(3) & 55.0 & 300 & 1550 & 1.2 & 35 & 36 & 38 \\
$Zn$(5) & 46.4 & 370 & 1640 & 1.2 & 26 & 26 & 30 \\
$Y$(0.3) & 66.0 & 210 & 4530 & 1.75 & 31 & 51 & 77 \\
$Y$(0.43) & 56.0 & 290 & 7170 & 1.45 & 14 & 28 & 40 \\
$Y$(0.08) & 91.5 & 186 & 1240 & 1.7 & 87 & 88 & 98 \\
$Y$(0.12) & 87.9 & 186 & 1565 & 1.8 & 75 & 82 & 97 \\
$Y$(0.16) & 83.7 & 177 & 1557 & 1.9 & 83 & 89 & 108 \\
$Y$(0.21) & 73.4 & 216 & 2559 & 2.1 & 47 & 59 & 73 \\
$Y$(0.23) & 67.9 & 215 & 2630 & 2.3 & 46 & 58 & 73 \\
$Y$(0.26) & 63.8 & 202 & 2740 & 2.0 & 48 & 60 & 83 \\
$Y$(0.3) & 60.0 & 210 & 2880 & 1.75 & 43 & 54 & 77 \\
$Y$(0.35) & 58.0 & 204 & 3890 & 1.6 & 35 & 50 & 82 \\
$Y$(0.4) & 56.0 & 229 & 4320 & 1.5 & 28 & 42 & 65 \\
$Hg$(0.049) & 70.0 & 216 & 16200 & 9.2 & 23 & 60 & 115 \\
$Hg$(0.055) & 78.2 & 161 & 10300 & 8.2 & 43 & 92 & 206 \\
$Hg$(0.055) & 78.5 & 200 & 12600 & 8.2 & 28 & 69 & 134 \\
$Hg$(0.066) & 88.5 & 153 & 7040 & 6.85 & 56 & 105 & 229 \\
 \hline
\end{tabular}
\end{table}}
\begin{table}[tbp]
\caption{Mass enhancement in cuprates}
\begin{tabular}[t]{lll}
Compound & $m_{ab}$ & $m_{c}$ \\ \hline
$La$(0.2) & 22.1 & 3558 \\
$La$(0.15) & 15.0 & 2698 \\
$La$(0.1) & 11.3 & 1909 \\
$Y$(0.0) & 7.2 & 584 \\
$Y$(0.12) & 8.3 & 600 \\
$Y$(0.3) & 10.6 & 1994 \\ \hline
\end{tabular}
\end{table}
We compare two last expressions with the experimental $T_{c}$ of more than $%
30$ different cuprates, for which both $\lambda _{ab}$ and $\lambda
_{c}$ are measured along with $R_{H}(T_{c}+0)$ in Table 2 and
Fig.18. The Hall ratio has a strong temperature dependence above
$T_{c}$. Therefore, we use the experimental Hall ratio just above
the transition. In a few cases (mercury compounds), where
$R_{H}(T_{c}+0)$ is unknown, we take the inverse chemical density of
carriers (divided by $e$) as $R_{H}$. For almost all samples the
theoretical $T_{c}$ fits experimental values within an experimental
error bar for the penetration depth (about $\pm 10\%$). There are a
few ${Zn}$ doped YBCO samples, Fig.18, whose critical temperature is
higher than the theoretical estimate. If we assume that the
degeneracy of the bipolaron spectrum is removed by the random potential of $%
Zn$, then the theoretical $T_{c}$ would be almost the same as the
experimental values for these samples as well.

One can argue that due to a large anisotropy cuprates may belong to a 2D `$%
XY $' universality class with the Kosterlitz-Thouless \ (KT)
critical temperature $T_{KT}$ of preformed bosons \cite{ald,poc} or
the Cooper pairs \cite{kiv}. Should it be the case, one would hardly
discriminate the Cooper pairs with respect to bipolarons. KT
critical temperature is expressed through the in-plane penetration
depth alone as \cite{kiv}
\begin{equation}
k_{B}T_{KT}\approx {\frac{0.9d\hbar ^{2}}{{16\pi e^{2}\lambda
_{ab}^{2}}}}.
\end{equation}
It appears significantly higher than the experimental values in many
cases
(see  Table 2). There are also quite a few samples with about the same $%
{\lambda _{ab}}$ and the same $d$, but with very different values of
$T_{c},$ which proves that the phase transition is not the KT
transition. On the contrary, our parameter-free fit of the
experimental critical temperature and the critical behavior (see
below) favor $3D$ Bose-Einstein condensation of charged bosons as
the mechanism of high T$_{c}$ rather than any low-dimensional
phase-fluctuation scenario. The fluctuation theory \cite{schn}
further confirms the three-dimensional character of the phase
transition in cuprates. However, it does not mean that all
high-temperature superconductors are in the BEC regime with charged
bosons as supercarriers. Some of them, in particular, electron-doped
cuprates, $MgB_2$ and doped fullerenes might be in the BCS or
intermediate regime, which makes the BCS-BEC crossover problem to be
relevant. Starting from the pioneering works by Eagles \cite{eag2}
and Legget \cite{leg} this problem received particular attention in
the framework of a {\it negative} Hubbard $U$ model \cite{noz,micr}.
Both analytical (diagrammatic \cite{dia}, path integral
\cite{alerub}) and numerical \cite {rand} studies have addressed the
intermediate coupling regime beyond a variational approximation
\cite{noz}, including 2D systems \cite {var2,rand,tra}. However, in
using the negative Hubbard $U$ model, we have to realize that this
model, which predicts a smooth BCS-BEC crossover, cannot be applied
to the BCS-bipolaron crossover. The essential effect of the polaron
band-narrowing (section 2) is missing in the negative (and positive)
Hubbard $U$ model. The polaron collapse of the bandwidth is mainly
responsible for high $T_{c}$. It strongly affects the BCS-BEC
crossover significantly reducing the crossover region.

It is interesting to estimate the effective mass tensor using the
penetration depth and the Hall ratio. These estimates for in-plane
and out-of-plane boson masses are presented in Table 3. They agree
with the inter-site bipolaron mass (section 3). We notice, however,
that the absolute value of the effective mass in terms of the free
electron mass does not describe the actual band mass renormalization
if the bare (band) mass is unknown. Nevertheless an assumption
\cite{kiv} that the number of carriers is determined by the
Luttinger theorem (i.e. $n \approx 1$) would lead to much heavier
carriers with $m^{\ast }$ about $100 m_{e}.$

\subsection{Isotope effect on $T_{c}$ and on supercarrier mass}
The advances in the fabrication of the isotope substituted samples
made it possible to measure a sizable isotope effect , $\alpha
=-d\ln T_{c}/d\ln M$ in many high-$T_{c}$ oxides. This led to a
general conclusion that phonons are relevant for high $T_{c}$.
Moreover the isotope effect in cuprates was found to be quite
different from the BCS prediction, $\alpha =0.5$ (or less). Several
compounds showed $\alpha > 0.5$ \cite{cra}, and a small negative
value of $\alpha $ was found in $Bi-2223$ \cite{bor}.

Essential features of the isotope effect, in particular large values in low $%
T_{c}$ cuprates, an overall trend to lower value as $T_{c}$
increases \cite {fra}, and a small or even negative $\alpha $ in
some high $T_{c}$ cuprates were understood in the framework of the
bipolaron theory \cite{aleiso}. With increasing ion mass the
bipolaron mass increases and the Bose-Einstein condensation temperature $%
T_{c}$ decreases in the bipolaronic superconductor. On the contrary
in polaronic superconductors an increase of the ion mass leads to a
band narrowing and to an enhancement of the polaron density of
states, and to an increase of $T_{c}$. Hence the isotope exponent in
$T_{c}$ can distinguish the BCS like polaronic superconductivity
with $\alpha < 0$ , and the Bose-Einstein condensation of small
bipolarons with $\alpha > 0$. Moreover, underdoped cuprates, which
are definitely in the BEC regime, could have $\alpha
>0.5,$  as observed.

The isotope effect on $T_{c}$ is linked with the  isotope effect on
the carrier mass, $\alpha_{m^*}$, as \cite{aleiso}
\begin{equation}
\alpha=-d\ln T_c/d\ln M=\alpha_{m^*}[1-Z/(\lambda-\mu_c)],
\end{equation}
where $\alpha_{m^*}= d \ln m^*/d \ln M$ and $Z=m/m^* \ll 1$. In
ordinary metals, where the Migdal approximation is believed to be
valid, the renormalized effective mass of electrons is independent
of the ion mass $M$ because the electron-phonon interaction constant
$\lambda$ does not depend on $M$. However, when the e-ph interaction
is sufficiently strong, the electrons form polarons dressed by
lattice distortions, with an effective mass $m^{\ast} = m \exp
(\gamma E_p/\hbar\omega)$. While $E_p$ in the above expression does
not depend on the ion mass, the phonon frequency does.  As a result,
there is a large isotope effect on the carrier mass in polaronic
conductors, $\alpha_{m^*} = (1/2)\ln (m^*/m)$ \cite{aleiso}, in
contrast to the zero isotope effect in ordinary metals.

Such an effect was observed in cuprates  in the London penetration
depth of isotope-substituted samples \cite{zhao}. The carrier
density is unchanged with the isotope substitution of $O^{16}$ by
$O^{18}$,  so that the isotope effect on $\lambda _{ab}$ measures
directly the isotope effect on the carrier mass. In particular, the
carrier mass isotope exponent $\alpha
_{m^*}$ was found as large as $\alpha _{m^*}=0.8$ in $%
La_{1.895}Sr_{0.105}CuO_{4}$. Then the polaron mass enhancement
should be $m^{\ast \ast }/m\approx 5$ in this material. Using
Eq.(57) we obtain the in-plane bipolaron mass as large as $m^{\ast
\ast }\approx 10m_{e}$ with the bare hopping integral $T(NNN)=0.2$
eV. The in-plane magnetic field penetration depth, calculated with
this mass is $\ $ $\lambda _{ab}=[m^{\ast \ast }/8\pi
ne^{2}]^{1/2}\approx 316$nm, where $n$ is the hole density. It
agrees well with the experimental one, $\lambda
_{ab}\simeq 320$nm. Using the measured values of $\ \lambda _{ab}=320$ nm, $%
\lambda _{c}=4160$ nm, and of $R_{H}=4\times 10^{-3}$ $cm^{3}/C$
(just above $T_{c}$) we obtain $T_{c}=31K$ from Eq.(100) in
astonishing agreement with the experimental value $T_{c}=30$ $K$  in
this compound. More recent high resolution angle resolved
photoemission spectroscopy studies \cite{LAN} provided further
compelling evidence for strong e-ph interaction in the cuprates.
They revealed a fine phonon structure in the electron self-energy of
underdoped La$_{2-x}$Sr$_x$CuO$_4$ samples  and a complicated
isotope effect in the electron spectral function of Bi2212 that
depended on the electron energy and momentum.

\subsection{Specific heat anomaly}

Bose liquids (or more precisely $He^{4}$) show the characteristic
$\lambda $ -point singularity of their specific heat, but superfluid
Fermi liquids  like BCS superconductors exhibit a sharp second order
phase transition accompanied by a finite jump in the specific heat.
It was established beyond doubt \cite{fish,lor,ind,jun,schM} that
the anomaly in high $T_{c}$ cuprates differs qualitatively from the
BSC prediction. As was stressed by Salamon et al.\cite{sal} the heat
capacity is logarithmic near the transition, and consequently,
cannot be adequately treated by the mean-field BCS theory even
including the gaussian fluctuations. In particular, estimates using
the gaussian fluctuations yield an unusually small coherence volume,
Table 4, comparable with the unit cell volume \cite{lor}.
\begin{table}[tbp]
\caption{Coherence volume $\Omega $ in $\AA ^{3}$, the in-plane $\protect\xi %
_{ab}$ and out- of- plane $\protect\xi _{c}$ coherence lengths
derived from a Ginzburg-Landau analysis of the specific heat
\cite{lor}}
\begin{tabular}[t]{llll}
\mbox{}$Compound$ & $\Omega$ & $\xi^{2}_{ab}$,$(\AA^{2})$ & $\xi_{c}$, ($\AA$%
) \\ \hline
$YBa_{2}Cu_{3}O_{7}$ & 400 & 125 & 3.2 \\
$YBa_{2}Cu_{3}O_{7-0.025}$ & 309 & 119 & 2.6 \\
$YBa_{2}Cu_{3}O_{7-0.05}$ & 250 & 119 & 2.1 \\
$YBa_{2}Cu_{3}O_{7-0.1}$ & 143 & 119 & 1.2 \\
$Ca_{0.8}Y_{0.2}Sr_{2}Tl_{0.5}Pb_{0.5}Cu_{2}O_{7}$ & 84 & 70 & 1.2 \\
$Tl_{1.8}Ba_{2}Ca_{2.2}Cu_{3}O_{10}$ &  & 40 & $<0.9$%
\end{tabular}
\end{table}
The magnetic field dependence of the anomaly \cite{jun0} is also
unusual, but it can be described by the bipolaron model
\cite{alekablia,ZAV}. Calculations of the specific heat of charged
bosons in a magnetic field require an analytical DOS, $N(\epsilon
,B)$ of a particle, scattered by other particles and/or by a random
potential of impurities. One can use DOS in the magnetic field with
an impurity scattering calculated in the non-crossing approximation
\cite{aleH}. The specific heat coefficient
\begin{equation}
\frac{C(T,B)}{T}=\frac{d}{TdT}\int d\epsilon \frac{N(\epsilon ,B)\epsilon }{%
\exp [(\epsilon -\mu )/T]-1},
\end{equation}
calculated with this DOS and with $\mu $ determined from $n_{b}=\int
d\epsilon N(\epsilon ,B)f(\epsilon )$, is shown in Fig.20. The broad
maximum at $T\approx T_{c}$ is practically the same as in the ideal
Bose gas without scattering. It barely shifts in the magnetic field.
However, there is the second anomaly at lower temperatures, which is
absent in the ideal gas. It shifts with the magnetic field, tracing
precisely the resistive transition, as clearly seen from the
difference between the specific heat in the field and zero-field
curve, Fig.20b.
\begin{figure}
\begin{center}
\includegraphics[angle=-0,width=0.65\textwidth]{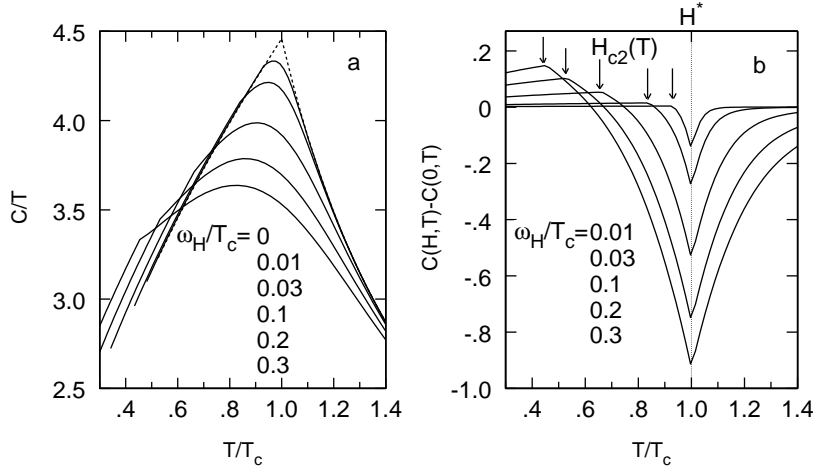}
\vskip -0.5mm \caption{Temperature dependence of the specific heat
divided by temperature (arb. units) of  the charged Bose-gas
scattered by impurities for several fields ($\omega_{H}=2eB/m^{\ast
\ast })$. (b) shows two anomalies, the lowest one traces resistive
transition, while the highest anomaly is the normal state feature.}
\end{center}
\end{figure}
The specific heat, Fig. 20, is in striking resemblance with the
Geneva group's experiments on $DyBa_{2}Cu_{3}0_{7}$ and on
$YBa_{2}Cu_{3}O_{7}$ \cite{jun0}, where both anomalies were
observed. Within the bipolaron model, when the magnetic field is
applied, it hardly changes the temperature dependence of the
chemical potential near the zero field $T_{c}$ because the energy
spectrum of thermally excited bosons is practically unchanged. This
is because their characteristic energy (of the order of $T_{c}$)
remains huge compared with the magnetic energy of the order of
$2eB/m^{\ast \ast }$. In contrast, the energy spectrum of low energy
bosons is strongly perturbed even by a weak magnetic field. As a
result the chemical potential `touches' the band edge at lower
temperatures, while having almost the same `kink'-like temperature
dependence around $T_{c}$ as in zero field. While the lower anomaly
corresponds to the true long-range order due to the Bose-Einstein
condensation, the higher one is just a `memory' about the zero-field
transition. This microscopic consideration shows that a genuine
phase transition into a superconducting state is related to a
resistive transition, and to the lower specific heat anomaly, while
the broad higher anomaly is a normal state feature of the bosonic
system in the external magnetic field. Different from the BCS
superconductor these two anomalies are well separated in the bosonic
superconductor at any field but zero.

\subsection{Universal upper critical field}
The upper critical field, $H_{c2}(T)=\Phi _{0}/2\pi \xi (T)^{2}$, is
very different in the BCS superconductor  and in the charged
Bose-gas  (CBG). While $H_{c2}(T)$ is linear in temperature near
$T_{c}$ in the Landau theory of second-order phase transitions, it
has a positive curvature $H_{c2}(T)\sim (T_{c}-T)^{3/2}$ in CBG
\cite{aleH}. Also at zero temperature $H_{c2}(0)$ is normally below
the Pauli pair-breaking limit given by $H_{p}\simeq 1.84T_{c}$ (in
Tesla) in the BCS theory, but the limit can be exceeded by many
times in CBG.

In cuprates \cite{buc,mac0,boz,lawrie,gan,alezav,ZAV}, spin-ladders
\cite {spin} and organic superconductors \cite{org} high magnetic
field studies revealed a non-BCS upward curvature of resistive
$H_{c2}(T)$. When measurements were performed on low-T$_{c}$
unconventional superconductors \cite{mac0,boz,lawrie,spin,org}, the
Pauli limit was exceeded by several times. A non-linear temperature
dependence in the vicinity of $\ T_{c}$ was unambiguously observed
in a few samples \cite{alezav,lawrie,gan,ZAV}. Importantly, a
thermodynamically determined $H_{c2}$ turned out much higher than
the resistive $H_{c2}$ \cite{wen} due to contrasting magnetic field
dependencies of the specific heat anomaly and of resistive
transition.

I believe that many unconventional superconductors are in the
`bosonic' limit of preformed real-space bipolarons, so their
resistive $H_{c2}$ is actually a critical field of the Bose-Einstein
condensation of charged bosons \cite{aleH}. Calculations above
carried out for the heat capacity of CBG lead to the conclusion that
the resistive $H_{c2}$ and the thermodynamically determined $H_{c2}$
are very different in bosonic superconductors. While the magnetic
field destroys the condensate of ideal bosons, it hardly shifts the
specific heat anomaly as observed.

A comprehensive scaling of resistive $H_{c2}$ measurements in
unconventional superconductors is shown in Fig.21 \cite{ZAV} in the
framework of the microscopic model of charged bosons scattered by
impurities. An expression for $H_{c2}(T)$ accounting for a
temperature dependence of the number of delocalized bosons,
$n_{b}(T),$ can be written as \cite{ZAV}
\begin{equation}
H_{c2}(T)=H_{0}\left[ \frac{n_{b}(T)}{tn_{b}(T_{c})}-t^{1/2}\right]
^{3/2},
\end{equation}
where $T_{c}$ is the zero-field critical temperature, and
$t=T/T_{c}$. Here the scaling constant $H_{0}$ depends on the
mean-free path $l$ , $H_{0}=\Phi _{0}/2\pi \xi _{0}^{2}$, with the
characteristic (coherence) length $\xi _{0}\simeq
(l/n_{b}(T_{c}))^{1/4}$. In the vicinity of $T_{c}$ one obtains the
parameter-free $H_{c2}(T)\propto (1-t)^{3/2}$ using this equation,
but the low-temperature behavior depends on a particular scattering
mechanism, and a detailed structure of the density of localized
states. As suggested by the normal state Hall measurements in
cuprates  $n_{b}(T)$ can be parameterized as $%
n_{b}(T)=n_{b}(0)+constant\times T$ (see also Eq.(62)), so that
$H_{c2}(T)$ is described by a single-parameter expression as
\begin{equation}
H_{c2}(T)=H_{0}\left[ \frac{b(1-t)}{t}+1-t^{1/2}\right] ^{3/2}.
\end{equation}

\begin{figure}
\begin{center}
\includegraphics[angle=-0,width=0.55\textwidth]{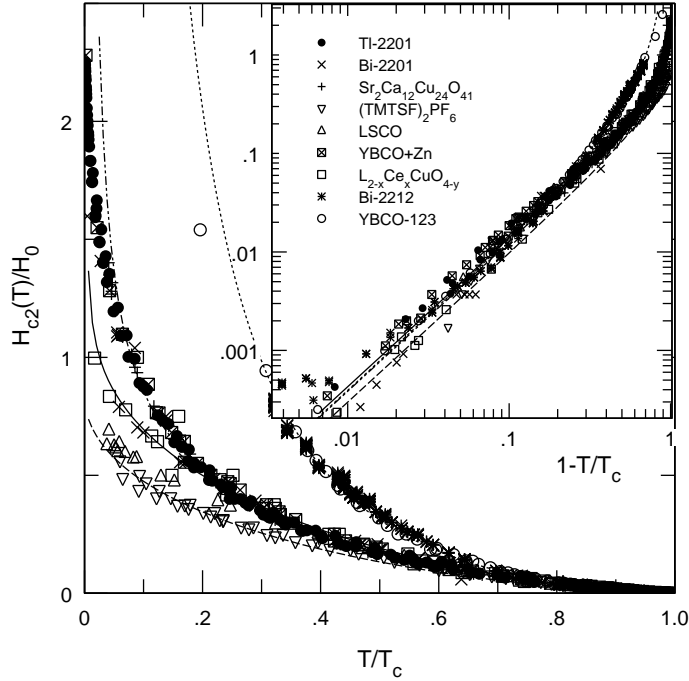}
\vskip -0.5mm \caption{Resistive upper critical field (determined at
50\% of the transition) of cuprates, spin-ladders and organic
superconductors scaled according to Eq.(105) \cite{ZAV}. The
parameter b is 1 (solid line), 0.02 (dashed-dotted line), 0.0012
(dotted line), and 0 (dashed line). The inset shows a universal
scaling of the same data near $T_{c}$ on the logarithmic scale.}
\end{center}
\end{figure}
The parameter $b$ is proportional to the number of delocalized
bosons at zero temperature. We expect that this expression is
applied in the whole temperature region except ultra-low
temperatures, where the non-crossing approximation fails
\cite{alebeerkab2}. Exceeding the Pauli pair-breaking limit readily
follows from the fact, that the singlet-pair binding energy is
related to the normal-state pseudogap temperature $T^{\ast }$,
rather than to $T_{c}$. $T^{\ast }$ is higher than $T_{c}$ in
bosonic superconductors and in cuprates. The universal scaling of
$H_{c2}$ near $T_{c}$ is confirmed by resistive measurements of the
upper critical field of many cuprates, spin-ladders, and organic
superconductors, as shown in Fig.21. All measurements reveal a
universal $(1-t)^{3/2}$ behavior in a wide temperature region
(inset) near $T_c$. The low-temperature behavior of $%
H_{c2}(T)/H_{0}$ is not universal, but well described using Eq.(105)
with the single fitting parameter, $b$. The parameter is close to $%
1 $ in high quality cuprates with a very narrow resistive
transition.
 It naturally becomes rather small in overdoped cuprates
where randomness is more essential.

\subsection{Symmetry and space modulations of the order parameter}
Independent observations of normal state pseudogaps in a number of
magnetic and kinetic measurements, and the unusual critical
behavior, discussed above, tell us that many cuprates may not be
$BCS$ superconductors. Indeed their superconducting state is as
anomalous as the normal one. In particular, there is strong evidence
for a $d$-like order parameter (changing sign when the $CuO_{2}$
plane is rotated by $\pi /2$) in cuprates \cite{ann}. A number of
phase-sensitive experiments \cite{pha} provide unambiguous evidence
in this direction; furthermore, the low temperature magnetic
penetration depth \cite{bonn,xia} was found to be linear in a few
cuprates as expected for a d-wave BCS superconductor. However, $SIN$ and $%
SIS $ tunnelling studies, the $c$-axis Josephson tunnelling
\cite{klem} and some high-precision magnetic measurements \cite
{mulsym} show a more usual $s$-like symmetry or even reveal an
upturn in the temperature dependence of the penetration depth below
some characteristic temperature \cite{wal}. Also both angle-resolved
photoemission (ARPES) \cite{shen} and scanning tunnelling microscopy
(STM) \cite{ren} have shown that the maximum energy gap and $2\Delta
/T_{c}$ ratio is several times larger  than expected in the
weak-coupling BCS theory, or in its intermediate-coupling Eliashberg
generalization. Strong deviations from the Fermi/BCS-liquid behavior
are suggestive of a new electronic state in cuprates, which is a
charged Bose liquid of bipolarons \cite{aleedw}.

Actually there are more complicated
  deviations from the conventional Fermi/BCS-liquid behavior than the normal state
  pseudogaps. Recent studies of the gap function revealed  two distinctly
 different gaps with different magnetic field and temperature
 dependence \cite{deu,kras}, and
 the checkerboard spatial modulations of the tunnelling DOS,
  with \cite{hoff} and without \cite{hoff2,kapit} applied
magnetic fields. We have proposed a simple phenomenological model
\cite{aleand} explaining two different gaps in  cuprates. The main
assumption, supported by a parameter-free estimate  of the Fermi
energy (section 3.5), is
 that the attractive potential is large compared with the renormalized Fermi
 energy, so that the ground state is the Bose-Einstein condensate of
 tightly bound real-space pairs. Here I
present an explanation of the symmetry of the order parameter and
real-space modulations of tunnelling DOS \cite{alesym} in the
framework of the bipolaron theory.

The anomalous Bogoliubov-Gor'kov average
$\cal{F}(\mathbf{r,r^\prime})$
 depends on the relative coordinate  of two electrons (holes)
 and on the center-of-mass coordinate, ${\bf \rho =r-r^\prime},%
{\bf R}=({\bf r + r^\prime})/2$. Its Fourier transform, $f({\bf %
k,K})$, depends on the relative momentum ${\bf k}$ and on the
center-of-mass momentum ${\bf K.}$ In the BCS theory ${\bf K=}0$,
and the Fourier transform of the order parameter is
proportional to the gap in the quasi-particle excitation spectrum, $f({\bf k,K%
})\sim \Delta _{{\bf k}}$. Hence the symmetry of the order parameter
and the symmetry of the gap are the same in the
weak-coupling regime. Under the rotation of the coordinate system, $\Delta _{%
{\bf k}}$ changes its sign, if the Cooper pairing appears in the
d-channel.

Real-space pairs might also have an unconventional symmetry due to a
specific symmetry of the pairing potential as in the case of the
Cooper pairs, but in any case the ground state and DOS are
homogeneous, if pairs are condensed with ${\bf K}=0$. On the other
hand,  the symmetry of the order parameter could be different from
the `internal' symmetry of the pair wave function, and from the
symmetry of a single-particle excitation gap in the strong-coupling
regime \cite{alebook1}. If the pair band dispersion has its minima
at finite ${\bf K}$ in the center-of-mass BZ, the Bose condensate is
inhomogeneous. In particular,   the center-of-mass bipolaron energy
bands could have their minima at the Brillouin zone boundaries at
${\bf K}=(\pi,0)$ and  three other equivalent momenta \cite{alesym}
(the lattice constant is taken as $a=1$). These four states are
degenerate, so that
the condensate wave function $\psi({\bf m})$ in the real (Wannier) space, ${\bf m}%
=(m_{x},m_{y}),$ is their superposition,
\begin{equation}
\psi({\bf m})=\sum_{{\bf K}=(\pm \pi ,0),(0,\pm \pi )}b_{{\bf K}%
}e^{-i{\bf K\cdot m}},
\end{equation}
where $b_{{\bf K}}=\pm \sqrt{n_{c}}/2$ are $c$-numbers,  and
$n_c(T)$ is the  atomic density of the Bose-condensate. The
superposition, Eq.(106), respects the time-reversal and parity
symmetries, if
\begin{equation}
\psi ({\bf m})=\sqrt{n_{c}}\left[ \cos (\pi m_{x})\pm \cos (\pi
m_{y})\right] .
\end{equation}
The order parameter, Eq.(107), has $d$-wave symmetry  changing sign
in the real space, when the lattice is rotated by $\pi /2$. This
symmetry is
entirely due to the pair-band energy dispersion with four minima at ${\bf %
K} \neq 0$, rather than due a  specific pairing potential. It
reveals itself as a {\it checkerboard} modulation of the hole
density with two-dimensional patterns, oriented along the diagonals.
From this insight  one can expect  a fundamental connection between
stripes detected by different techniques \cite{bia}  and the
symmetry of the order parameter in cuprates \cite{alesym}.

Importantly, even if preformed singlet bipolarons  are condensed at
$\Gamma$ point of their center-of-mass BZ (i.e. with ${\bf K}=0$),
the superconducting order-parameter could be $d-wave$ due to the
"orientation" degeneracy of inter-site pairs, Fig.10. As proposed by
Andreev \cite{and} the orientation  degeneracy of inter-site pairs
provides d-wave symmetry also irrespective of the symmetry of the
pairing potential, but without any space modulations of the hole
density. It could be relevant for overdoped cuprates (see below).

Now  let us take into account that in the superconducting state
($T<T_c$)  single-particle excitations
 interact with the pair condensate via the same  attractive potential, which forms the pairs \cite{aleand}.
 The Hamiltonian  describing
the interaction of  excitations with the pair Bose-condensate in the
Wannier representation is
\begin{equation}
 H = -\sum_{s,{\bf m,n}}[t({\bf m-n})+\mu \delta_{\bf m,n}]c^{\dagger}_{s \bf m}c_{s \bf
 n}
 + \sum_{\bf m}[\Delta({\bf m})c^{\dagger}_{\uparrow \bf m}c_{\downarrow \bf
 m}+H.c.],
 \end{equation}
 where $s= \uparrow,\downarrow$ is the
 spin,   and $\Delta({\bf m}) \propto \psi ({\bf
 m})$ is a coherent gap function. Applying  equations of motion for  the Heisenberg operators
 $\tilde{c}^{\dagger}_{s \bf m}(t)$ and $\tilde{c}_{s \bf
 m}(t)$, and the Bogoliubov transformation \cite{bog}
 \begin{equation}
\tilde{c}_{\uparrow \bf m}(t)=\sum_{\nu} [u_{\nu}({\bf m})
\alpha_{\nu} e^{-i\epsilon_{\nu}t} +v_{\nu}^* ({\bf
m})\beta_{\nu}^{\dagger} e^{i\epsilon_{\nu}t}],
\end{equation}
\begin{equation}
\tilde{c}_{\downarrow \bf m}(t)=\sum_{\nu} [u_{\nu}({\bf m})
\beta_{\nu} e^{-i\epsilon_{\nu}t} - v_{\nu}^* ({\bf
m})\alpha_{\nu}^{\dagger} e^{i\epsilon_{\nu}t}],
\end{equation}
one  obtains BdG equations describing  the single-particle
excitation spectrum,
\begin{equation}
\epsilon u({\bf m})=-\sum_{\bf n} [t({\bf m-n})+\mu \delta_{\bf
m,n}]u({\bf n}) + \Delta({\bf m})v ({\bf m}),
\end{equation}
\begin{equation}
-\epsilon v({\bf m})=-\sum_{\bf n} [t({\bf m-n})+\mu \delta_{\bf
m,n}]v({\bf n}) + \Delta({\bf m})u({\bf m}),
\end{equation}
where  excitation quantum numbers $\nu$ are omitted for
transparency. Different from the conventional BdG equations in the
weak-coupling limit, there is virtually no feedback  of  single
particle excitations on the off-diagonal potential, $\Delta({\bf
m})$, in the strong-coupling regime. The number of these excitations
is low at temperatures below $T^*\equiv \Delta_p$, so that the
coherent potential $\Delta ({\bf m})$ is an external (rather than a
self-consistent) field,  solely determined by the pair Bose
condensate \cite{aleand}. While the analytical solution is not
possible for any arbitrary off-diagonal interaction $\Delta({\bf
m})$, one can readily solve the infinite system of discrete
equations (111,112) for a periodic $\Delta({\bf m})$ with a period
commensurate with the lattice constant. For example,
\begin{equation}
\Delta({\bf m})= \Delta_c [e^{i\pi m_x} - e^{i\pi m_y}],
\end{equation}
 corresponds to the pair condensate at ${\bf K}=(\pm\pi,0)$
and $(0,\pm\pi)$, Eq.(107), with a temperature dependent (coherent)
$\Delta_c \propto \sqrt {n_c(T)}$. In this case the quasi-momentum
${\bf k}$ is the proper quantum number, $\nu= {\bf k}$, and the
excitation wave-function is a superposition of plane waves,
\begin{eqnarray}
u_{\nu}({\bf m}) &=&u_{{\bf k}}e^{i{\bf k}\cdot {\bf m}}+\tilde{u}_{{\bf k}}e^{i({\bf k-g})\cdot {\bf m}}, \\
v_{{\nu}}({\bf m}) &=&v_{{\bf k}}e^{i{({\bf k-g}_x})\cdot {\bf
m}}+\tilde{v}_{{\bf k}}e^{i({{\bf k-g}_y})\cdot {\bf m}}.
\end{eqnarray}
Here ${\bf g}_x=(\pi,0)$, ${\bf g}_y=(0,\pi)$, and ${\bf
g}=(\pi,\pi)$ are reciprocal doubled lattice vectors. Substituting
these equations into the BdG equations (111,112) one obtains four
coupled algebraic equations,
\begin{eqnarray}
\epsilon _{\bf k}u_{\bf k}&=&\xi_{\bf k}u_{\bf k}-\Delta_c (v_{\bf k}-\tilde{v}_{\bf k}), \\
\epsilon _{\bf k}\tilde{u}_{\bf k}&=&\xi_{\bf k-g}\tilde{u}_{\bf
k}+\Delta_c (v_{\bf k}-\tilde{v}_{\bf k}),
\\
-\epsilon_{\bf k}v_{\bf k}&=&\xi_{{\bf k-g}_x}v_{\bf k}+\Delta_c
(u_{\bf k}-\tilde{u}_{\bf k}), \\ -\epsilon_{\bf k}\tilde{v}_{\bf
k}&=&\xi_{{\bf k-g}_y}\tilde{v}_{\bf k}-\Delta_c (u_{\bf
k}-\tilde{u}_{\bf k}) ,
\end{eqnarray}
where $ \xi_{\bf k}=-\sum_{\bf n}t({\bf n}) e^{i{\bf k \cdot n}}
-\mu$. The determinant of the system (116-119) yields the following
equation for  the energy spectrum $\epsilon$:
\begin{eqnarray}
&&(\epsilon-\xi_{\bf k})(\epsilon-\xi_{\bf k-g})(\epsilon+\xi_{{\bf
k-g}_x})(\epsilon+\xi_{{\bf k-g}_y})\cr
&=&\Delta_c^2(2\epsilon+\xi_{{\bf k-g}_x}+\xi_{{\bf
k-g}_y})(2\epsilon-\xi_{\bf k}-\xi_{\bf k-g}).
\end{eqnarray}
Two positive  roots for $\epsilon$ describe the single-particle
excitation spectrum. Their calculation is rather cumbersome, but not
in the extreme strong-coupling limit, where the pair binding energy
$2\Delta_p$ is large compared with $\Delta_c$ and with the polaron
bandwidth. The
 chemical potential in this limit is pinned
 below a single-particle band edge, so  $\mu$ is negative, and
its magnitude is large compared with $\Delta_c$.
 Then the right
hand side in Eq.(120) is a perturbation, and the spectrum is
\begin{eqnarray}
\epsilon_{1\bf k}&\approx& \xi_{\bf k} -{\Delta_c^2\over{\mu}}, \\
\epsilon_{2\bf k}&\approx& \xi_{\bf k-g} -{\Delta_c^2\over{\mu}}.
\end{eqnarray}

If a metallic tip is placed at the point ${\bf m}$ above  the
surface of a sample, the STM current $I(V, {\bf m)}$ creates an
electron (or hole) at this point. Applying the Fermi-Dirac golden
rule and the Bogoliubov transformation, Eqs.(109,110), and assuming
that the temperature is much lower than $\Delta_p/k_B$ one readily
obtains the tunnelling conductance
\begin{equation}
\sigma (V, {\bf m})\equiv {dI(V, {\bf m)}\over {dV}}\propto
\sum_{\nu} | u_{\nu}({\bf m})|^2 \delta(eV-\epsilon_{\nu}),
 \end{equation}
 which is a local excitation DOS. The solution  Eq.(114) leads to a spatially modulated
 conductance,
\begin{equation}
\sigma(V, {\bf m)}= \sigma_{reg}(V) +\sigma_{mod} (V) \cos(\pi
m_x+\pi m_y).
\end{equation}
The smooth (regular) contribution is
\begin{equation}
\sigma_{reg}(V)=\sigma_0 \sum_{{\bf k},r=1,2}(u_{r{\bf k}}^2
+\tilde{u}_{r {\bf k}}^{2}) \delta(eV-\epsilon_{r{\bf k}}),
\end{equation}
and the amplitude of the modulated contribution is
\begin{equation}
\sigma_{mod}(V)= 2\sigma_0 \sum_{{\bf k},r=1,2}u_{r{\bf
k}}\tilde{u}_{r {\bf k}} \delta(eV-\epsilon_{r{\bf k}}),
\end{equation}
where $\sigma_0$ is a constant. Conductance modulations reveal a
checkerboard pattern, as the Bose condensate itself,
\begin{equation}
{\sigma-\sigma_{reg}\over{\sigma_{reg}}}=A \cos(\pi m_x+\pi m_y),
\end{equation}
where
 \begin{eqnarray}
 &&A=2 \sum_{\bf k} \left[u_{1
\bf k}\tilde{u}_{1\bf k} \delta (eV-\epsilon_{1 \bf k})+ u_{2\bf
k}\tilde{u}_{2 \bf k}  \delta (eV-\epsilon_{2 \bf k})\right]/\cr &&
\sum_{\bf k} \left [(u_{1 \bf k}^2 +\tilde{u}_{1 \bf k}^2) \delta
(eV-\epsilon_{1 \bf k})+(\tilde{u}_{2 \bf k}^2+u_{2 \bf k}^2) \delta
(eV-\epsilon_{2 \bf k})\right] \nonumber
\end{eqnarray}
is the amplitude  of  modulations  depending on the voltage $V$ and
temperature. An  analytical result can be obtained in the
strong-coupling limit with the excitation spectrum  given by Eqs.
(121,122) for the voltage  near the threshold, $eV\approx \Delta_p$.
In this case only  states near bottoms of each excitation band
contribute to the integrals in Eq.(125), so that
\begin{equation}
\tilde{u}_{1\bf k}={\xi_{\bf k}-\epsilon_{1 \bf k}\over{\epsilon_{1
\bf k}-\xi_{\bf k-g}}}u_{1\bf k} \approx -u_{1\bf
k}{\Delta_c^2\over{ \mu w}}\ll {u}_{1\bf k},
\end{equation}
and
\begin{equation}
u_{2\bf k}={\xi_{\bf k-g}-\epsilon_{2 \bf k}\over{\epsilon_{2 \bf
k}-\xi_{\bf k}}}\tilde{u}_{2\bf k} \approx -\tilde{u}_{2\bf
k}{\Delta_c^2\over{ \mu w}}\ll \tilde{u}_{2\bf k}.
\end{equation}
Substituting  these expressions into $A$, Eq.(127), yields in the
lowest order of $\Delta_c$,
\begin{equation}
A\approx -{2\Delta_c^2\over{ \mu w}}.
\end{equation}
The result, Eq.(127), is  reminiscent of STM data
\cite{hoff,hoff2,kapit,fu}, where  spatial checkerboard modulations
of $\sigma$ were observed in a few cuprates. Both commensurate and
incommensurate
 modulations were found depending on sample composition. In our model
the period is determined by the center-of mass wave vectors ${\bf
K}$ of the Bose-condensed preformed pairs. While the general case
has to be solved numerically, the perturbation result, Eq.(127) is
qualitatively applied for any ${\bf K}$ at least close to $T_{c}$,
where the coherent gap is small, if one replaces $\cos(\pi m_x+\pi
m_y)$ by $\cos(K_{x} m_x+K_{y} m_y)$. Different from any other
scenario, proposed so far, the hole density, which is about twice of
the condensate density at low temperatures, is spatially modulated
with the  period determined by the inverse wave vectors
corresponding to the center-of-mass pair band-minima. This 'kinetic'
interpretation of charge modulations in cuprates was originally
proposed \cite{alesym} before STM results became available. It could
account for those DOS modulations in  superconducting samples, which
disappear above $T_c$ because  the coherent gap $\Delta_c(T)$
vanishes, so that $A=0$ above $T_c$ in Eq.(127). Indeed some
inelastic neutron scattering experiments show that
incommensurate inelastic peaks are observed $only$ in the $%
superconducting$ state of high-$T_c$ cuprates \cite{bou}. The
vanishing at $T_{c}$ of incommensurate  peaks is inconsistent with
any other stripe picture, where a characteristic distance needs to
be observed in the normal state as well.  On the other hand some STM
studies (see, for example \cite{ver}) report incommensurate and
commensurate DOS modulations somewhat above $T_c$, in particular, in
heavily underdoped cuprates \cite{hancond}. I believe that those
modulations are due to a single-particle band structure and impurity
states near the top of the valence band in doped charge-transfer
insulators, rather than a signature of any cooperative phenomenon.

In this way the strong-coupling Fr\"ohlich-Coulomb model, Eq.(2),
 links charge heterogeneity, pairing, and pseudo-gaps as
manifestations of the strong electron-phonon attractive
 interaction in narrow bands of doped Mott-Hubbard insulators.

\section{Overdoped cuprates: Boson-Fermion mixtures}

It has been mentioned in section 1.1 that the chemical potential
 could enter the oxygen band in overdoped samples \cite{alebook1}  as a result of the
overlap of  bipolaron and polaron bands, so the Fermi-level crossing
could be seen in ARPES, and  pseudogaps gradually vanish, Fig.22. If
the number of states below the Fermi level in the polaron band is
less than the number of holes, bipolarons remain stable mobile
quasi-particles because the Pauli exclusion principle prevents their
decay into single polarons, Fig.22.  A comprehensive analysis by
Kornilovitch \cite{kor2} of a two-body problem on the square lattice
with the nearest-neighbor attraction strongly supports such a model
of overdoped cuprates.   It has been found that the stability of
pairs increases with their momentum. The pairs are formed easier
along the ($\pi,0$) direction than along the ($\pi, \pi$) direction.
This might lead to the appearance of "hot pairing spots" on the
$K_x$ and $K_y$ axes, while other regions of BZ remain unpaired.
\begin{figure}
\begin{center}
\includegraphics[angle=-90,width=0.55\textwidth]{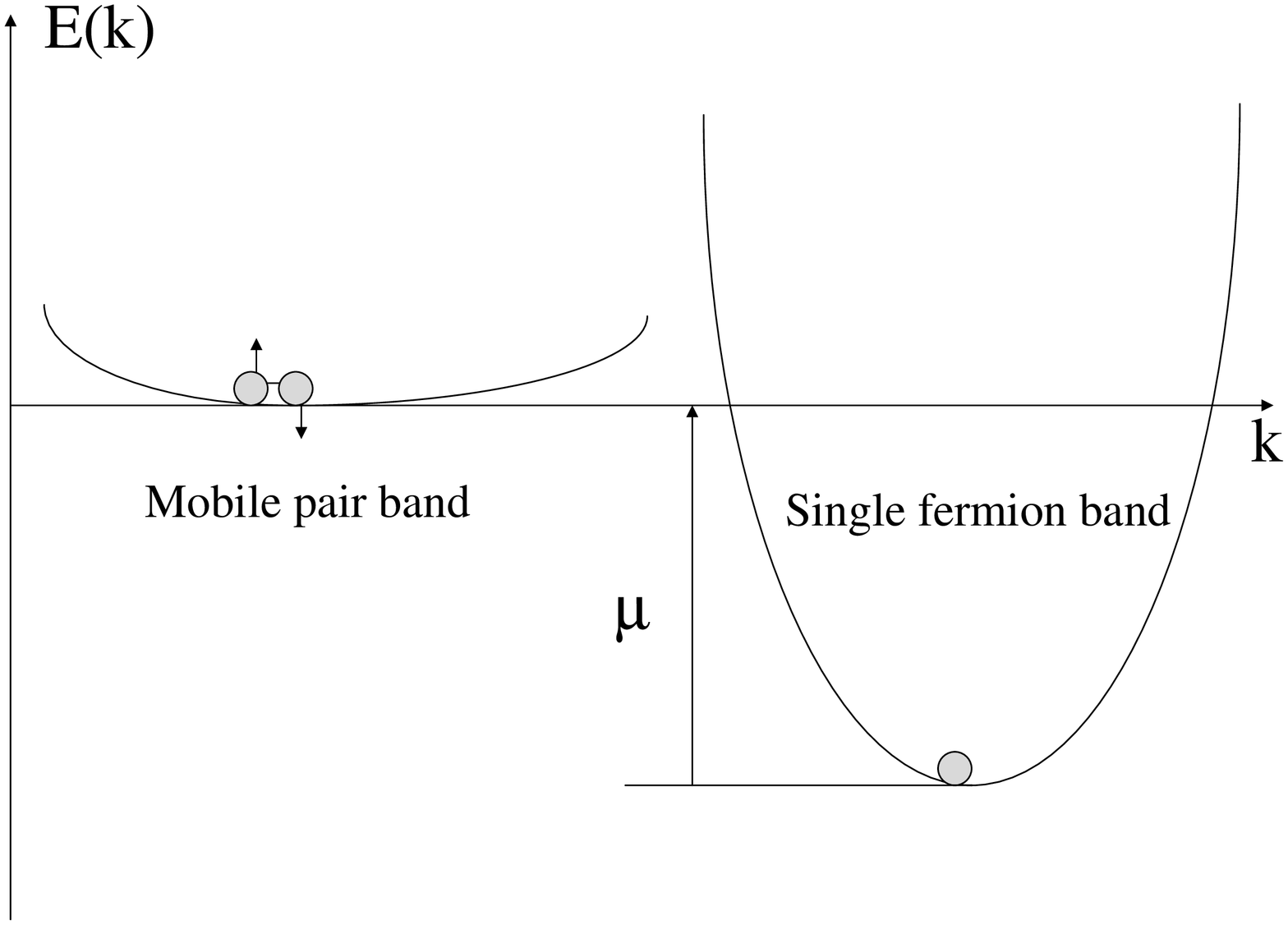}
\vskip -0.5mm \caption{The Pauli exclusion principle prevents a
decay of mobile bipolarons into single polarons in overdoped
cuprates because all states below the Fermi level $\mu$ are
occupied.}
\end{center}
\end{figure}

Such a system represents a mixture of mobile 2e-charged bosons and
e-charged degenerate fermions, first considered by us before the
discovery \cite{ale3}, somewhat similar to  neutral He$^3$-He$^4$
mixtures. While the normal state kinetic properties are dominated by
a lighter polaronic Fermi liquid, the superconducting state is the
Bose-Einstein condensate (BEC) of heavier bipolarons in the mixture.
As we discuss in the remaining part, it is valid even in the extreme
case of intrinsically immobile bipolarons, if they are hybridized
with mobile polarons, irrespective to the strength of the
hybridization interaction.

\subsection{Mobile fermions hybridized with immobile bosons: Boson-Fermion model}
As an alternative to the picture of HTS with mobile bipolarons,
Figs.12,22, some authors \cite{lee,ran,ran0,kos,lar,dam,dam2,mic}
proposed a so-called Boson-Fermion model (BFM), where intrinsically
immobile pairs are coupled with the Fermi sea of itinerant single
fermions by the hybridization interaction. Soon after Anderson and
Street and Mott \cite{ander} introduced localized pairs in amorphous
semiconductors, such two component model of negative $U$ centers
coupled with the Fermi sea of itinerant fermions was employed to
study superconductivity in disordered metal-semiconductor alloys
\cite{sim,tin}. When the attractive potential $U$ is large, the
model is reduced to localized hard-core bosons spontaneously
decaying into itinerant electrons and vice versa, different from a
non-converting mixture of mobile charged bosons and fermions
\cite{ale3,and}.

This boson-fermion model (BFM) was applied more generally to
describe pairing electron processes with localization-delocalization
\cite{ion}. The model attracted more attention in connection with
high-temperature superconductors
\cite{lee,ran,ran0,kos,lar,ale,cris,dam,dam2,mic}.  In particular,
Refs. \cite{dam,dam2} claimed that 2D BFM with immobile hard-core
bosons is capable to reproduce some physical properties and the
phase diagram of cuprates. BFM has been also adopted for a
description of superfluidity of atomic fermions scattered into bound
(molecular) states \cite{chio}.

Most studies of BFM  below its transition into a low-temperature
condensed phase applied a
 mean-field approximation (MFA), replacing  zero-momentum boson operators by c-numbers
 and neglecting the boson self-energy  in the density sum rule \cite{lee,ran,kos,lar,dam,dam2,mic,chio}.
  When the bare boson energy is well above the chemical potential, the BCS ground state
  was
  found  with bosons being only virtually excited \cite{lee,ran}.
  MFA led to a conclusion
 that BFM exhibits features compatible with BCS characteristics
 \cite{kos}, and describes a crossover from the BCS-like to a local pair  behavior
 \cite{mic}.  The transition was found  more mean-field-like than
 the usual Bose condensation, i.e. characterized by a relatively
 small value of the fluctuation parameter $Gi$ \cite{lar}.

 At the same time our previous study of BFM \cite{ale}  beyond MFA revealed a crucial
effect of the boson self-energy on the normal state boson spectral
function and the transition temperature $T_{c}$. Ref.\cite{ale}
proved that the Cooper pairing of fermions via virtual bosonic
states  is impossible in any-dimensional BFM. It occurs only
simultaneously with the Bose-Einstein condensation of real bosons.
The origin of this simultaneous condensation lies in a softening of
the boson mode at $T=T_c$ caused by its hybridization with fermions.
The energy of zero-momentum bosons is renormalized down to
\emph{zero}  at $T=T_c$, no matter how weak the boson-fermion
coupling and how large the bare boson energy are
 \cite{ale}.  One can also expect that the boson self-energy should qualitatively
modify the phase transition and the
 condensed phase of BFM below $T_c$.

\subsection{Absence of BCS-BEC crossover in BFM}
Let us examine  the phase transition and the condensed state of
 BFM  beyond the ordinary mean-field
approximation in two (2D) and three (3D)dimensions. It appears  that
$T_{c}=0$ K in the two-dimensional model,  even in the absence of
any Coulomb repulsion,  and the  phase
 transition is never a BCS-like second-order phase transition even in 3D BFM
 because of the complete boson
softening.

The 2D BFM is defined by the Hamiltonian,
\begin{eqnarray}
H &=&\sum_{{\bf k},\sigma =\uparrow ,\downarrow }\xi _{{\bf k}}c_{{\bf k}%
,\sigma }^{\dagger }c_{{\bf k},\sigma }+E_{0}\sum_{{\bf q}}b_{{\bf q}%
}^{\dagger }b_{{\bf q}}+ \\
&&{\rm g}N^{-1/2}\sum_{{\bf q,k}}\left( \phi _{{\bf k}}b_{{\bf
q}}^{\dagger }c_{-{\bf k}+{\bf q}/2,\uparrow }c_{{\bf k}+{\bf
q}/2,\downarrow }+H.c.\right) ,  \nonumber
\end{eqnarray}
where $\xi _{{\bf k}}=-2t(\cos k_{x}+\cos k_{y})-\mu $ is the 2D
energy spectrum of fermions, $E_{0}\equiv \Delta _{B}-2\mu $ is the
bare boson energy with respect to their chemical potential $2\mu $,
${\rm g}$ is the
magnitude of the anisotropic hybridization interaction, $\phi _{{\bf k}%
}=\phi _{-{\bf k}}$ is the anisotropy factor, and $N$ is the number
of cells. Here and further I take $\hbar=c=k_B=1$ and the lattice
constant $a=1$. Ref. \cite{dam} argued that 'superconductivity is
induced in this
model from the anisotropic charge-exchange interaction (${\rm g}\phi _{{\bf k%
}}$) between the conduction-band fermions and the immobile hard-core
bosons', and 'the on-site Coulomb repulsion  competes with this
pairing' reducing the critical temperature $T_{c}$ less than by
25\%. Also it has been argued \cite{dam2}, that the calculated upper
critical field of the model fits well the experimental results in
cuprates.

Here I show that $T_{c}=0$ K in the two-dimensional model, Eq.(131),
even in the absence of any Coulomb repulsion,  and the mean-field
approximation is meaningless for any-dimensional BFM because of the
complete boson softening. Replacing boson operators by $c$-numbers
for ${\bf q}=0$ in Eq.(131) one obtains  a linearized BCS-like
equation for the fermion order-parameter (the gap function) $\Delta
_{{\bf k}}$,
\begin{equation}
\Delta _{{\bf k}}=\frac{{\rm \tilde{g}}^{2}\phi _{{\bf k}}}{E_{0}N}\sum_{%
{\bf k}^{\prime }}\phi _{{\bf k}^{\prime }}{\frac{\Delta _{{\bf
k}^{\prime
}}\tanh (\xi _{{\bf k}^{\prime }}/2T)}{{2\xi _{{\bf k}^{\prime }}}},%
}
\end{equation}
with the coupling constant  ${\rm \tilde{g}}^{2}={\rm
g}^{2}(1-2n^{B})$, renormalized by the hard-core effects. Using a
two-particle fermion vertex part in the Cooper channel one can prove
that this equation is perfectly correct even beyond the conventional
non-crossing approximation \cite{ale}. The problem with MFA does not
stem  from this BSC-like equation, but from an incorrect definition
of the bare boson energy with respect to the chemical potential,
$E_{0}(T)$. This energy is determined by the atomic density of
bosons ($n_{b}$)  as (Eq.(9) in Ref. \cite{dam})
\begin{equation}
\tanh \frac{E_{0}}{2T}=1-2n_{b}.
\end{equation}
While Eq.(132) is perfectly correct, Eq.(133) is incorrect because
the boson self-energy $\Sigma _{b}({\bf q},\Omega)$ due to the same
hybridization  interaction is missing. At first sight \cite{dam} the
self-energy  is small in comparison to the kinetic energy of
fermions, if ${\rm g}$ is small. However $\Sigma _{b}(0,0)$ diverges
logarithmical at zero temperature \cite{ale}, no matter how week the
interaction is. Therefore it should be kept in the density sum-rule,
Eq.(133). Introducing the boson Green's function
\begin{equation}
D({\bf q},\Omega)=\frac{1-2n_b}{i\Omega -E_{0}-\Sigma _{b}({\bf q}%
,\Omega)}
\end{equation}
one must replace incorrect Eq.(133) by

\bigskip
\begin{equation}
-{\frac{T}{{N}}}\sum_{{\bf q},n}e^{i\Omega \tau }D({\bf q}%
,\Omega)=n_{b},
\end{equation}
where $\tau =+0$, and $\Omega =2\pi Tn$ ($n=0,\pm 1,\pm 2...$).

The divergent (cooperon) contribution to $\Sigma _{b}({\bf
q},\Omega)$ is given by \cite{ale},
\begin{eqnarray}
&&\Sigma _{b}({\bf q},\Omega)=-\frac{{\rm \tilde{g}}^{2}}{2N}\sum_{{\bf %
k}}\phi _{{\bf k}}^{2}\times  \\
&&\frac{\tanh [\xi _{{\bf k-q}/2}/(2T)]+\tanh [\xi _{{\bf k+q}/2%
}/(2T)]}{\xi _{{\bf k-q}/2}+\xi _{{\bf k+q}/2}-i\Omega}, \nonumber
\end{eqnarray}
so that one obtains
\begin{equation}
\Sigma _{b}({\bf q},0)=\Sigma _{b}(0,0)+\frac{q^{2}}{2M^{\ast }}+{\cal O}%
(q^{4})
\end{equation}
for small ${\bf q}$ and any anisotropy factor compatible with the
point-group symmetry of the cuprates. Here $M^{\ast }$ is the boson
mass, calculated analytically in Ref.\cite{ale} for the isotropic
exchange
interaction and parabolic fermion band dispersion (see also Ref.\cite{cris}%
). The BCS-like equation (132) has a nontrivial solution for $\Delta
_{{\bf k}} $ at $T=T_c$, if
\begin{equation}
E_{0}=-\Sigma _{b}(0,0).
\end{equation}
Substituting Eqs.(137,138)  into the sum-rule, Eq.(135), one obtains
a logarithmical divergent integral with respect to ${\bf q}$, and
\begin{equation}
T_{c}=\frac{const}{\int_{0}dq/q}=0.
\end{equation}
The devastating result, Eq.(139) is a direct consequence of the
well-known theorem, which states that BEC is impossible in 2D. One
may erroneously believe that MFA results \cite{dam,dam2} are still
applied in three-dimensions, where BEC is possible. However,
increasing dimensionality does not make MFA a meaningful
approximation for the boson-fermion model. This approximation leads
to a  conclusion that a BCS-like superconducting state
 occurs below the
 critical temperature   $T_{c}\simeq \mu \exp\left( -{%
E_{0}/z_c}\right) $ via fermion pairs being \emph{virtually} excited
into
 $unoccupied$ bosonic states \cite{lee,ran}.  Here $z_c=\tilde{g}^{2}N(0)$ and
$N(0)$ is the density of states (DOS) in the fermionic band near the
Fermi level $\mu $. However,  the Cooper pairing of fermions
 is impossible via virtual unoccupied bosonic states also in 3D BFM. Indeed,
Eqs.(132,138) do not depend on the dimensionality, so that the
analytical continuation of Eq.(132) to real frequencies $\omega$
yields the partial boson DOS as $\rho(\omega)=(1-2n_b)
\delta(\omega)$ at $T=T_c$ and ${\bf q}=0$ in any-dimensional BFM.
The Cooper pairing may occur only simultaneously with the
Bose-Einstein condensation of real bosons in the exact theory of 3D
BFM \cite{ale}. The origin of the simultaneous condensation of the
fermionic and bosonic fields in 3D BFM lies in the  softening of the
boson mode at $T=T_c$ caused by its hybridization with fermions.

Taking into account the boson damping and dispersion shows that the
boson spectrum significantly changes for all momenta. Continuing the
self-energy, Eq.(136) to real frequencies yields  the damping (i.e.
the imaginary part of the self-energy) as \cite{ale}
\begin{equation}
\gamma({\bf q},\omega)={\pi z_c\over{4q\xi}} \ln
\left[{\cosh(q\xi+\omega/(4T_{c}))\over{\cosh(-q\xi+\omega/(4T_{c}))}}\right],
\end{equation}
where $\xi=v_F/(4T_{c})$ is a coherence length, and $v_F$ is the
Fermi velocity. The damping is significant when $q\xi<<1$. In this
region $\gamma({\bf q},\omega)=\omega\pi z_{c}/(8T_{c})$ is
comparable or even larger than  the boson energy $\omega$. Hence
bosons look like overdamped diffusive modes, rather than
quasiparticles in the long-wave limit \cite{ale,cris}, contrary to
the erroneous conclusion of Ref.\cite{ran0}, that there is 'the
onset of coherent free-particle-like motion of the bosons' in this
limit. Only outside the long-wave   region, the damping becomes
small. Indeed, using Eq.(138) one obtains $\gamma({\bf
q},\omega)=\omega \pi z_{c}/(2qv_F)<< \omega$, so that bosons at
 $q >>1/\xi$ are well defined quasiparticles
 with a logarithmic dispersion, $\omega(q)=z_c \ln(q
\xi)$ \cite{ale}.  Hence the boson energy disperses over the whole
energy interval from zero up to $E_0$.

The main mathematical problem with MFA in 3D also stems from the
density sum rule, Eq.(135) which determines the chemical potential
of the system and consequently the bare boson energy $E_{0}(T)$ as a
function of temperature. In the framework of MFA one takes the bare
boson energy   in Eq.(132) as a temperature independent parameter,
$E_0=\tilde{g}^2N(0)\ln (\mu/T_c)$ \cite{lar}, or determines it from
the conservation of the total number of particles  neglecting the
boson self-energy, Eq.(133) \cite{ran,dam,mic,chio}. Then Eq.(132)
looks like the conventional linearized Ginzburg-Landau-Gor'kov
equation \cite{gor2}  with a negative coefficient $\alpha \propto
T-T_c$ at $T<T_c$ in the linear term. Then one concludes that the
phase transition is almost the conventional BCS-like transition, at
least at $E_0\gg T_c$ \cite{lee,ran,lar}. These findings are
mathematically and physically erroneous. Indeed, the term of the sum
in Eq.(135) with $\Omega_n=0$ is given by the integral
\begin{equation}
T\int {d{\bf q}\over{2\pi^3}}{1\over{E_0+\Sigma_b({\bf q},0)}}.
\end{equation}
 The integral converges, if and
only if $ E_0 \geq -\Sigma_b(0,0)$. In fact,
\begin{equation}
E_0+\Sigma_b(0,0)=0
\end{equation}
 is strictly zero in the
Bose-condensed state, because $\mu_b=-[E_0+\Sigma_b(0,0)]$
corresponds to the boson chemical potential relative to the lower
edge of the boson energy spectrum. More generally, $\mu_b=0$
corresponds to the appearance of the Bogoliubov-Goldstone mode due
to a broken symmetry  below $T_c$. This exact result makes the BSC
equation (132) simply an identity  with  $\alpha(T) \equiv 0$ at any
temperature below $T_c$. On the other hand, MFA
 violates the density
sum-rule, predicting the wrong negative $\alpha(T)$ below $T_c$.
Since $\alpha(T)=0$, one may expect that the conventional upper
 critical field, $H_{c2}(T)$ is zero in BFM. To determine
 $H_{c2}(T)$ and explore the condensed phase of 3D BFM, one can
  apply the Gor'kov formalism \cite{gor2}, as described below.

\subsection{Normal and anomalous Green's functions of 3D BFM: pairing
of bosons} Let us now explore a simplified version of 3D BFM  in an
external magnetic field ${\bf B= \nabla \times A}$ neglecting the
hard-core effects \cite{alepair},
\begin{eqnarray}
H&=&\int d{\bf r}\sum_s \psi _{s}^{\dagger }({\bf r})\hat{h}({\bf
r})\psi _{s}({\bf r}) +g[\phi({\bf r})\psi _{\uparrow }^{\dagger
}({\bf r})\psi _{\downarrow }^{\dagger }({\bf
r})+H.c.]\nonumber\\
&+&E_0\phi^{\dagger}({\bf r})\phi({\bf r}),
\end{eqnarray}
where $\psi_{s}({\bf r})$ and $\phi({\bf r})$ are fermionic and
bosonic fields, $s=\uparrow, \downarrow$ is the spin, $\hat{h}({\bf
r)=}-[\nabla +ie{\bf A(r)]}^{2}/(2m)-\mu$ is the fermion kinetic
energy operator. Here  the volume of the system is taken as $V=1$.

The Matsubara field operators, $Q=\exp (H\tau )Q({\bf r)}\exp
(-H\tau ), \bar{Q}=\exp(H\tau) Q^{\dagger}({\bf r)}\exp (-H\tau )$
($Q\equiv\psi_s, \phi$) evolve with the imaginary time $-1/T\leq
\tau \leq 1/T$ as

\begin{eqnarray}
-\frac{\partial \psi _{\uparrow }({\bf r},\tau )}{\partial \tau } &=&\hat{h}(%
{\bf r)}\psi _{\uparrow }({\bf r},\tau )+g \phi({\bf r},
\tau)\bar{\psi} _{\downarrow }({\bf r,}\tau ), \\
\frac{\partial \bar{\psi} _{\downarrow }({\bf r,}\tau )}{\partial
\tau } &=&\hat{h}^{\ast }({\bf r)}\bar{\psi} _{\downarrow }({\bf
r,}\tau )-g \bar{\phi}({\bf r},\tau)\psi _{\uparrow }({\bf r,}\tau
), \\ -\frac{\partial \phi ({\bf r,}\tau )}{\partial \tau} &=& E_0
\phi({\bf r,}\tau )+g\psi _{\downarrow }({\bf r,}\tau )\psi
_{\uparrow }({\bf r,}\tau ).
\end{eqnarray}
The  theory of the condensed state can be formulated with the
 normal and anomalous fermion GFs \cite{gor2}, $ {\cal
G}({\bf r,r}^{\prime },\tau)=- \langle T_{\tau
}\psi _{s}({\bf r},\tau )\bar{\psi} _{s}({\bf r}^{\prime }{\bf ,}%
0)\rangle$, ${\cal F}^{+}({\bf r,r}^{\prime },\tau)= \langle T_{\tau
}\bar{\psi} _{\downarrow }({\bf r,}\tau) \bar{\psi} _{\uparrow
}({\bf r}^{\prime
},0)\rangle$, respectively, where the operation $%
T_{\tau }$ performs the time ordering. Fermionic and bosonic fields
condense simultaneously \cite{ale}. Following Bogoliubov \cite{bog}
the bosonic  condensate is described by separating a large matrix
element $\phi_{0} ({\bf r})$ in $\phi({\bf r},\tau) $ as a number,
while the remaining part $\tilde{\phi}({\bf r},\tau)$ describes a
supracondensate field, $ \phi ({\bf r},\tau)=\phi _{0}({\bf
r})+\tilde{\phi}({\bf r},\tau)$. Then using Eq.(146) one obtains
\begin{equation}
g\phi_0({\bf r})= \Delta({\bf r})\equiv-{g^2\over{E_0}}{\cal F}({\bf
r,r},0+),
\end{equation}
where ${\cal F}({\bf r,r}^{\prime },\tau)= \langle T_{\tau }\psi
_{\downarrow }({\bf r,}\tau) \psi _{\uparrow } ({\bf r}^{\prime
},0)\rangle$. The equations for  GFs are obtained by using Eqs.
(144-146) and the  diagrammatic technique \cite{abr} in the
framework of the non-crossing approximation \cite{ref}, as shown in
Fig.23 and Fig.24.

An important novel feature of BFM is a pairing of supracondensate
bosons, caused by their hybridization with the fermionic condensate,
as follows from the last diagram in Fig.24. Hence, one has to
introduce an \emph{anomalous} supracondensate boson GF, ${\cal
B}^{+}({\bf r,r}^{\prime },\tau)=\langle T_{\tau
}\bar{\tilde{\phi}}({\bf r,}\tau) \bar{\tilde {\phi}}({\bf
r}^{\prime },0)\rangle$ along with the normal boson GF, ${\cal
D}({\bf r,r}^{\prime },\tau)=- \langle T_{\tau
}\tilde{\phi} ({\bf r},\tau )\bar{\tilde{\phi}}({\bf r}^{\prime }{\bf ,}%
0)\rangle$.

\begin{figure}
\begin{center}

\includegraphics[angle=-90,width=0.75\textwidth]{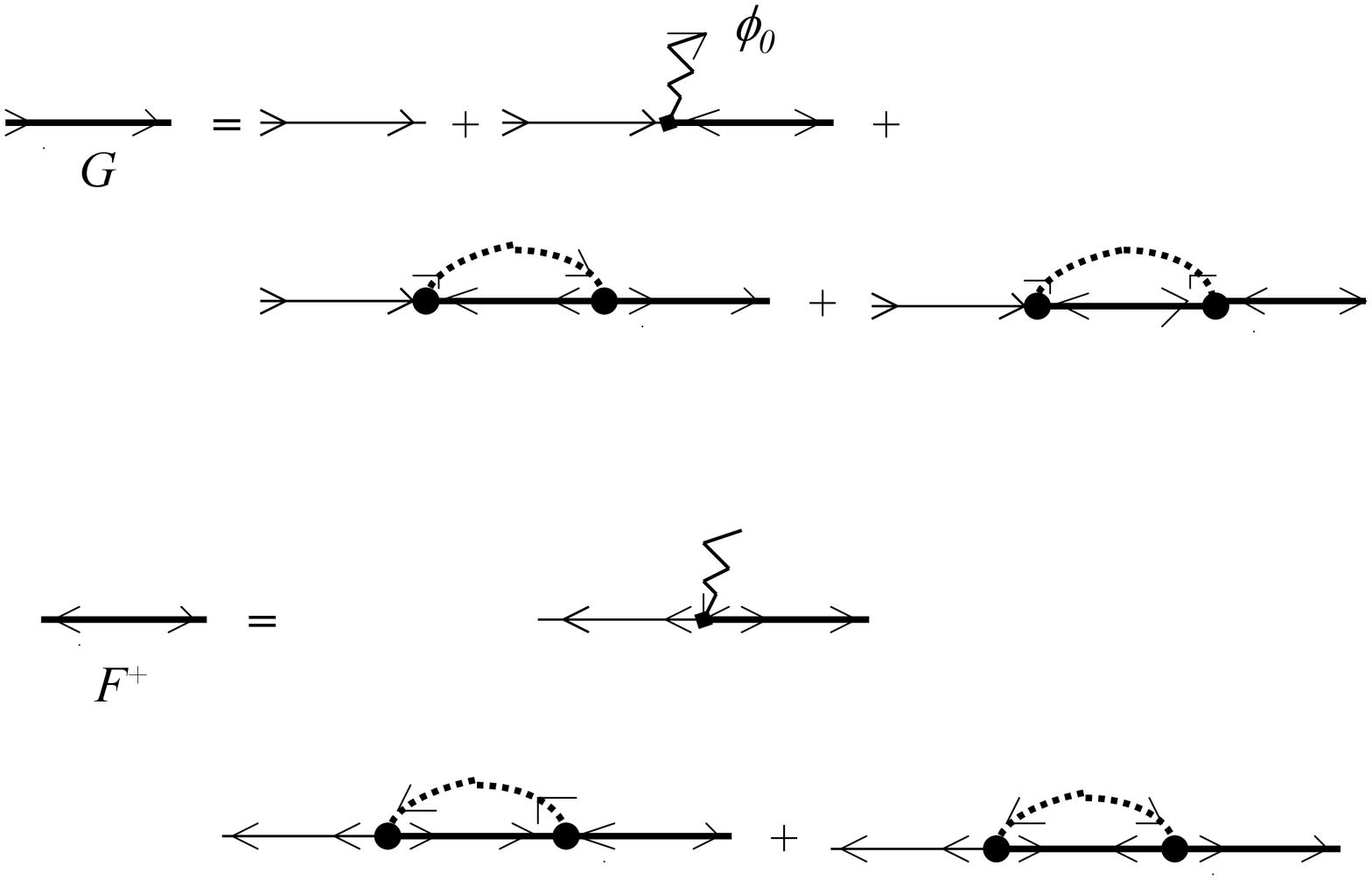}
\caption{Diagrams for the normal and anomalous fermion GFs. Zig-zag
arrows represent the single-particle Bose condensate $\phi_0$,
dotted lines are the boson GFs, solid lines are the
 fermion GFs. Vertex (dot) corresponds to the hybridization
 interaction.}
\end{center}
\end{figure}
The diagrams, Fig.23 and Fig.24, are transformed into analytical
equations for the time Fourier-components of the fermion GFs with
the Matsubara frequencies $\omega=\pi T (2n+1)$ ($n=0, \pm 1,\pm
2,...$) as
\begin{eqnarray}
&&[i\omega -\hat{h}({\bf r)}]{\cal G}_{\omega}({\bf r,r}^{\prime })
 = \delta ({\bf r-r}^{\prime })-\Delta ({\bf r)}{\cal
F}_{\omega}^{+}({\bf r,r}^{\prime } ) \cr
 &-& g^2 T
\sum_{\omega^{\prime}} \int d{\bf x}{\cal G}_{-\omega^{\prime}}({\bf
x,r}){\cal D}_{\omega-\omega^{\prime}}({\bf r,x}){\cal
G}_{\omega}({\bf x,r}^{\prime})
 \cr &-& g^2T
 \sum_{\omega^{\prime}} \int d{\bf
x}{\cal F}^{+}_{\omega'}({\bf r,x}){\cal B}_{\omega+\omega^{\prime}}
({\bf r,x}){\cal F}^{+}_{\omega}({\bf x,r^{\prime}}),
\end{eqnarray}

\begin{eqnarray}
&&[-i\omega -\hat{h}^{\ast}({\bf r)}]{\cal F}^{+}_{\omega}({\bf
r,r}^{\prime }) = \Delta^{\ast} ({\bf r)}{\cal G}_{\omega}({\bf
r,r}^{\prime } ) \cr
 &-& g^2 T
\sum_{\omega^{\prime}} \int d{\bf x}{\cal G}_{\omega^{\prime}}({\bf
r,x}){\cal D}_{\omega^{\prime}-\omega}({\bf x,r}){\cal
F}^{+}_{\omega}({\bf x,r}^{\prime}) \cr &+& g^2T
 \sum_{\omega^{\prime}} \int d{\bf
x}{\cal F}_{-\omega'}({\bf r,x}){\cal
B}^{+}_{-\omega-\omega^{\prime}} ({\bf r,x}){\cal G}_{\omega}({\bf
x,r^{\prime}}), \nonumber
\end{eqnarray}
 and
\begin{eqnarray}
&&(i\Omega -E_0) {\cal D}_{\Omega}({\bf r,r}^{\prime }) = \delta
({\bf r-r}^{\prime })\cr &-& g^2 T \sum_{\omega^{\prime}} \int d{\bf
x}{\cal G}_{\omega^{\prime}}({\bf r,x}){\cal
G}_{\Omega-\omega^{\prime}}({\bf r,x}){\cal D}_{\Omega}({\bf
x,r}^{\prime}) \cr &-& g^2T
 \sum_{\omega^{\prime}} \int d{\bf
x}{\cal F}_{\omega'}({\bf r,x}){\cal F}_{\Omega-\omega^{\prime}}
({\bf r,x}){\cal B}^{+}_{\Omega}({\bf x,r}^{\prime}),
\end{eqnarray}
\begin{eqnarray}
&&(-i\Omega -E_0){\cal B}^{+}_{\Omega}({\bf r,r}^{\prime }) = \cr &&
g^2T
 \sum_{\omega^{\prime}} \int d{\bf
x}{\cal F}_{-\omega'}^{+}({\bf r,x}){\cal
F}_{-\Omega+\omega^{\prime}}^{+}({\bf r,x}){\cal D}_{\Omega}({\bf
x,r^{\prime}}) \cr &-& g^2 T \sum_{\omega^{\prime}} \int d{\bf
x}{\cal G}_{-\omega^{\prime}}({\bf x,r}){\cal
G}_{\omega^{\prime}-\Omega}({\bf x,r}){\cal B}^{+}_{\Omega}({\bf
x,r}^{\prime}). \nonumber
\end{eqnarray}
for the boson GFs  with  ${\cal B}({\bf r,r}^{\prime },\tau)=\langle
T_{\tau }\tilde{\phi}({\bf r,}\tau) \tilde {\phi}({\bf r}^{\prime
},0)\rangle$.

\begin{figure}
\begin{center}

\includegraphics[angle=-90,width=0.75\textwidth]{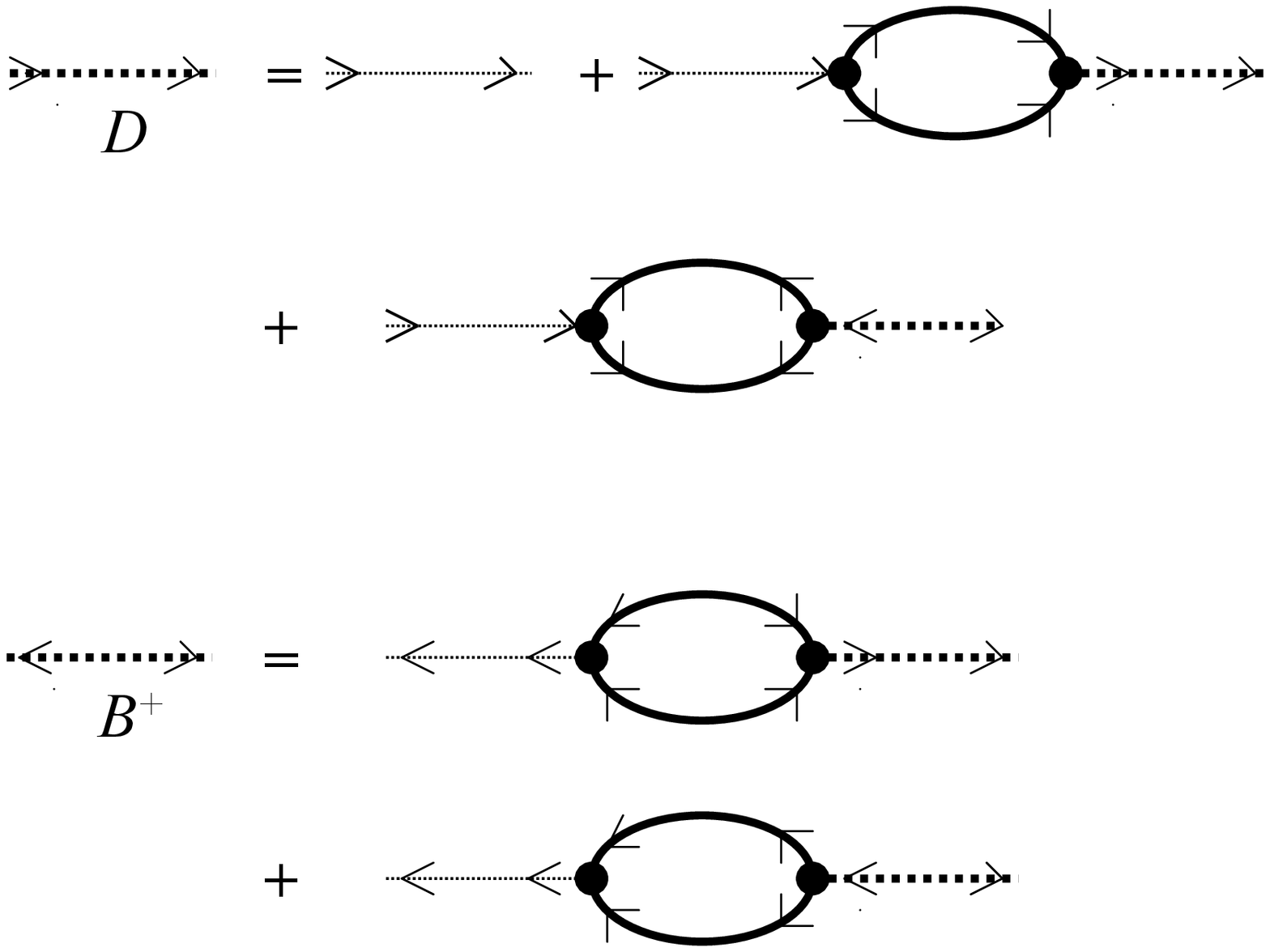}
 \caption{Diagrams for the supracondensate boson
 GFs. The Cooper-pairing of fermions  leads to the Cooper-pair-like boson condensate,
 described by the boson anomalous GF, ${\cal
B}^{+}$.}
\end{center}
\end{figure}

\subsection{Gor'kov expansion}
These equations can be formally solved in the homogeneous case
without the external field, ${\bf A}=0$. Transforming into the
momentum space yields GFs' time-space Fourier components as
\begin{eqnarray}
{\cal G}({\bf k},\omega) &=-&{\frac{i\tilde{\omega}^{\ast} +\xi
_{\bf k}}{ |i\tilde{\omega} -\xi _{\bf
k}|^2+|\tilde{\Delta}({\bf k}, \omega)|^{2}}},\\
{\cal F}^{+}({\bf k}, \omega) &=&{\frac{\tilde{\Delta}^{\ast}({\bf
k}, \omega)}{ |i\tilde{\omega} -\xi _{\bf k}|^2+|\tilde{\Delta}({\bf
k}, \omega)|^{2}}} ,
\end{eqnarray}
and
\begin{eqnarray}
{\cal D}({\bf q},\omega) &=&-{\frac{i\tilde{\Omega}^{\ast} +E_0}
{|i\tilde{\Omega} -E_0|^2+|\Gamma({\bf q}, \Omega)|^{2}}},
 \\
{\cal B}^{+}({\bf q}, \omega) &=&{\frac{\Gamma^{\ast}({\bf q},
\Omega) }{ |i\tilde{\Omega} -E_0|^2+|\Gamma({\bf q}, \Omega)|^{2}}},
\end{eqnarray}
where $\tilde{\omega}\equiv\omega+i\Sigma_f({\bf k}, \omega)$, $
\tilde{\Omega}\equiv\Omega+i\Sigma_b({\bf q}, \Omega)$, and $\xi
_{\bf k}=k^{2}/(2m)-\mu$.  The fermionic order parameter,
renormalized with respect to the mean-field $\Delta$ due to the
formation of the boson-pair condensate, is given by
\begin{equation}
\tilde{\Delta}({\bf k}, \omega)=\Delta + g^2T
 \sum_{\omega^{\prime}}\int {d {\bf q}\over{2\pi^3}} {\cal F}^{+}({\bf k}-{\bf q},\omega'
){\cal B}({\bf q},\omega+\omega^{\prime}),
\end{equation}
and the boson-pair order parameter, generated by the hybridization
with the fermion Cooper pairs, is
\begin{equation}
 \Gamma({\bf
q}, \Omega)= g^2T \sum_{\omega^{\prime}}\int {d {\bf
k}\over{2\pi^3}} {\cal F}({\bf k},\omega'){\cal F}({\bf q}-{\bf
k},\Omega-\omega^{\prime}).
\end{equation}
  Hence,
there are three coupled condensates in the model described by the
 off-diagonal fields $g\phi_0$, $\tilde{\Delta}$, and
$\Gamma$, rather than two, as in MFA. At low temperatures all of
them have about the same magnitude, as the fermion and boson
self-energies,
\begin{equation}
\Sigma_f({\bf k}, \omega)=-g^2T
 \sum_{\omega^{\prime}}\int {d {\bf q}\over{2\pi^3}} {\cal G}({\bf q}-{\bf k},-\omega'){\cal
D}({\bf q},\omega-\omega^{\prime}),
\end{equation}
\begin{equation}
\Sigma_b({\bf q}, \Omega)=-g^2T
 \sum_{\omega^{\prime}}\int {d {\bf q}\over{2\pi^3}} {\cal G}({\bf k},\omega'){\cal
G}({\bf q}-{\bf k},\Omega-\omega^{\prime}),
\end{equation}
 respectively.

On the other hand, when the temperature is   close to $T_c$ (i.e.
$T_c-T \ll
 T_c$),
   the boson pair condensate  is weak compared with the single-particle boson and the Cooper pair condensates.
  In this temperature range
$\Gamma$, Eq.(155) is of the second order in $\Delta$, $\Gamma
\propto \Delta^{2}$, so that the anomalous boson GF can be
neglected, since $\Delta$ is small. The fermion self-energy,
Eq.(156) is a regular function of $\omega$ and ${\bf k}$, so  it can
be absorbed in the renormalized fermion band dispersion. Then the
fermion normal and anomalous GFs, Eqs.(150,151) look like the
familiar GFs of the BCS theory, and one can apply the Gor'kov
expansion \cite{gor2}  in powers of $\Delta ({\bf r)}$ to describe
the condensed phase of BFM in the magnetic field near the
transition. Using Eq.(147)  one obtains
 to the terms linear in $\Delta $
\begin{equation}
\Delta^{\ast} ({\bf r)} ={g^2\over{E_0}}T\sum_{\omega _{n}}\int d{\bf x}{\cal G}%
_{-\omega _{n}}^{(n)}({\bf x,r})\Delta ^{\ast }({\bf x)}{\cal
G}_{\omega _{n}}^{(n)}({\bf x,r}).
\end{equation}
The spatial variations of the vector potential  are small near the
transition.  If ${\bf A}({\bf r})$ varies slowly, the normal state
GF, ${\cal G}_{\omega }^{(n)}({\bf r,r}^{\prime })$ differs from the
zero-field normal state GF, ${\cal G}_{\omega
}^{(0)}(\bf{r-r^{\prime}} )$ only by a phase \cite{gor2} ${\cal
G}_{\omega }^{(n)}({\bf r,r}^{\prime })=\exp [-ie{\bf A(r)}\cdot
(\bf{r-r}^{\prime })]{\cal G}_{\omega }^{(0)}(\bf{r-r}^{\prime } )$.
Expanding all quantities  near the point ${\bf x=r}$ in Eq.(158) up
to the second order in ${\bf x-r} $ inclusive, one obtains the
linearized  equation for the fermionic order parameter as
\begin{equation}
 \gamma [\nabla -2ie{\bf
A(r)]}^{2}\Delta({\bf r})=\alpha \Delta({\bf r}),
\end{equation}
where
\begin{equation}
\alpha = 1+{\Sigma_b(0,0)\over{E_0}}\approx 1-
{g^2N(0)\over{E_0}}\ln {\mu\over{T}},
\end{equation}
and $\gamma \approx 7\zeta (3)v_{F}^{2}g^2N(0)/(48\pi^2 T^{2}E_0)$.

 The coefficient $\alpha(T)$ disappears in Eq.(159), since $E_{0}=-\Sigma _{b}(0,0)$ at and \emph{below} $T_c$.
It means that the  phase
 transition is never a BCS-like second-order phase transition
 even at large $E_0$ and small $g$. In fact, the
 transition  is driven by the Bose-Einstein condensation of \emph{
 real} bosons with ${\bf q}=0$, which occur  due to the complete
 softening of their spectrum at  $T_c$.
 Remarkably, the conventional upper critical field, determined as the field, where a non-trivial
 solution of the \emph{linearized} Gor'kov equation (159)  occurs, is
 zero in BFM, $H_{c2}(T)=0$. It is not  a finite $H_{c2}(T)$
 found in Ref. \cite{dam} using MFA. The qualitative failure of MFA  might be rather unexpected, if one
believes that bosons in Eq.(131) play the same role as phonons in
the BCS superconductor. This is not the case for two reasons. The
first one is  the density sum-rule, Eq.(135), for bosons which is
not applied to phonons. The second being that the boson self-energy
is given by the divergent (at $T=0$) Cooperon diagram, while the
self-energy of phonons is finite at small coupling.

 Even
 at  temperatures well below $T_c$ the condensed state is fundamentally
 different from the MFA ground state, because of the pairing of
 bosons. The latter is similar to the Cooper-like pairing of
 supracondensate $^{4}He$ atoms \cite{pas}, proposed  as an explanation of the small density of the single-particle
 Bose condensate in superfluid Helium-4.   The pair-boson condensate
 should significantly
modify the thermodynamic
  properties of the condensed BFM
  compared with the MFA predictions.
  Hence the common wisdom
 that at weak coupling  the boson-fermion model is adequately described by the BCS
 theory is  negated by our analysis beyond MFA. There is no BCS-BEC
 crossover in BFM, and the phase transition is of the BEC-type at any parameters.

\section{Conclusion}

Here I have argued that  attractive electron correlations,
prerequisite to HTS, are caused by an almost unretarded
electron-phonon interaction sufficient to overcome the direct
Coulomb repulsion in cuprates. Low energy physics of
high-temperature superconductors is that of lattice polarons and of
real-space hole pairs dressed by phonons, i.e. mobile lattice
bipolarons. Single polarons are thermal excitations in underdoped
cuprates, but they could be degenerate in the boson-fermion mixture
in overdoped cuprates. The superconducting state is the
Bose-Einstein 3D condensate of bipolarons at any doping including
the overdoped domain, irrespective to a possible hybridization of
single polarons and bipolarons. Our multi-polaron approach to the
HTS problem accounts for many normal state properties of
superconducting cuprates including the temperature-dependent spin
susceptibility, nonlinear in-plane and thermally activated
out-of-plane resistivities, the temperature-dependent  Hall effect,
the  Nernst signal and
 diamagnetism  above $T_c$, spin and charge pseudogaps. It provides a parameter-free
fit of experimental $T_c$ and describes isotope effects, specific
heat anomalies, resistive upper critical fields, and the symmetry
and space modulations of the order parameter.

  I thank A.M. Bratkovsky, C.J. Dent, P.P. Edwards, J.P. Hague, V.V. Kabanov, P.E.
  Kornilovitch, W.Y. Liang,
J.H. Samson, P.E. Spencer, and V.N. Zavaritsky for a long-standing
collaboration and valuable discussions. I highly appreciate
enlightening   discussions with A.F. Andreev, I. Bozovic, L.P.
Gor'kov, J.E. Hirsch, A.P. Levanyuk, R. Micnas, D. Mihailovic, and
S. Robaszkiewicz. The work was supported by EPSRC (UK) (grant
EP/C518365/1), the Leverhulme Trust (UK),  the Royal Society, and by
NATO.

\newpage


\begin{thebibliography}{200}
\bibitem{AKM}  Here the polaron DOS is  energy independent different from Eq.(8) in Ref.\cite{alekabmot}
with a one-dimensional DOS.
\bibitem{ref} The crossing diagrams are small  as $g/\mu
\ll 1$ \cite{ale}.
\end{thebibliography}
\end{document}